\documentclass[11pt,a4paper]{article}
\usepackage{jcappub}
\usepackage[font=footnotesize,up]{caption}
\usepackage{wrapfig}
\usepackage{multirow}
\usepackage{booktabs}
\usepackage{array}
\usepackage{tabulary}
\usepackage{rotating}
\usepackage{xcolor}
\usepackage{multicol}
\usepackage{colortbl}
\usepackage{hhline}
\newcommand{\st}{\textsl}
\newcommand{\bt}{\textbf}
\newcommand{\tit}{\textit}
\newcommand{\rt}{\mathrm}
\newcommand{\bs}{\boldsymbol}

\newcommand{\sq}{\scalebox{0.75}}
\newcommand{\su}{\scalebox{1.10}}
\renewcommand{\deg}{\displaystyle^{\circ}}

\newcommand{\m}{m_{\chi}}
\newcommand{\td}{\tau_{\chi}}
\newcommand{\sv}{\langle\sigma_{\rt{ann}}v\rangle}

\newcommand{\svu}{~\rt{cm}^{\sq{3}}~\rt{s}^{-{\sq{1}}}}

\newcommand{\Aeff}{A_{\rt{eff}}}
\renewcommand{\th}{\theta^{2}}
\newcommand{\teff}{{\su{\tit{t}}}_{\sq{eff}}}
\newcommand{\UL}{^{\scalebox{0.7}{UL}}}
\newcommand{\Non}{N_{\scalebox{0.7}{ON}}}
\newcommand{\Noff}{N_{\scalebox{0.7}{OFF}}}
\newcommand{\Nobs}{N_{\scalebox{0.7}{OBS}}}
\newcommand{\Nest}{N_{\scalebox{0.7}{EST}}}

\newcommand{\sNobs}{N_{\scalebox{0.5}{OBS}}}
\newcommand{\sNest}{N_{\scalebox{0.5}{EST}}}
\newcommand{\Pon}{P_{\scalebox{0.7}{ON}}}
\newcommand{\Poff}{P_{\scalebox{0.7}{OFF}}}
\newcommand{\Eth}{E_{\sq{th}}}
\renewcommand{\AA}{\st{A\&A }}

\newcommand{\AP}{\st{Astropart. Phys. }}                        
\newcommand{\APJ}{\st{Astrophys. J. }}

\newcommand{\MNRAS}{\st{Mon. Not. Roy. Astron. Soc. }}

\newcommand{\NPB}{\st{Nucl. Phys.} \bt{B}}
\newcommand{\PLB}{\st{Phys. Lett.} \bt{B}}
\newcommand{\PR}{\st{Phys. Rept. }}
\newcommand{\PRD}{\st{Phys. Rev.} \bt{D}}
\newcommand{\PRL}{\st{Phys. Rev. Lett. }}
\title{Optimized dark matter searches in deep observations of Segue~1
  with MAGIC}

\author [1, \star] {J.~Aleksi\'c}
\author [2] {S.~Ansoldi}
\author [3] {L.~A.~Antonelli}
\author [4] {P.~Antoranz}
\author [5] {A.~Babic}
\author [6] {P.~Bangale}
\author [6] {U.~Barres de Almeida}
\author [7] {J.~A.~Barrio}
\author [8,25] {J.~Becerra Gonz\'alez}
\author [9] {W.~Bednarek}
% ...
\author{(the MAGIC Collaboration)}
\author [33]{and A. Ibarra}

\affiliation [1] {IFAE, Campus UAB, E-08193 Bellaterra, Spain}
\affiliation [2] {Universit\`a di Udine, and INFN Trieste, I-33100 Udine, Italy}
\affiliation [3] {INAF National Institute for Astrophysics, I-00136 Rome, Italy}
\affiliation [4] {Universit\`a  di Siena, and INFN Pisa, I-53100 Siena, Italy}
\affiliation [5] {Croatian MAGIC Consortium, Rudjer Boskovic Institute, University of Rijeka and University of Split, HR-10000 Zagreb, Croatia}
\affiliation [6] {Max-Planck-Institut f\"ur Physik, D-80805 M\"unchen, Germany}
\affiliation [7] {Universidad Complutense, E-28040 Madrid, Spain}
\affiliation [8] {Inst. de Astrof\'isica de Canarias, E-38200 La Laguna, Tenerife, Spain}
\affiliation [9] {University of \L\'od\'z, PL-90236 Lodz, Poland}
% ...
\affiliation [25] {now at: NASA Goddard Space Flight Center, Greenbelt, MD 20771, USA and Department of Physics and Department of Astronomy, University of Maryland, College Park, MD 20742, USA}
% ...
\affiliation [33] {Physik-Department T30d, Technische Universit\"at
  M\"unchen, James-Franck-Stra\ss e, 85748 Garching, Germany}

\affiliation[\star]{corresponding author}
\emailAdd{jelena@ifae.es, jrico@ifae, saverio.lombardi@pd.infn.it}
\abstract{We present the results of stereoscopic observations of the
  satellite galaxy Segue~1 with the MAGIC Telescopes, carried out
  between 2011 and 2013. With almost 160 hours of good-quality data,
  this is the deepest observational campaign on any dwarf galaxy
  performed so far in the very high energy range of the
  electromagnetic spectrum. We search this large data sample for
  signals of dark matter particles in the mass range between 100 GeV
  and 20 TeV. For this we use the full likelihood analysis method,
  which provides optimal sensitivity to characteristic gamma-ray
  spectral features, like those expected from dark matter annihilation
  or decay. In particular, we focus our search on gamma-rays produced
  from different final state Standard Model particles, annihilation
  with internal bremsstrahlung, monochromatic lines and box-shaped
  signals. Our results represent the most stringent constraints to the
  annihilation cross-section or decay lifetime obtained from
  observations of satellite galaxies, for masses above few hundred
  GeV. In particular, our strongest limit (95\% confidence level)
  corresponds to a $\sim$500 GeV dark matter particle annihilating
  into $\tau^{+}\tau^{-}$, and is of order $\sv \simeq$
  1.2$\times$10$^{-\sq{24}} \svu$ --- a factor $\sim$40 above the
  $\sv$ thermal value.}

\begin{document}
\maketitle
\flushbottom

\section{Introduction}
\label{I}

Dark matter (DM) is an, as yet, unidentified type of matter, which
accounts for about 85\% of the total mass content and 26\% of the
total energy density of the Universe \cite{I_Planck}. Despite the
abundant evidence on all astrophysical length scales implying its
existence, the nature of DM is still to be determined. Observations
require the DM particles to be electrically neutral, non-baryonic and
stable on cosmological time scales. Furthermore, in order to allow the
small-scale structures to form, these particles must be ``cold''
(i.e. non-relativistic) at the onset of structure formation. However,
a particle fulfilling all those requirements does not exist within the
Standard Model (SM); thus, the existence of DM inequivocally points to
new physics. A particularly well motivated class of cold DM candidates
are the weakly interacting massive particles (WIMPs,
\cite{I_WIMP}). WIMPs are expected to have a mass in the range between
$\sim$10 GeV and a few TeV, interaction cross-sections typical of the
weak scale and they naturally provide the required relic density
(``WIMP miracle'', see, e.g., \cite{I_DMreview}). Several extensions
of the SM include WIMP candidates, most notably Supersymmetry (see,
e.g., \cite{I_SUSY}), as well as theories with extra dimensions
\cite{I_UED}, minimal SM extensions \cite{I_MDM}, and others (for a
review see, e.g., \cite{I_DMreview}).

The search for WIMPs is conducted on three complementing fronts,
namely: production in particle accelerators, direct detection in
underground experiments and indirect detection. The latter implies the
searches, by ground- and space-based observatories, of the SM
particles presumably produced in WIMP annihilations or
decays. Accelerator and direct detection experiments are most
sensitive to DM particles with mass below a few hundred GeV. Positive
results for signals of $\sim$10 GeV DM reported by some
direct-searches experiments \cite{I_DAMA, I_CoGeNT, I_CDMS,
  I_CDMSlite} could not be confirmed by other detectors, and are in
tension with results obtained by XENON100 \cite{I_XENON100si,
  I_XENON100sd} and LUX \cite{I_LUX}. On the other hand, the rather
heavy Higgs boson \cite{I_HiggsBoson} and the lack of indications of
new physics at the Large Hadron Collider, strongly constrain the
existence of a WIMP at the electroweak scale. Therefore, the current
status of these experimental searches strengthens the motivation for
DM particles with masses at the TeV scale or above --- the mass range
best (and sometimes exclusively) covered by the Imaging Air Cherenkov
Telescopes (IACTs). For this reason, IACT observations in the very
high energy domain (E $\gtrsim100$ GeV) provide extremely valuable
clues to unravel the nature of the DM. Such searches are the primary
scope of this work.

Among the best-favored targets for indirect DM detection with
gamma-ray observatories are dwarf spheroidal galaxies (dSphs). The
dSph satellites of the Milky Way are relatively close-by (less than a few
hundred kpc away), and in general less affected by contamination from
gamma rays of astrophysical origin than some other DM targets, like
the Galactic Center (GC) and galaxy clusters (see, e.g,
\cite{I_GC_Astro, I_Cluster_Astro}). Furthermore, given the low
baryonic content and large amounts of DM expected in these kind of
galaxies, dSphs are considered highly promising targets for indirect
DM searches. Over the last decade, a number of dSphs have been
observed by the present generation of IACTs: MAGIC
\cite{I_MAGIC_Draco, I_MAGIC_Willman, I_MAGIC_Segue},
H.E.S.S. \cite{I_HESS_CanisM, I_HESS_Sagittarius,
  I_HESS_SculptorCarina} and VERITAS \cite{I_VERITAS,
  I_VERITAS_Segue}, as well as by the Large Area Telescope (LAT) on
board the \tit{Fermi} satellite \cite{I_Fermi_dSph}. 

In this work we present the results of deep observations of the dSph
galaxy Segue~1 with MAGIC. Discovered in 2006 in the imaging data from
the Sloan Digital Sky Survey \cite{I_Segue_Discovery}, Segue~1 is
classified as an ultra-faint dSph, of absolute magnitude $M_{\rm{V}} =
-$1.5$^{+{\rm{0.6}}}_{-\rm{0.8}}$. With a mass-to-light ratio estimated
to be $\sim$3400 $M_{\odot}/L_{\odot}$ \cite{I_Segue_Properties}, this
is the most DM-dominated object known so far. Furthermore, given its
relative closeness (23$\pm$2 kpc), lack of backgrounds of conventional
origin, high expected DM flux and its favorable position in the
Northern hemisphere and outside of the Galactic plane (RA, DEC =
10.12$^{\sq{h}}$, 16.08$\deg$), Segue~1 has been selected as an
excellent target for DM searches with MAGIC.

We present the results of a three-year long (2011--2013) observational
campaign on Segue~1 carried out with MAGIC in stereoscopic mode. With
almost 160 hours of good-quality data, this is the deepest exposure of
any dSph by any IACT so far. No significant gamma-ray signal is
found. The gathered data are used to set constraints on various models
of DM annihilation and decay, providing the most sensitive indirect DM
search on dSphs for the WIMP mass range between few hundred GeV and 10
TeV. In particular, we improve our previous limits, obtained from
$\sim$30 hours of Segue~1 observations in the single telescope mode
\cite{I_MAGIC_Segue}, by one order of magnitude. This improvement is
achieved through the increased sensitivity of the MAGIC stereo system,
the deep exposure, and the use of the \tit{full likelihood analysis}
\cite{I_FL} --- a method optimized for searches of characteristic,
DM-induced gamma-ray spectral features.

This paper is structured as follows: first, we describe the
observations of Segue~1 with MAGIC, data reduction and standard
analysis procedures (section \ref{O}). Then, in section \ref{FL}, we
describe the basics of the full likelihood method, used for the
analysis of our data, and the particular choices we have made
regarding the likelihood construction. In section \ref{DM} we give
details on the expected photon flux from DM annihilation and decay,
with accent on the spectral shapes of the considered models and the
calculations of the astrophysical term of the flux. Section \ref{R}
presents our results --- the upper limits on annihilation cross section
and lower limits on the DM particle lifetime for the studied channels,
and in section \ref{D} those are put into context and compared with
constraints from other gamma-ray experiments. Lastly, section \ref{C}
summarizes this paper and our conclusions.
\section{Observations and conventional data analysis}
\label{O}

\begin{table}[t]
  \begin{center}
    \setlength{\extrarowheight}{1pt}
    {\small
      \begin{tabulary}{0.99\textwidth}{ L C C C C }
        \toprule[0.1em]
        & \bt{Sample A} & \bt{Sample B1} & \bt{Sample B2} & \bt{Sample C} \\\midrule[0.1em]
        Readout & DRS2 & DRS4 & DRS4 & DRS4\\
        MAGIC-I camera & old & old & old & new \\\midrule
        Obs. period & Jan--May 2011 & Jan--Feb 2012 & Mar--May 2012 & Nov 2012--Feb 2013\\
        Obs. time [h] & 64 & 24.28 & 59.77 & 55.05 \\
        \tit{Zd} range [deg] & 13--33.7 & 13--32.5 & 13--35.7 & 13--37 \\
        \tit{Az} range [deg] & 104.8--250.2 & 120.2--252.0 & 115.4--257.2 & 103.8--259.4\\\midrule
        Wobble around & dummy & dummy & dummy & Segue~1\\
        Wobble offset [deg] & 0.29 & 0.29 & 0.29 & 0.40 \\\midrule
        W1 $\teff$ [h] & 22.66 & 6.07 & 25.02 & 23.71 \\
        W2 $\teff$ [h] & 24.35 & 6.20 & 26.11 & 23.80 \\ 
        ${\bs{\teff}}$ [h] & 47.00 & 12.26 & 51.13 & 47.51 \\           
        \bottomrule[0.1em]
        \vspace{0.05pt}\bt{Total} ${\bs{\teff}}$ [h] & & & & \vspace{0.05pt}\bt{157.9}\\\bottomrule[0.1em]
      \end{tabulary}
    }
    \vspace{-10pt}
  \end{center}
  \caption{Basic details of the Segue~1 observational campaign with
    MAGIC. Refer to the main text for additional explanations.}
  \label{Table1}
  \vspace{-5pt}
\end{table}
The \tit{Florian Goebel} Major Atmospheric Gamma-ray Imaging Cherenkov
(MAGIC) Telescopes are located at the Roque de los Muchachos
Observatory (28.8$\deg$ N, 17.9$\deg$ W; 2200 m a.s.l.) on the Canary
Island of La Palma, Spain. The system consists of two, 17 m diameter
telescopes with fast imaging cameras of 3.5$\deg$ field of view. The
first telescope, MAGIC-I, has been operational since 2004, and in 2009
it was joined by MAGIC-II. The trigger threshold of the system is
$\sim$50 GeV for the standard observations; the integral sensitivity
above 250 GeV, for 5$\sigma$ detection in 50 hours, is $\sim$0.7\% of
the Crab Nebula flux, with an angular resolution better than
0.07$\deg$ (39\% containment radius, \cite{O_MAGIC_performance}).

Observations of Segue~1 were performed between January 2011 and
February 2013. During this period, the MAGIC Telescopes underwent a
series of important hardware changes, aimed at the homogenization and
improvement of the performance of both instruments
\cite{O_MAGIC_performance}. First, at the end of 2011, the readout
electronics of the telescopes were upgraded to the Domino Ring Sampler
version 4 (DRS4)-based readouts, thus reducing the dead time of 0.5~ms
(introduced by the previously used DRS2-based readout electronics in
MAGIC-II) to 26~$\mu$s \cite{O_MAGIC_readout}. Second, by the end of
2012, the camera of MAGIC-I was exchanged with a replica of that of
MAGIC-II, with uniform pixel size and extended trigger area with
respect to the old camera \cite{O_MAGIC_camera}.

\begin{figure}[t!]
  \centering
  \vspace{-5pt}\hspace{5pt}
  \includegraphics[trim=5 5 20 20,clip=true,width=0.75\textwidth]{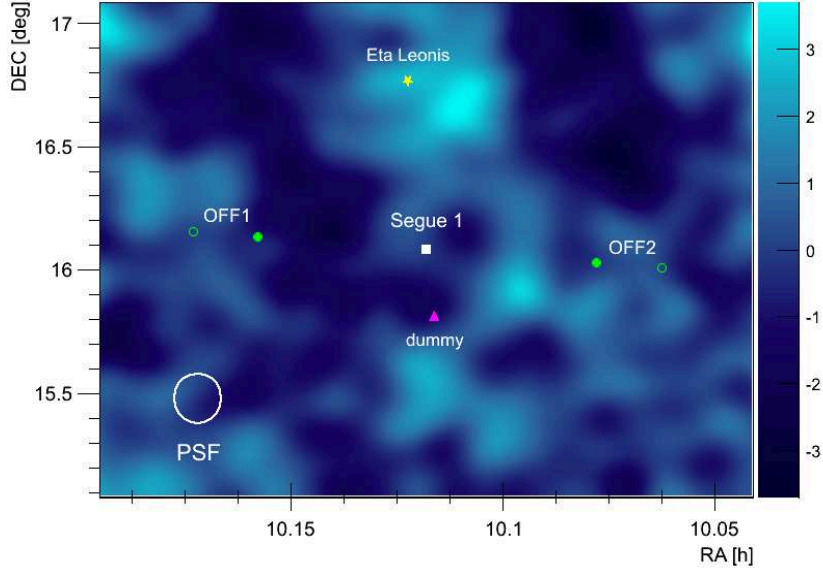}
  \vspace{0pt}
  \caption{Segue~1 significance skymap above 
    150 GeV, from 157.9 hours of observations. 
    Nominal positions of Segue~1 and $\eta$ 
    Leo are marked with the square and the star, 
    respectively. Also shown is the wobbling 
    scheme used in different samples: for 
    periods A, B1 and B2 the wobbling was 
    performed with respect to the 'dummy' 
    position (triangle), and around Segue~1 for 
    period C, while the centers of the \st{OFF} 
    regions were used for background estimation 
    at the position of the source (full circles for 
    periods A, B1 and B2, and empty circles for 
    period C). Positions of the camera center 
    are found at an equal distance from Segue~1 and the respective
    \st{OFF} positions, and are not shown here. See the main text for more details.}
  \label{Fig1}
\end{figure}
Due to the upgrade, the response of the telescopes varied throughout
the Segue~1 campaign. Consequently, we have defined four different
observational samples, such that for each of them the performance and
response of the instruments are considered stable. Sample A
corresponds to the beginning of the campaign in 2011, before the
upgrade started. Samples B1 and B2 refer to the first half of 2012,
when both telescopes were operating with new, DRS4-based readouts. B1
corresponds to the commissioning phase after the first upgrade, and is
affected by several faulty readout channels, which were fixed for
period B2. Finally, sample C was obtained at the end of 2012 and
beginning of 2013, with the final configuration of the system
including the new camera of MAGIC-I. Table \ref{Table1} summarizes the
relevant observational parameters of each of the four periods. Each
sample has been processed separately, with the use of contemporaneous
Crab Nebula observations (for optimization of the analysis cuts and
validation of the analysis procedures) and dedicated Monte Carlo
productions (for evaluation of the response of the instruments).

\begin{figure}[t]
  \centering
  \vspace{-5pt}\hspace{5pt}
  \includegraphics[trim=5 15 20 20,clip=true,width=0.75\textwidth]{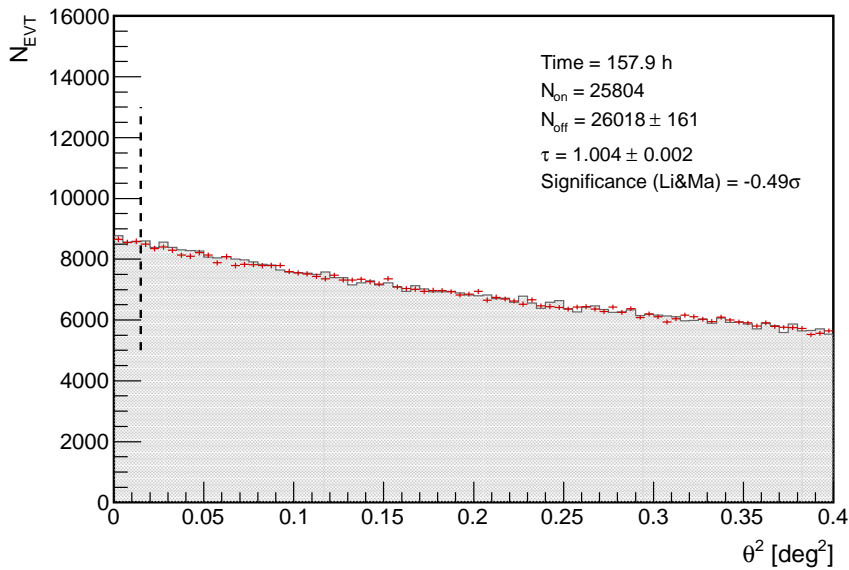}
  \vspace{0pt}
  \caption{Segue~1 $\th$-distribution above 60 
    GeV, from 157.9 hours of observations. The 
    signal (\st{ON} region) is presented by red 
    points, while the background (\st{OFF} region) 
    is the shaded gray area. The \st{OFF} sample 
    is normalized to the \st{ON} sample in the 
    region where no signal is expected, for $\th$ 
    between 0.15 and 0.4 deg$^{2}$. The vertical 
    dashed line shows the $\th_{\rt{cut}}$ value.}
  \label{Fig2}
\end{figure}

The data taking itself was performed in the so-called wobble mode
\cite{O_wobble}, using two pointing (wobble) positions. The residual
background from Segue~1 (\st{ON} region) was estimated for each of the
two pointings (W1 and W2) separately, using an \st{OFF} region placed
at the same relative location with respect to the center of the camera
as the \st{ON} region, but from the complementary wobble
observation. To ensure a homogeneous \st{ON}/\st{OFF} acceptance,
special care was taken during the data taking to observe both wobble
positions for similar amounts of time within similar azimuthal
(\tit{Az}) ranges. For periods A, B1 and B2, in order to exclude the
nearby, 3.5-apparent magnitude star $\eta$~Leo from the trigger area
of MAGIC-I old camera, the wobbling was done with respect to a 'dummy'
position located 0.27$\deg$ away from Segue~1 and on the opposite side
with respect to $\eta$~Leo (figure \ref{Fig1}). With wobbling done at
an offset of 0.29$\deg$, in a direction approximately perpendicular to
the one joining $\eta$~Leo and Segue~1, the pointing positions were
$\sim$0.4$\deg$ and $\sim$1$\deg$ away from the potential source and
the star, respectively. For period C, with the new MAGIC-I camera, the
star could no longer be excluded from its extended trigger area, so
the standard observational scheme was used instead: the wobble
positions were chosen at the opposite sides of Segue~1, at an
0.4$\deg$ offset and in a direction parallel to the one used for the
A, B1 and B2 samples.

Data reduction was performed using the standard MAGIC analysis
software MARS \cite{O_MARS}. Data quality selection was based mainly
on stable trigger rates, reducing the total sample of 203.1 hours of
observations to 157.9 hours of good quality data (see table
\ref{Table1} for total and effective observation times ($\teff$) of
each of the four samples). For the data analysis, after the image
cleaning, we retain only events with a \st{size} (total signal inside
the shower) larger than 50 photo-electrons in each telescope and
reconstructed energy greater than 60 GeV. Furthermore, for every
event, we calculate the angular distance $\theta$ between the
reconstructed arrival direction and the \st{ON} (or \st{OFF}) center
position, as well as the signal-to-noise maximization variable
\st{hadronness} ($h$) by means of the Random Forest method
\cite{O_MAGIC_RF}, and select events with $\th < \th_{\rt{cut}}$ and
$h < h_{\rt{cut}}$. The values of $\th_{\rt{cut}}$ and $h_{\rt{cut}}$
are optimized on the contemporaneous Crab Nebula data samples, by
maximizing the expected integral flux sensitivity (according to Li \&
Ma \cite{O_LiMa}) in the whole energy range. For all four of the
considered data samples, we obtain $\th_{\rt{cut}}$ = 0.015
deg$^{\sq{2}}$ \footnote{We note that the $\th_{\rt{cut}}$ is
  optimized for a point-like source, whereas Segue~1 is slightly
  extended for MAGIC (see section \ref{DM:J}). We have confirmed that
  the sensitivity could be improved by up to 10\% for $\th_{\rt{cut}}$
  values slightly larger than the selected ones. This, however, would
  have been at the expense of background modeling (the number of
  available modeling zones would be reduced, see section \ref{FL}),
  thus this option was disregarded.}, whereas for $h_{\rt{cut}}$ we
get 0.30 for samples A and B1, and 0.25 for samples B2 and C.

Additionally, the effect that $\eta$~Leo has on the data was also
taken into account. Namely, pixels illuminated by the star produce
background triggers with a rate larger than those due to atmospheric
showers. To avoid the saturation of the data acquisition with such
events, the thresholds of the affected pixels are automatically
increased/decreased with the rotation of the star in the camera. As a
consequence, a region is created around the $\eta$~Leo position where
the efficiency for shower detection is reduced with respect to the nominal
one. This causes differences between the spatial distributions of
events with respect to the \st{ON} and \st{OFF} positions, which could
introduce inhomogeneities in the instrument response function across
the field of view. In our Segue~1 observations, this effect is only
significant for sample C (because the star is closer to the camera
center). We restore the \st{ON}-\st{OFF} symmetry by removing those
events of reconstructed energy lower than 150 GeV and for which the
center of gravity of the shower image, in either one of the cameras,
lies less than 0.2$\deg$ away from the position of $\eta$~Leo (or from
a position at the same relative location with respect to the \st{OFF}
region as the star is from the \st{ON} region). The optimization of
these energy and angular distance cuts was done by imposing agreement
between \st{ON} and \st{OFF} $\th$-distributions for events with $0.35
< h < 0.60$ (i.e. excluding those selected as the signal
candidates). The corresponding Monte Carlo production was treated with
the same star-related cuts.

Figure \ref{Fig1} represents the skymap centered at Segue~1: no
significant gamma-ray excess can be seen at its nominal position. The
same conclusion is made from the overall $\th$-distribution (figure
\ref{Fig2}). Consequently, we proceed to calculate the differential
and integral flux upper limits for the gamma-ray emission from the
potential source, by means of the \tit{conventional} analysis approach
(for details and formalism, see \cite{I_FL}), currently standard for
IACTs. However, as it has been shown that the conventional method is
suboptimal to the full likelihood analysis for spectral emissions with
known characteristic features (as expected from gamma-rays of DM
origin, \cite{I_FL}), those upper limits are quoted for completeness
and cross-checking purposes only and can be found in Appendix
\ref{A}. The actual analysis of the Segue~1 data proceeds with the
full likelihood method, and more details on it are provided in the
following section.

\section{Full likelihood analysis}
\label{FL}

As shown by Aleksi\'{c}, Rico and Martinez in \cite{I_FL}, the full
likelihood approach maximizes the sensitivity to gamma-ray signals of
DM origin by including the information about the expected spectral
shape (which is fixed for a given DM model) into the calculations. The
sensitivity improvement obtained by the use of this method is
predicted to be a factor \mbox{$\sim$(1.5--2.5)} with respect to the
conventional analysis (depending on the spectral form of the searched
signal).

In this work, we present the results of the full likelihood analysis
applied to our Segue~1 observations with MAGIC. We follow the
formalism and nomenclature defined in \cite{I_FL} (for the Reader's
convenience, also included below), and address our specific choices
for the different terms entering the likelihood in more detail.

The basic concept behind the method is the comparison of the
\tit{measured} and \tit{expected} spectral distributions; that is, we
have to model the emission expected from the \st{ON} region. For a
given DM model, $M$, the spectral shape is known (see section
\ref{DM}), thus the only free parameter is the intensity of the
gamma-ray signal ($\theta$). The corresponding likelihood function has
the following form:
\begin{equation}
\mathcal{L}(\Nest, M(\theta) | \Nobs, E_1, ..., E_{\sNobs}) 
= \frac{{\Nest}^{\sNobs}}{\Nobs!}e^{-\sNest} \times
\prod\limits_{i=1}^{\sNobs}\mathcal{P}(E_{i}; M(\theta)),
\label{FL:eq1}
\end{equation}
with $\Nobs ( = \Non + \Noff)$ and $\Nest$ denoting the total number
of observed and estimated events, respectively, in \st{ON} and
\st{OFF} regions. $\mathcal{P}(E_{i}; M(\theta))$ is the value of the
probability density function of the event $i$ with \tit{measured}
energy $E_{i}$: 
\begin{equation}
\mathcal{P}(E; M(\theta)) = \frac{P (E;
  M(\theta))}{\int\limits_{E_{\rt{min}}}^{E_{\rt{max}}} P (E; M(\theta)) dE},
\label{FL:eq2}
\end{equation}
where $E_{\rt{min}}$ and $E_{\rt{max}}$ are the lower and upper limits
of the considered energy range. $P (E; M(\theta))$ represents the
differential rate of the events, such that:
\begin{equation}
P(E; M(\theta)) =
\begin{cases} 
\Poff(E_i), & i \in OFF \\
\Pon(E_i; M(\theta)), & i \in ON
\end{cases},
\label{FL:eq3}
\end{equation}
where $\Poff (E)$ and $\Pon (E; M(\theta))$ are the expected
differential rates from the \st{ON} and \st{OFF} regions. In this
work, $\Poff (E)$ is determined from the data (see bellow), whereas
$\Pon (M(\theta); E)$ is calculated as:
\begin{equation}
\Pon(E; M(\theta)) = \frac{1}{\tau} \Poff(E) 
+ \int\limits_{0}^{\infty} \frac{d\Phi(M(\theta))}{dE'}R(E; E')dE'.
\label{FL:eq4}
\end{equation}
True energy is denoted with $E'$; $d\Phi(M(\theta))/dE'$ is the
predicted differential gamma-ray flux, and $R (E; E')$ is the response
function of the telescope. Lastly, $\tau$ refers to the normalization
between \st{OFF} and \st{ON} regions.

Thus, in practice, for the construction of the full likelihood
function, we need the instrument response to gamma-ray events, a
model of the background differential rate, and an assumed
signal spectrum:

\begin{itemize}
\item the instrumental response function ($R(E; E')$) can be described
  by the effective collection area ($\Aeff$) and by the energy
  dispersion function. The latter is well approximated, for a given
  $E'$, with a Gaussian, whose mean and width will be referred to as
  the energy bias and resolution, respectively. For each of the four
  observational periods, the response functions are independently
  determined from the corresponding Monte Carlo simulations;
\item the model of the background differential rate ($\Poff(E)$) is
  obtained, for each period and each wobble pointing, directly from
  the Segue~1 observations at the complementary wobble position. For
  each pointing we select four model regions that have similar
  exposure as the \st{OFF} region, and we define them to be of the
  same angular size and at the same angular distance from the camera
  center as the corresponding \st{OFF} region, and adjacent to
  it. Then, by the means of the Kolmogorov-Smirnov statistics
  \mbox{(K-S)}, we compare the energy distribution of events from each
  of the modeling zones (individually and combined) to that from the
  \st{OFF} region. The modeling region(s) providing the highest K-S
  probability is (are) selected, and its (smoothed) measured
  differential energy distribution is used as the background model in
  the full likelihood (for more details, refer to
  \cite{FL_Teza}). 
  The statistical and systematic uncertainties introduced by this
  procedure in our final results are estimated by comparing the limits
  obtained using the selected modeling region(s) with those that we
  would obtain if the average of all four zones was used instead. Our
  constraints on DM properties are found to vary by up to 10\% for the
  considered range of DM particle masses.

\item the signal spectral function ($d\Phi(M(\theta))/dE'$) is known
  and fixed for a given DM model. In this work, we consider several
  channels of photon production from DM annihilation or decay:
  secondary photons from SM final states, gamma-ray lines, virtual
  internal bremsstrahlung (VIB) photons and gamma-ray boxes. More
  details on these final states are provided in section
  \ref{DM:Phot}. 
\end{itemize}

For each of the two pointing positions of each of the four defined
observational periods, an individual likelihood function is
constructed using the corresponding background model, plus the signal
spectral function convoluted with the appropriate response of the
telescopes. The global likelihood, encompassing the entire Segue~1
data sample, is obtained as a product of those individual ones
(eq. (5.1) in \cite{I_FL}). It is maximized with the gamma-ray signal
intensity as a free parameter, while $\Nest$ and $\tau$ of each
individual sample are treated as nuisance parameters with Poisson and
Gaussian distributions, respectively. The free parameter is bounded to
the physical region during the likelihood maximization: that is, the
signal intensity is not allowed to take negative values. We note that
the results obtained this way are conservative (i.e. they may have a
slight over-coverage, see \cite{O_Rolke}), since negative fluctuations
cannot produce artificially constraining limits.

The full likelihood calculations are performed for the 95\% confidence
level (CL) and one-sided confidence intervals ($\Delta \log
\mathcal{L}$ = 1.35) using the \texttt{TMinuit} class of ROOT
\cite{FL_TMinuit}.

\section{Expected dark matter flux}
\label{DM}

Before proceeding to the results of the full likelihood analysis of
our Segue~1 sample, let us first comment on how the limits on the
DM-induced gamma-ray signal are translated to limits on DM properties.

The prompt gamma-ray flux produced in annihilation or decay of DM
particles is given as a product of two terms:
\begin{equation}
  \frac{d\Phi(\Delta\Omega)}{dE'} = \frac{d\Phi^{\rt{\sq{PP}}}}{dE'}\times J(\Delta\Omega). 
  \label{eq1}
\end{equation}
The particle physics term, $d\Phi^{\rt{\sq{PP}}}/dE'$, solely depends
on the chosen DM model --- it is completely determined for the given
theoretical framework and its value is the same for all sources. The
astrophysical term, $J({\Delta\Omega})$, on the other hand, depends on
the observed source (its distance and geometry), the DM distribution
at the source region and the properties of the instrument and the
analysis. 

In the case of \st{annihilating} DM, the particle physics term takes
the form:
\begin{equation}
  \frac{d\Phi^{\rt{\sq{PP}}}}{dE'} = \frac{1}{4\pi}\frac{\langle\sigma_{\rt{ann}}v\rangle}
  {2m^{\sq{2}}_{\chi}}\frac{dN}{dE'},
  \label{eq2}
\end{equation}
where $\m$ is the DM particle mass, $\sv$ is the thermally averaged
product of the total annihilation cross-section and the velocity of
the DM particles, and $dN/dE' = \sum_{i=1}^{n}\rt{Br}_{i} dN_{i}/dE'$
is the differential gamma-ray yield per annihilation, summed over all
the $n$ possible channels that produce photons, weighted by the
corresponding branching ratio Br. All the information regarding the
spectral shape is contained in the $dN/dE'$ term.

On the other hand, the astrophysical factor ($J_{\rt{ann}}$) is the
integral of the square of the DM density ($\rho$) over the line of
sight (\tit{los}) and the solid angle covered by the observations
($\Delta\Omega$), i.e.:
\begin{equation}
  J_{\rt{ann}}(\Delta\Omega) = \int_{\Delta\Omega}\int_{los}\rho^{\sq{2}}(l,\Omega)dld\Omega.
  \label{eq3}
\end{equation}

For the case of \st{decaying} DM, the particle physics term depends on
the lifetime of the particle $\td$, and its form is obtained by
replacing the $\sv / 2\m$ contribution with $1/\td$ in
eq. (\ref{eq2}). As for the astrophysical term ($J_{\rt{dec}}$), it
scales linearly with the DM density ($\rho^{\sq{2}}\rightarrow\rho$ in
eq. (\ref{eq3})).

We express the results of our DM searches as upper limits to $\sv$ (for the
annihilation scenarios) or as lower limits to $\td$ (for the decaying
DM). In the full likelihood analysis, $\sv$ or $\td$ play the role of
the free parameter. 

\subsection{Considered spectral shapes}
\label{DM:Phot}

\begin{figure}[t!]
  \centering
  \vspace{-5pt}\hspace{0pt}
  \includegraphics[trim=20 15 20 20,clip=true,width=0.45\textwidth]{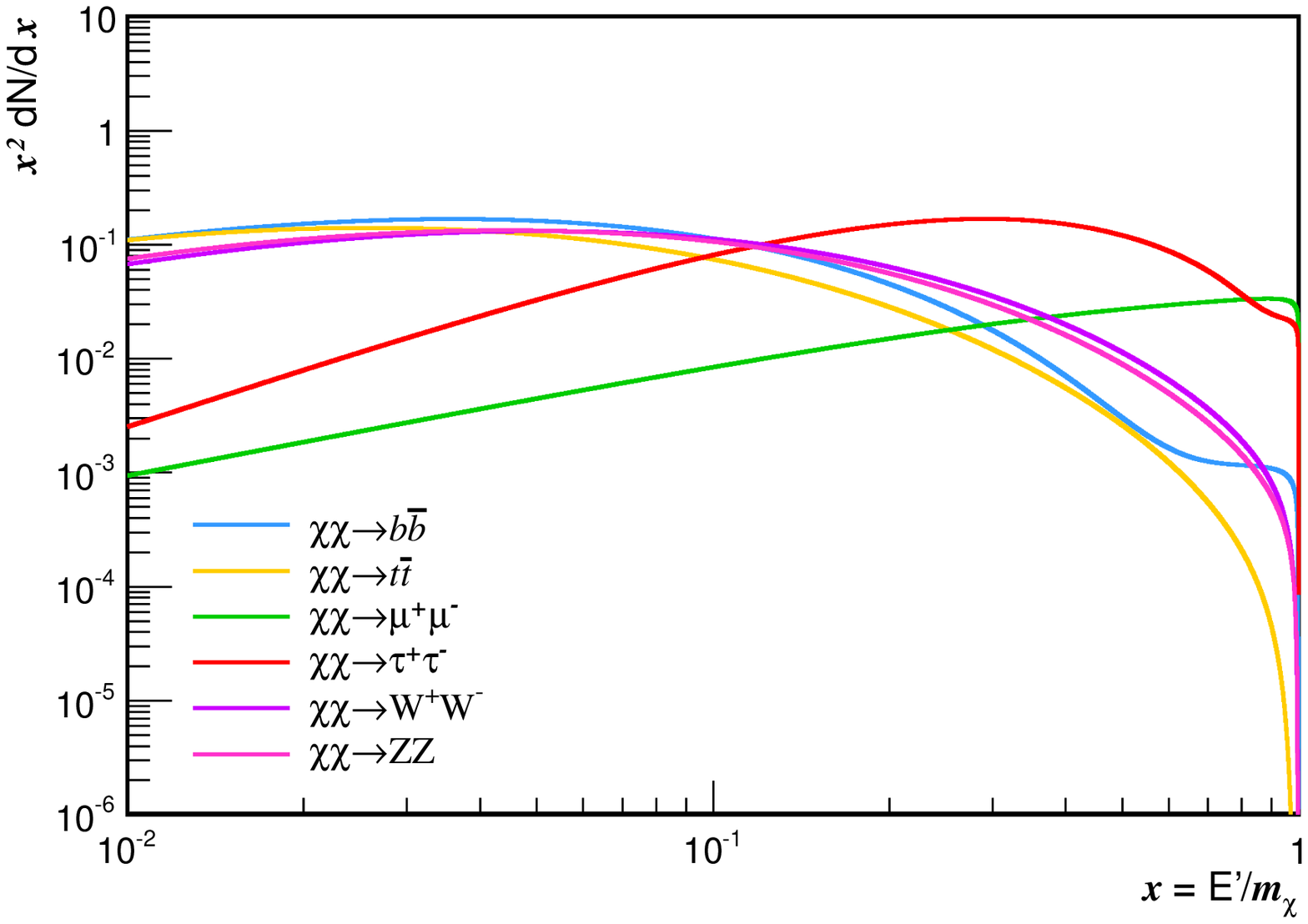}
  \includegraphics[trim=20 15 20 20,clip=true,width=0.45\textwidth]{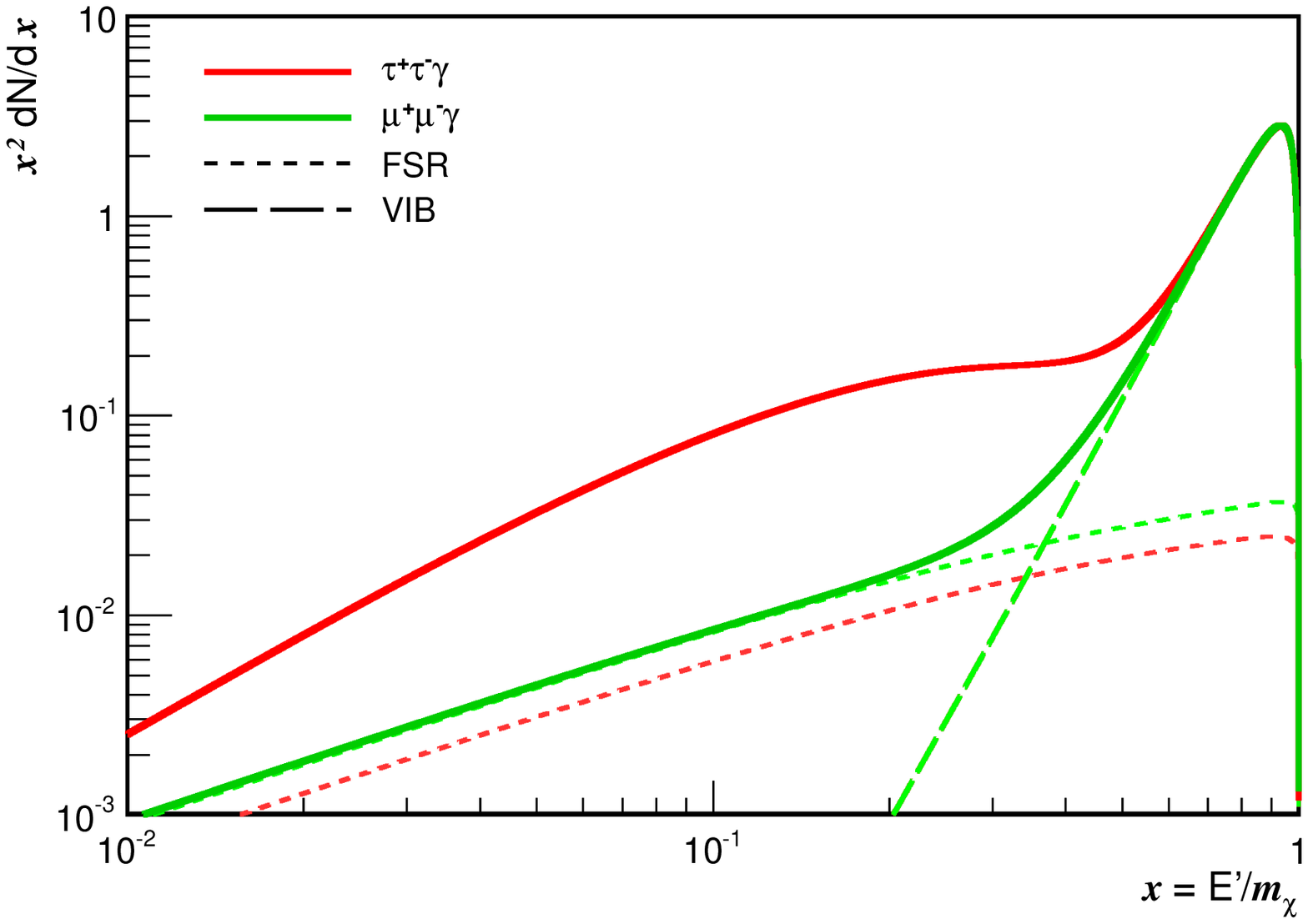}
  \vspace{0pt}
  \caption{Gamma-ray spectrum for DM 
    annihilation into different final states. 
    (\bt{Left}) Secondary photons (when 
    applicable, the FSR is included in the 
    spectrum). Modeling is done according 
    to the fits provided in \cite{DM_SecFit}. 
    (\bt{Right}) Spectral distribution from 
    annihilation into leptonic three-body 
    final states (solid lines), with the 
    contributions from FSR and VIB photons 
    (dashed and long-dashed lines, 
    respectively). The assumed 
    mass-splitting parameter value is 
    $\mu$ = 1.1.}
  \label{Fig3}
\end{figure}

As already mentioned in section \ref{FL}, in this work we search for
DM annihilating or decaying into different final states. In
particular, we consider the following channels: $b\bar{b}$,
$t\bar{t}$, $\mu^{+}\mu^{-}$, $\tau^{+}\tau^{-}$, $W^{+}W^{-}$ and
$ZZ$. The resulting spectra from secondary photons are continuous and
rather featureless, with a cutoff at the kinematical limit $E' = \m$
(figure \ref{Fig3}-left). In our analysis, we use the parametrization
presented in \cite{DM_SecFit}. When applicable, the final state radiation (FSR)
contribution is included in those fits. 

We also analyze final states leading to sharp spectral
features. First, we consider the direct annihilation into two photons
($\chi\chi\rightarrow\gamma\gamma$), or a photon and a boson
($\chi\chi\rightarrow\gamma Z/h$). Although loop-suppressed ($\sim
\mathcal{O}(1/\alpha^{2})$), such processes are of great interest, as
they would result in a sharp, monochromatic line-like feature in the
photon spectrum --- a feature whose detection would represent the
smoking gun for DM searches. Such a line is described as:
\begin{equation}
  \frac{dN}{dE'} = N_{\gamma}\delta (E'-E_{0}),
  \label{eq4}
\end{equation}
where $E_{0}=\m$ and $N_{\gamma} =$ 2 for annihilation into
$\gamma\gamma$, while $E_{0} = \m (1-M_{Z/h}^{\sq{2}}/4\m^{\sq{2}})$
and $N_{\gamma} =$ 1 for annihilation into $\gamma Z/h$. In the latter
case, there is also a contribution to the gamma-ray spectrum
originated in the fragmentation and decay of the $Z$ and Higgs
bosons. As for the case of the decaying DM, line production is also a
possibility. In this work, we consider the case of two-body decay into
one monoenergetic photon ($\chi\rightarrow\gamma\nu$) for fermionic DM
particles. The spectral function of the resulting line is obtained by
simply making the substitution $\m\rightarrow\m/2$ in eq. (\ref{eq4}).

Another scenario resulting in sharp spectral features involves
emission of VIB photons: if the DM particle is a Majorana fermion,
which couples via a Yukawa coupling to a charged fermion ($f$) and a
scalar ($\eta$), the photon produced in the internal bremsstrahlung
process ($\chi\chi\rightarrow f\bar{f}\gamma$) will have a very
characteristic spectrum, displaying a salient feature close to the
kinematic endpoint and resembling a distorted gamma-ray line. The
exact expression of the differential gamma-ray spectrum of the
2$\rightarrow$3 process is given by eq. (2.8) in \cite{DM_VIB}. The
total spectral distribution also receives a contribution from FSR of
the nearly on-shell fermions produced in the 2$\rightarrow$2
annihilation ($\chi\chi\rightarrow f\bar{f}$), as well as from the
fragmentation or decay of the fermions produced both in the
2$\rightarrow$2 and in the 2$\rightarrow$3 processes (figure
\ref{Fig3}-right). The contribution from FSR becomes more and more
important as the mass splitting $\mu$ ($\equiv
m_{\eta}^{\sq{2}}/\m^{\sq{2}}$) between $\eta$ and the DM particle
increases, eventually erasing the strong gamma-ray feature from
internal bremsstrahlung. It can be verified that the gamma-ray feature
stands out in the total spectrum when $\mu\lesssim$ 2, which is the
case we assume in our analysis.

Lastly, a sharp feature might arise in scenarios where a DM particle
annihilates into an on-shell intermediate scalar $\phi$, which
subsequently decays in flight into two photons:
$\chi\chi\rightarrow\phi\phi\rightarrow\gamma\gamma\gamma\gamma$. In
the rest frame of $\phi$, photons are emitted isotropically and
monoenergetically; therefore, in the galactic frame, the resulting
spectrum will be box-shaped (for the exact expression for $dN/dE'$,
see eq. (2) in \cite{DM_Box}). The center and width of such a feature
are completely determined by the masses of the scalar ($m_{\phi}$) and
DM particle: the box is centered at $E' = \m/2$ and its width is
$\Delta E' = \sqrt{\m^{\sq{2}}-m_{\phi}^{\sq{2}}}$. For
$m_{\phi}\approx\m$, almost all of the DM particle energy is
transferred to the photons, and the resulting spectral shape is
intense and similar to the monochromatic line. On the other hand, for
$m_{\phi}\ll\m$ the box becomes wide and dim in amplitude; still, it
extends to higher energies and thus is not negligible as a
contribution to the signal spectrum.
\subsection{The astrophysical factor $J$ for Segue~1}
\label{DM:J}

The choice of density profile plays a crucial role in the calculation
of the astrophysical factor $J$, as it has direct implications on the
expected photon flux. This is particularly true for DM annihilation
(eq. (\ref{eq3})): as $J_{\rt{ann}}$ is proportional to
$\rho^{\sq{2}}$, cored central distributions (described by a constant
density value close to the center) will yield lower fluxes than the
cusped ones (described by a steep power law in the central region),
for the same total DM content. This dependence is less pronounced for
the decaying DM, since $J_{\rt{dec}} \propto \int\rho$. Motivated by
the numerical simulations, we model the DM density distribution
assuming the Einasto profile \cite{DM_Einasto}, with scale radius
$r_{s}$ = 0.15 kpc, scale density $\rho_{s}$ =
1.1$\times$10$^{\sq{8}}$ $M_{\odot}$ kpc$^{-3}$ and slope $\alpha$ =
0.30 \cite{DM_Segue_J, I_MAGIC_Segue}.

\begin{figure}[t!]
  \centering
  \vspace{-10pt}\hspace{5pt}
  \includegraphics[trim=0 25 0 20,clip=true,width=0.45\textwidth]{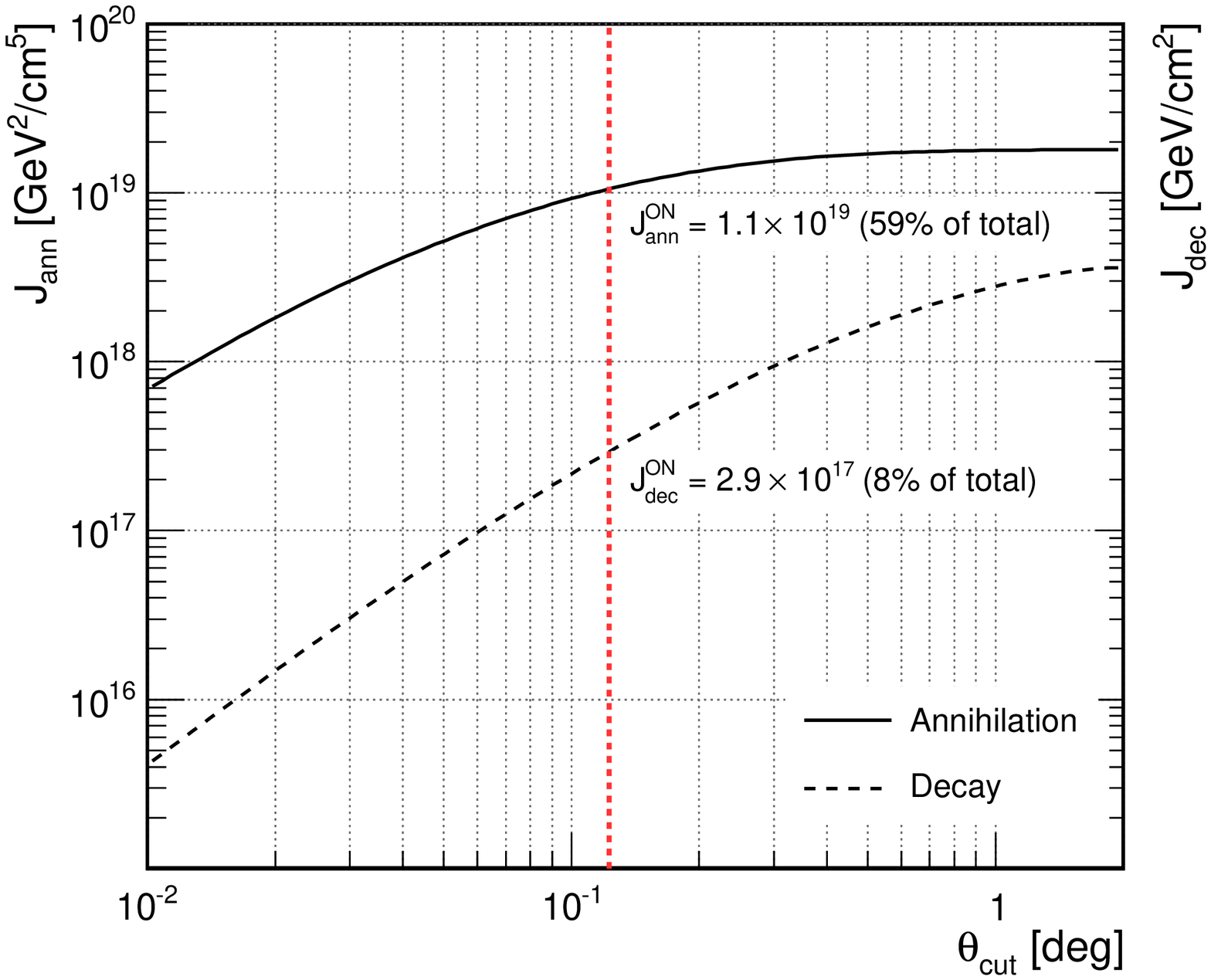}
  \hspace{5pt}
  \includegraphics[trim=0 25 0 20,clip=true,width=0.45\textwidth]{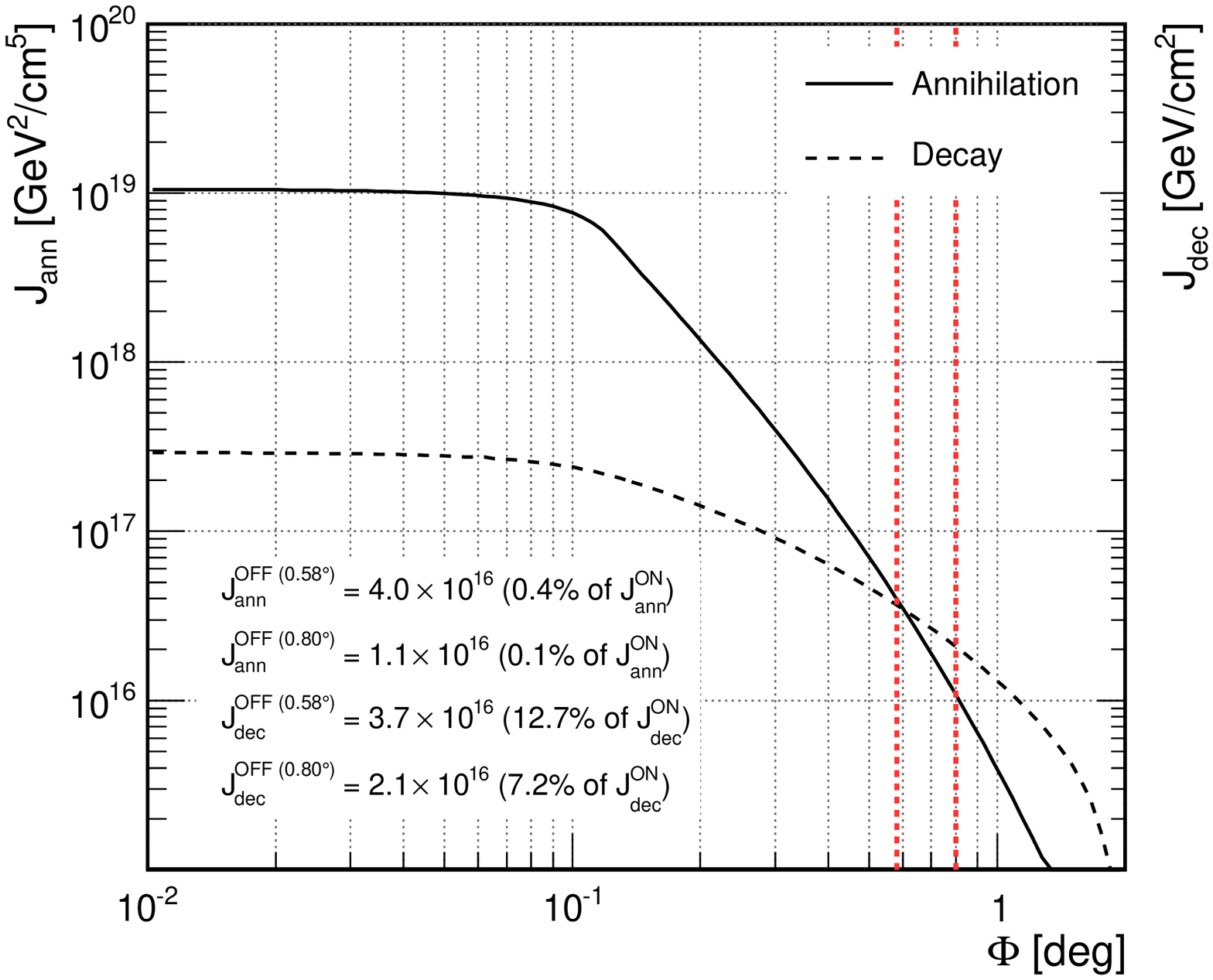}
  \vspace{0pt}
  \caption{ Astrophysical factor for Segue~1 for annihilating
    ($J_\rt{ann}$, solid line, left axis) and decaying ($J_\rt{dec}$,
    dashed line, right axis) DM, assuming the Einasto density profile.
    (\bt{Left}): as a function of the angular cut ($\th_{\rt{cut}}$,
    see section \ref{O}), for observations centered at the nominal
    position of Segue~1. The vertical dotted red line corresponds to
    the value used in this analysis, $\th_{\rt{cut}}$ =
    0.015$\deg$. (\bt{Right}) as a function of the angular distance
    from Segue~1 ($\Phi$), for a fixed anglar cut $\th_{\rt{cut}}$ =
    0.015$\deg$. The vertical dotted red lines correspond to the
    values of distance between the \st{ON} and \st{OFF} regions
    relevant in this analysis.}
  \label{Fig4}
\end{figure}
The value of $J$ is determined by the DM distribution within the
integrated solid angle $\Delta\Omega$ (eq. (\ref{eq3})), and hence by
the analysis angular cut $\th_{\rt{cut}}$ (figure \ref{Fig4}-left). In
addition, in order to compute the residual background in the \st{ON}
region, we measure the number of events acquired in the \st{OFF}
regions, defined by the same $\th_{\rt{cut}}$ and at an angular
distance from the position of Segue~1 of $\Phi$ = 0.58$\deg$ (for
samples A, B1 and B2) and $\Phi$ = 0.80$\deg$ (for sample C). The
\st{OFF} regions may contain non-negligible amounts of DM-induced
events (figure \ref{Fig4}-right), which are accounted in the analysis
as background, hence reducing the sensitivity for detection of signal
events in the \st{ON} region. We take this into account by using the
difference in $J$ between \st{ON} and \st{OFF} regions as
astrophysical factor, that is:
$J(\Delta\Omega) = J^\rt{ON}(\Delta\Omega) -
J^\rt{OFF}(\Delta\Omega)$. This correction is negligible (less than
1\%) for annihilation, but has an effect of $\sim$10\% for decay,
since in this case the abundant, although less concentrated quantities
of DM at large $\Phi$ contribute relatively more to the total expected
flux than in the case of annihilation. For the angular cut
$\th_{\rt{cut}}$ = 0.015$\deg$ and the used \st{OFF} positions, the
astrophysical factor for annihilating DM is $J_\rt{ann}$ =
1.1$\times$10$^{\sq{19}}$ GeV$^{\sq{2}}$ cm$^{-\sq{5}}$, and the
corresponding values for decay are $J_\rt{dec}$ =
2.5$\times$10$^{\sq{17}}$ GeV cm$^{-{\sq{2}}}$ for periods A, B1 and
B2, and $J_\rt{dec}$ = 2.7$\times$10$^{\sq{17}}$ GeV cm$^{-{\sq{2}}}$
for period C.

The dominating systematic uncertainty on $J$, resulting from the fit of
the Segue~1 DM distribution to the Einasto profile, is about a factor
of 4 at 1$\sigma$ level for $J_\rt{ann}$, and about a factor of 2 for
$J_\rt{dec}$ \cite{DM_Segue_J}. These uncertainties affect our $\sv$
and $\td$ limits linearly. A discussion about comparisons of $J$
uncertainties for different classes of objects is included in section
\ref{D:Exp}.

\section{Limits for dark matter annihilation and decay models}
\label{R}

\begin{figure}[t]
  \centering
  \includegraphics[trim=10 30 30 30,clip=true,width=0.32\textwidth]{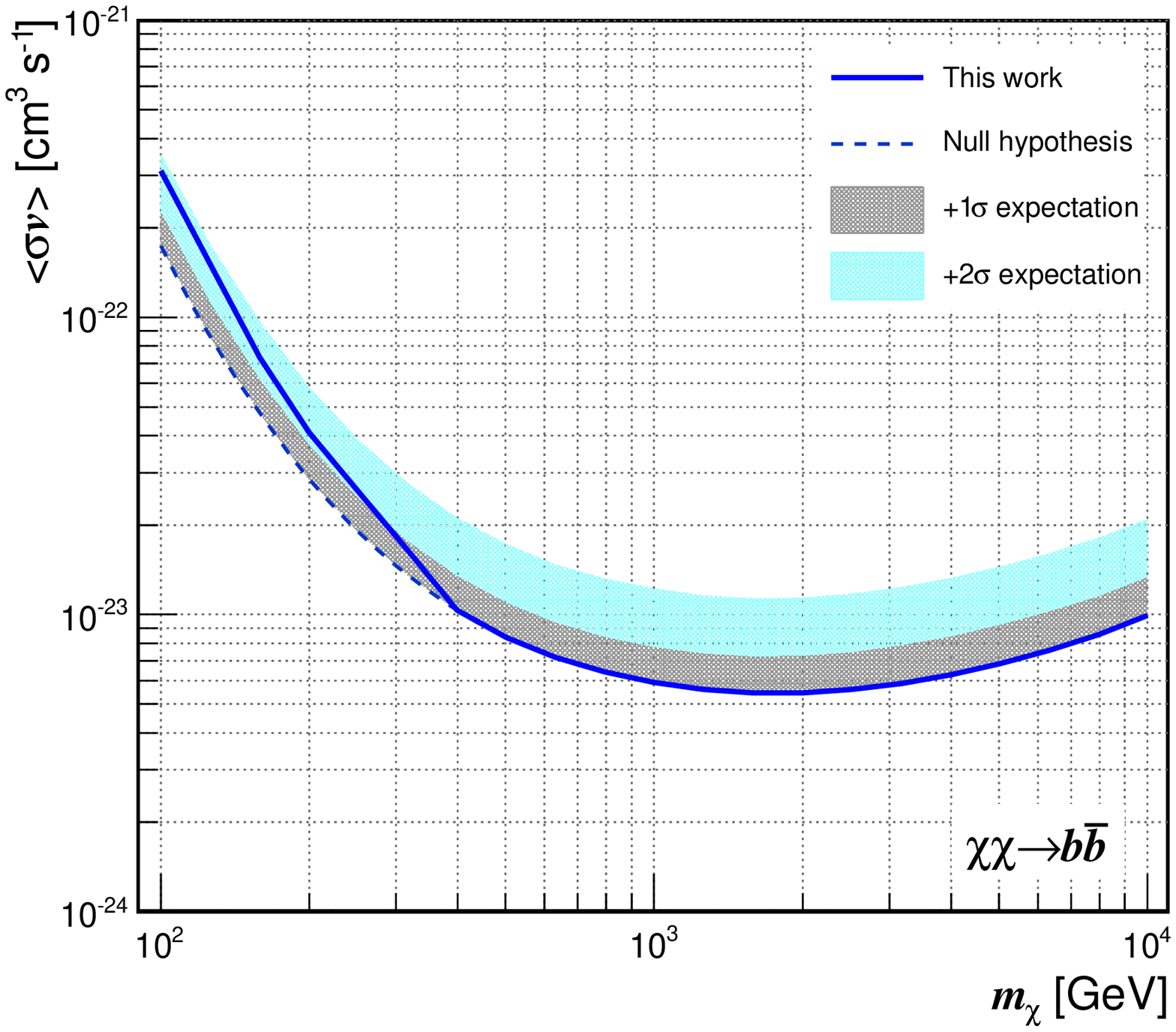}
  \includegraphics[trim=10 30 30 30,clip=true,width=0.32\textwidth]{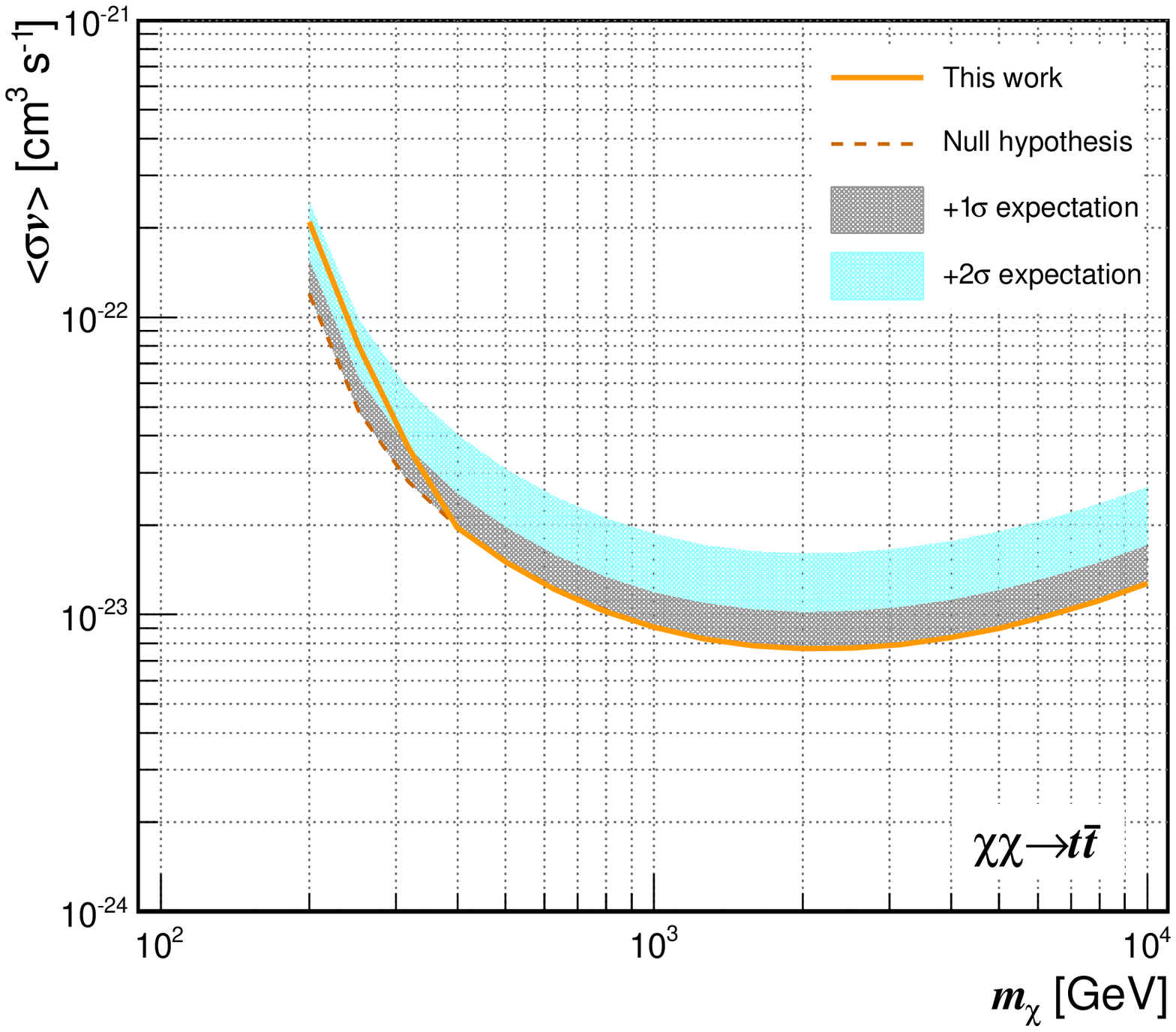}
  \includegraphics[trim=10 30 30 30,clip=true,width=0.32\textwidth]{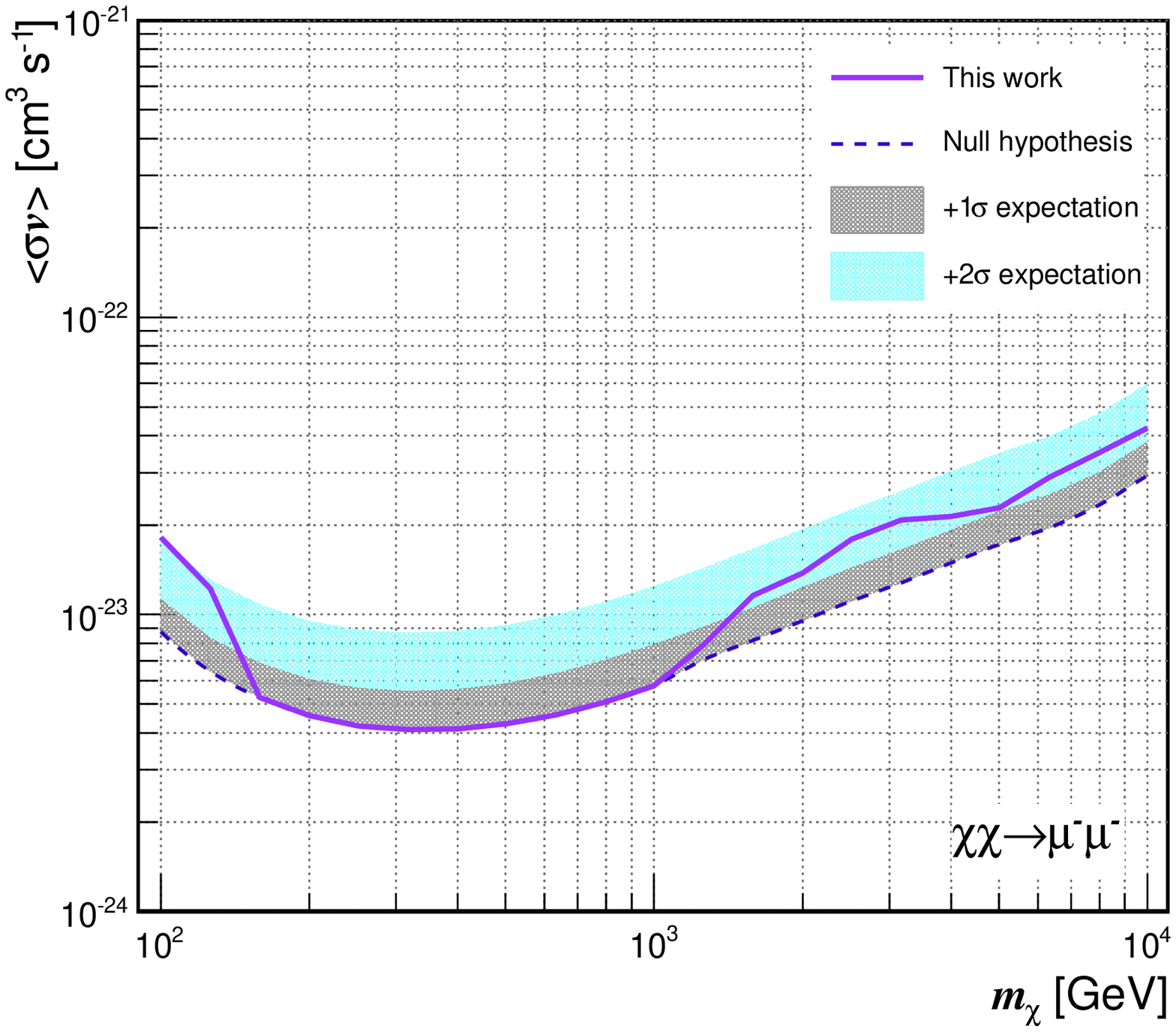}
  \includegraphics[trim=10 20 30 20,clip=true,width=0.32\textwidth]{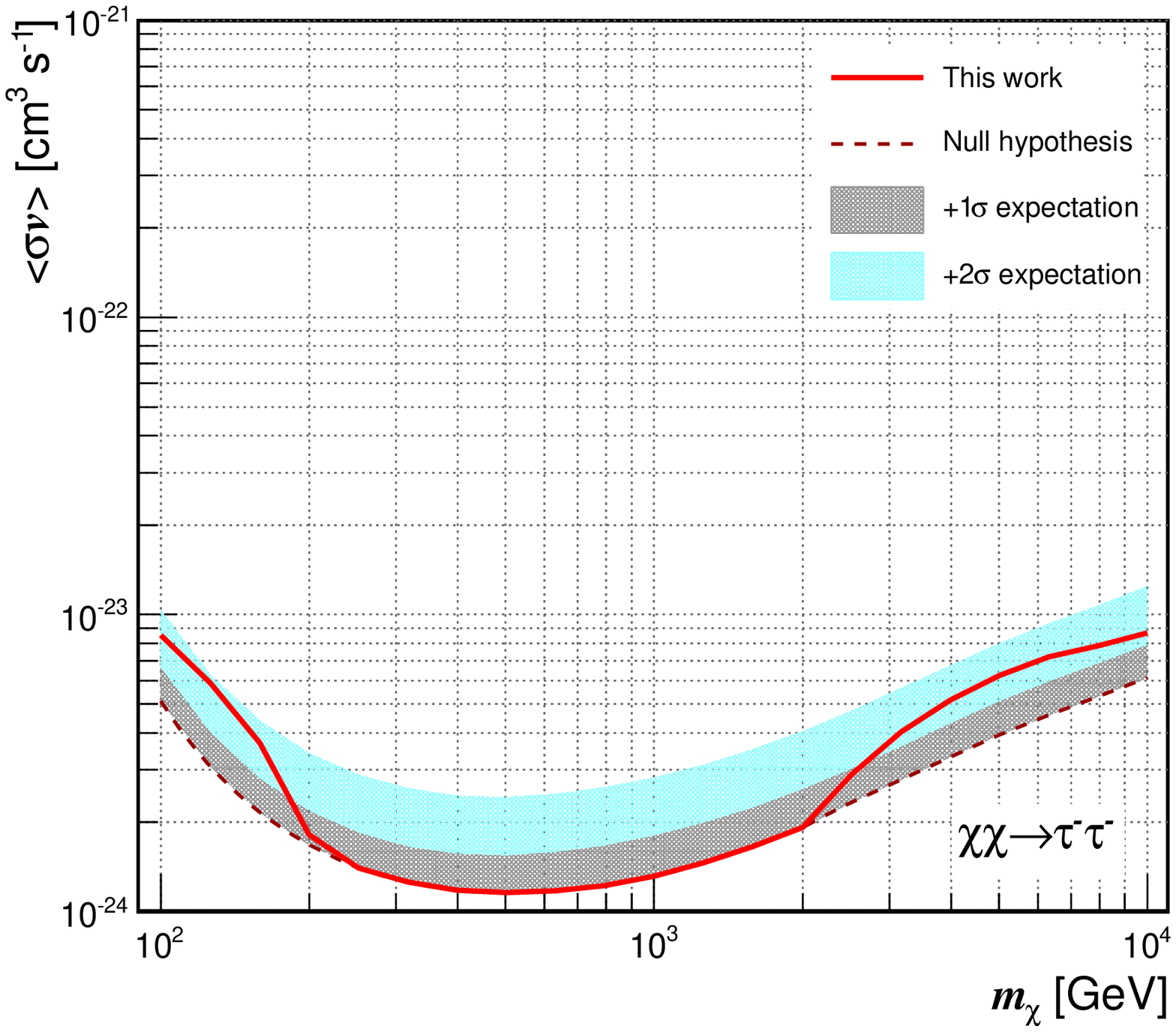}
  \includegraphics[trim=10 20 30 20,clip=true,width=0.32\textwidth]{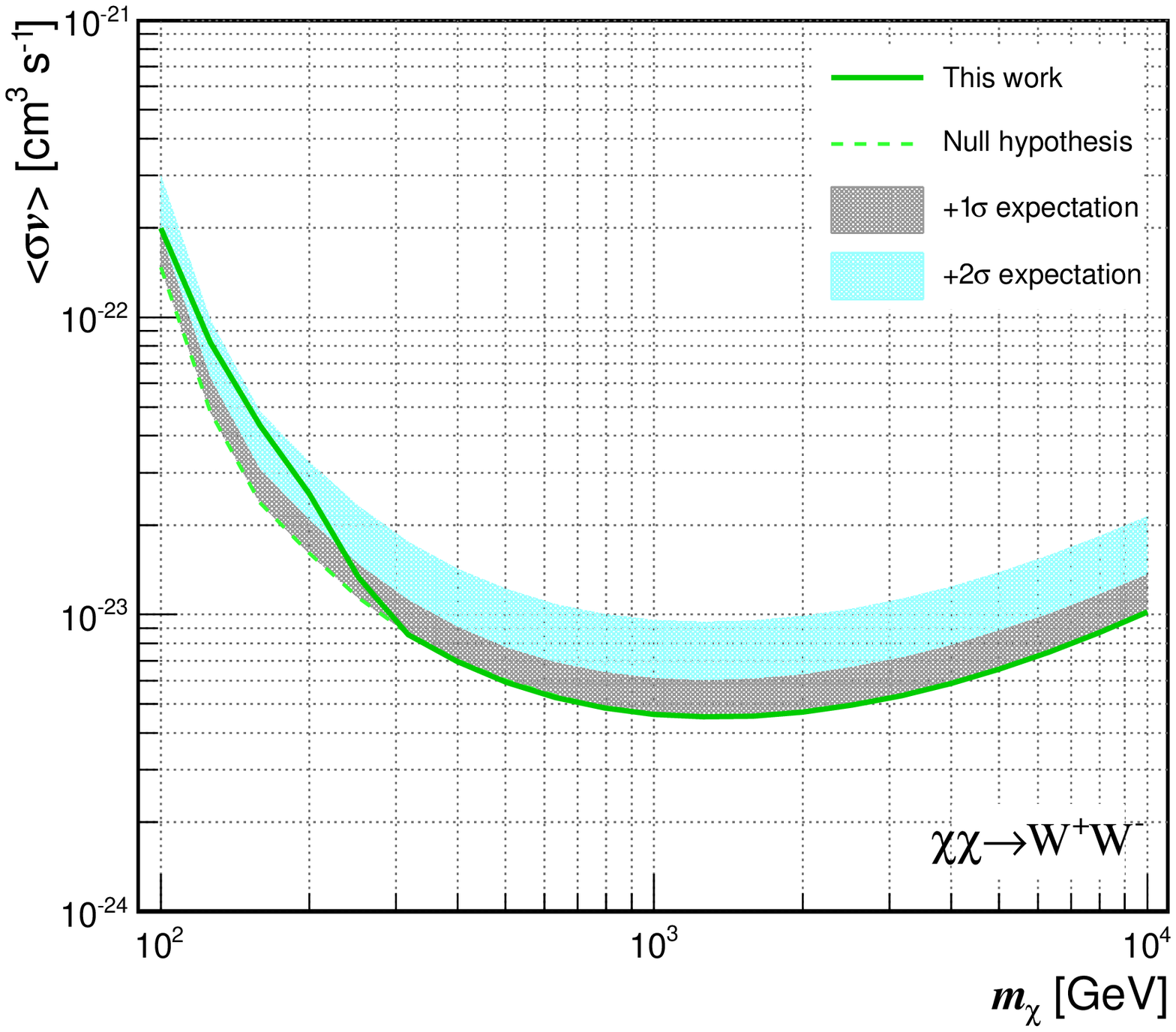}
  \includegraphics[trim=10 20 30 20,clip=true,width=0.32\textwidth]{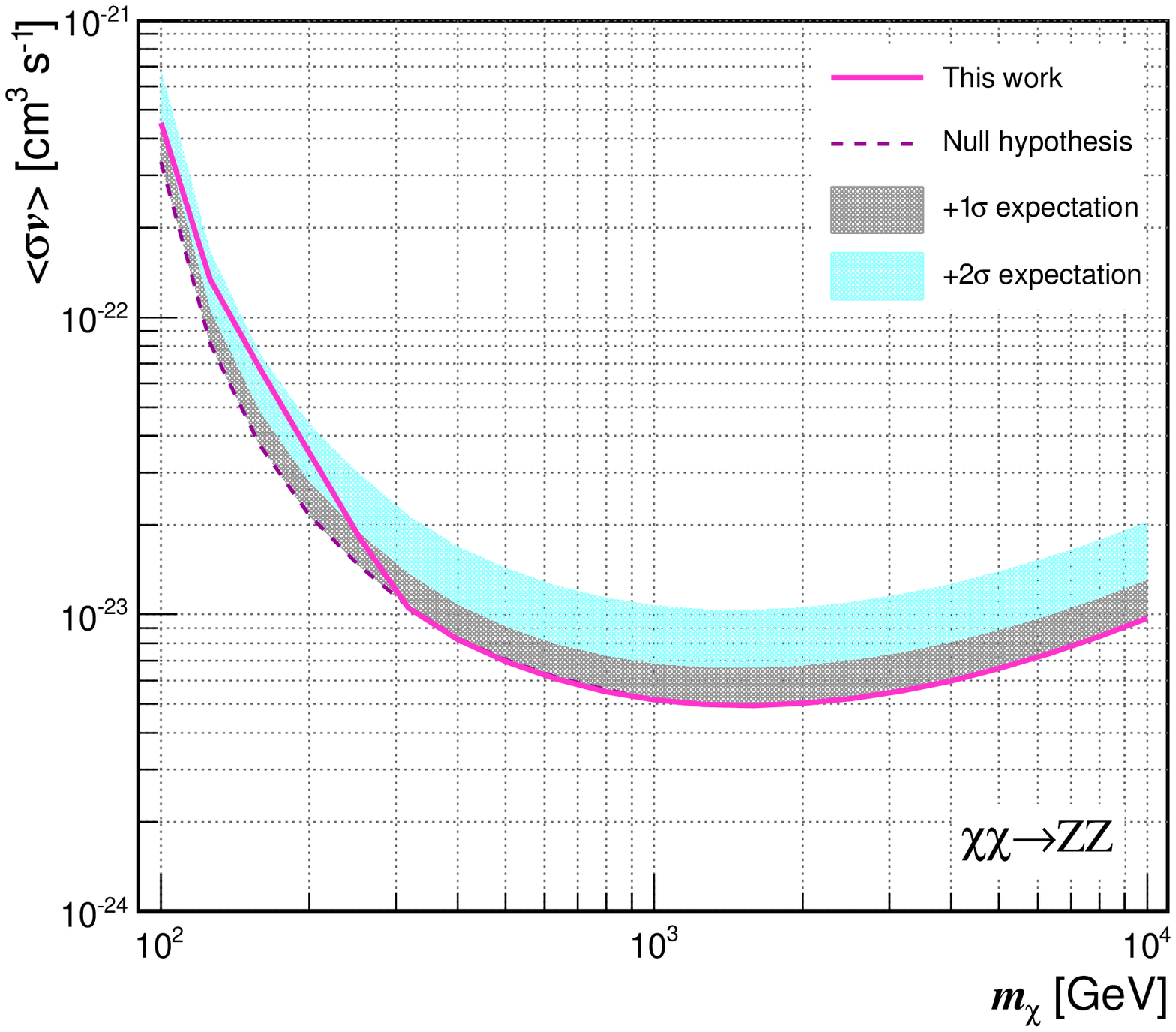}
  \caption{Upper limits on $\sv$ for different final state channels
    (from top to bottom and left to right): $b\bar{b}$, $t\bar{t}$,
    $\mu^{+}\mu^{-}$, $\tau^{+}\tau^{-}$, $W^{+}W^{-}$ and $ZZ$, from
    the Segue~1 observations with MAGIC. The calculated upper limit is
    shown as a solid line, together with the null-hypothesis
    expectations (dashed line), and expectations for 1$\sigma$ (shaded
    gray area) and $2\sigma$ (shaded light blue area) significant
    signal.}
\label{Fig5}
  \centering
  \vspace{10pt}
  \includegraphics[trim=10 20 30 20,clip=true,width=0.45\textwidth]{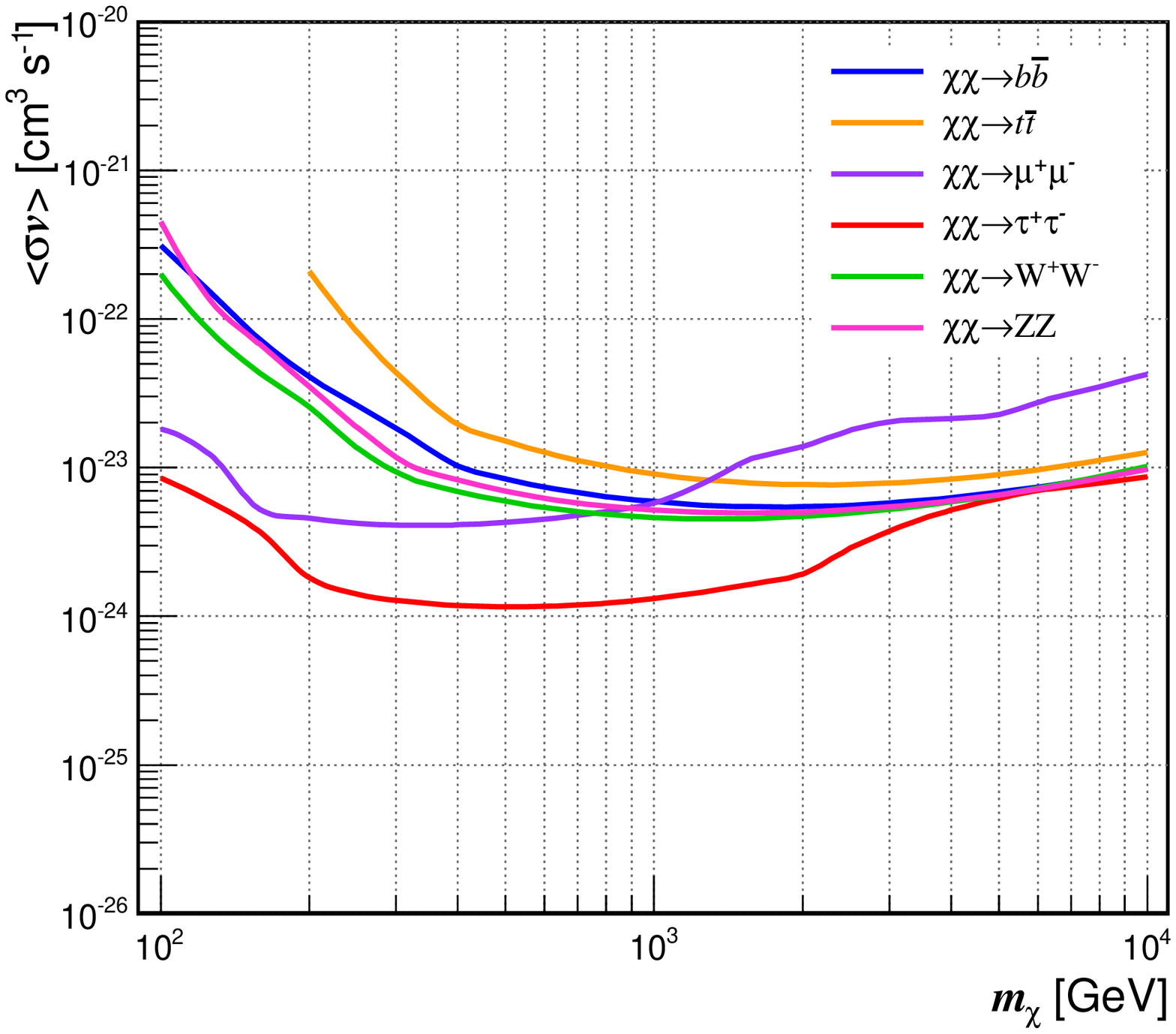}
  \includegraphics[trim=10 20 30 20,clip=true,width=0.45\textwidth]{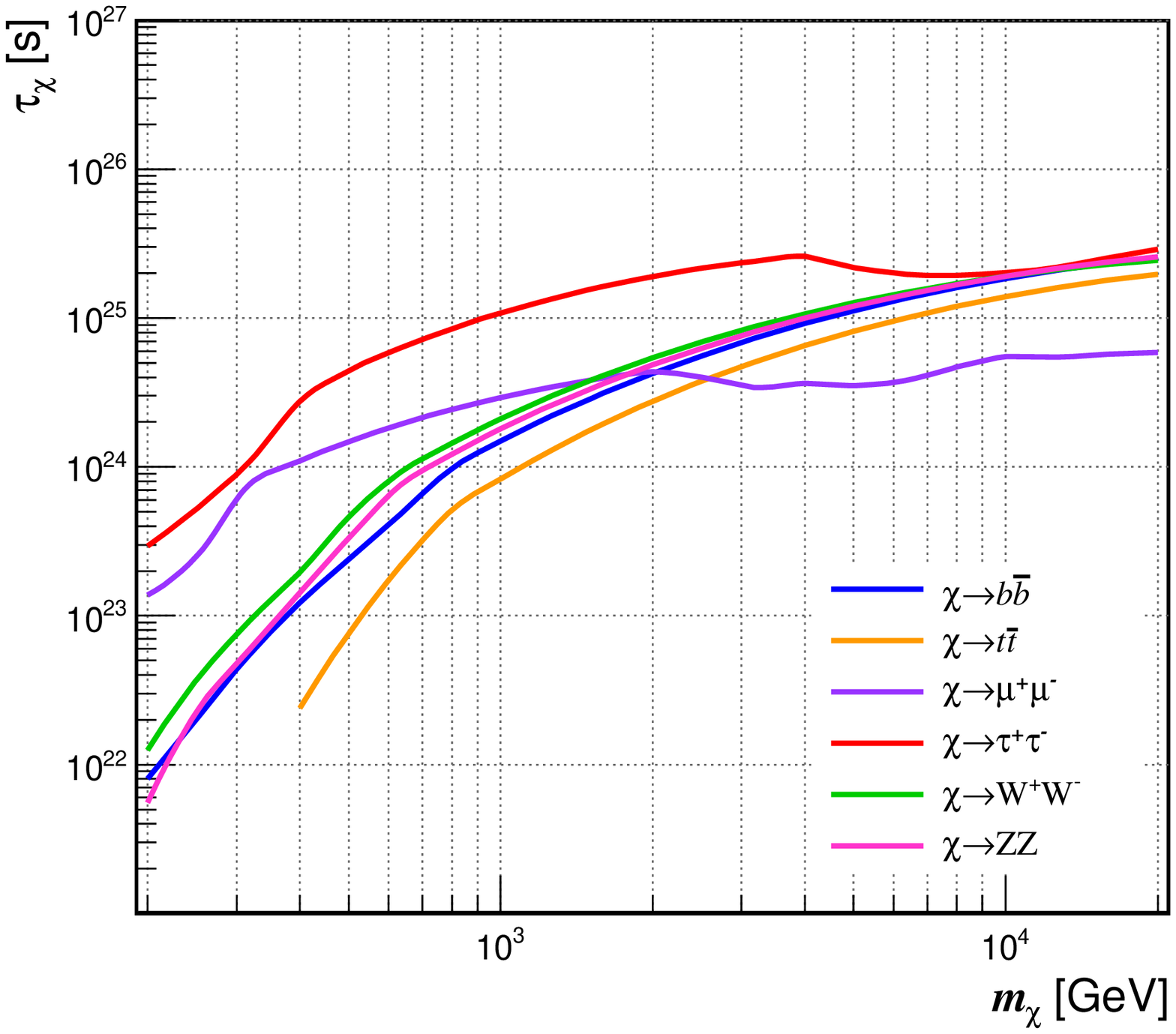}
  \vspace{0pt}
  \caption{Upper limits on $\sv$ (\bt{left}) and lower limits on $\td$
    (\bt{right}), for secondary photons produced from different final
    state SM particles, from the Segue~1 observations with MAGIC.}
  \label{Fig6}
\end{figure}
Here we present the results, in the context of indirect DM searches,
of 157.9 hours of selected data from the Segue~1 observations with
MAGIC, analyzed with the full likelihood approach. The results are
introduced in the following way: for each of the considered DM models,
a limit is set (95\% CL upper limit on $\sv$ or lower limit on $\td$)
by using the combined likelihood for the whole data sample. This
constraint is then compared to the expectations for the null
hypothesis (no signal), as well as for signals of 1$\sigma$ and
2$\sigma$ significances, estimated from the fast simulations
(comparison with negative signals is meaningless in this work, as the
free parameter is constrained to only have positive values, see
section \ref{FL}). In order to make the results as model-independent
as possible, in all the cases, the branching ratio is set to Br =
100\%. Considering the energy range for which the MAGIC telescopes are
sensitive to gamma rays, we search for DM particles of mass $\m$
between 100 GeV and 10 TeV for annihilation scenarios and between 200
GeV and 20 TeV for the decaying DM. Furthermore, all of the results
are produced without the assumptions of some additional boosts, either
from the presence of substructures \cite{R_substructures} or from
quantum effects \cite{R_Sommerfeld}.

\subsection{Secondary photons from final state Standard Model particles}
\label{R:Sec}

Figure \ref{Fig5} shows the upper limits on $\sv$, together with the
null hypothesis, 1$\sigma$ and 2$\sigma$ expectations, for
annihilation into six different final states: quarks ($b\bar{b}$,
$t\bar{t}$), leptons ($\mu^{+}\mu^{-}$, $\tau^{+}\tau^{-}$) and gauge
bosons ($W^{+}W^{-}$, $ZZ$). All bounds are consistent with the
no-detection scenario. For a more comprehensive overview, the $\sv$
upper limits for the considered final states are shown in figure
\ref{Fig6}-left.  A clear dependence between the shape of the expected
photon spectrum and the derived constraints can be noticed: the
strongest bound corresponds to the $\tau^{+}\tau^{-}$ channel
($\sv\sim$ 1.2$\times$10$^{-{\sq{24}}} \svu$), as it produces photons
whose spectrum is the hardest at energies for which the sensitivity of
MAGIC is at peak.

Similar considerations apply to the decaying DM scenarios: again, the
most constraining lower limit on $\td$ from Segue~1 observations is
obtained for the $\tau^{+}\tau^{-}$ channel, and is of the order of
$\td\sim$ 2.9$\times$10$^{\sq{25}}$ s.

\begin{figure}[t!]
  \centering
  \vspace{-5pt}\hspace{5pt}
  \includegraphics[trim=0 15 0 0,clip=true,width=0.55\textwidth]{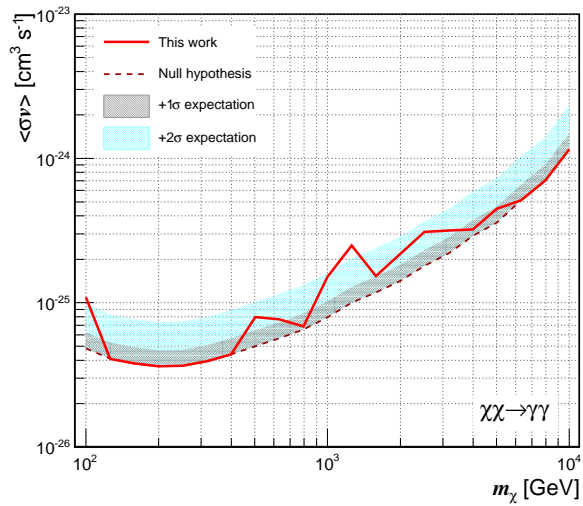}
  \vspace{0pt}
  \caption{Upper limits on $\sv$ for direct annihilation into two
    photons, as a function of $\m$, from the Segue~1 observations with
    MAGIC (solid line) and as expected for the case of no signal
    (dashed line), as well as for a signal of 1$\sigma$ or 2$\sigma$
    significance (gray and light blue shaded areas, respectively).}
  \label{Fig7}
\end{figure}
\subsection{Gamma-ray lines}
\label{R:Line}

Figure \ref{Fig7} shows the upper limits on $\sv$ for direct
annihilation of DM particles into two photons. For the considered $\m$
range, the constraints are set between 3.6$\times$10$^{-{\sq{26}}}$
and 1.1$\times$10$^{-{\sq{24}}} \svu$. In almost the entire considered
mass range, the upper limits are within 1$\sigma$ from the null
hypothesis; the largest deviation is observed at $\m\sim$ 1.3 TeV where
the signal is slightly larger than 2$\sigma$. The probability that
this is caused by random fluctuations of the background is relatively
large ($\sim$5\%) and hence not enough to be considered a hint of a
signal\footnote{The same reasoning applies to other excesses of
  similar magnitude calculated for other models studied in this
  section.}. On the other hand, should the excess be caused by
gamma-rays from DM annihilation or decay, a detection at a
5$\sigma$ significance level would require about 1000 hours of
observations with a sensitivity comparable to the ones used here.

Upper limits on $\sv$ from DM annihilation into photon and $Z$ boson,
for the considered DM masses, span the range between
7.8$\times$10$^{-{\sq{26}}} \svu$ and 2.3$\times$10$^{-{\sq{24}}}
\svu$. In the calculation of these limits we do not take into account
the contribution of secondary photons originating from fragmentation
and decay of $Z$, as the bound from the resulting continuous
contribution is expected to be negligible compared to that from the line
(figures \ref{Fig5} and \ref{Fig7}). Furthermore, due to the finite
width of the $Z$ boson, the gamma-ray line is not monochromatic. The
calculation of the consequent corrections to the $\sv$ upper limits
are beyond the scope of this paper; however, we note that, given the
energy resolution of MAGIC, the line broadening due to $Z$ width
($\Gamma\sim$ 2.5 GeV) is not expected to be of relevance in the
considered $\m$ range.

We also search for lines produced in DM decay. The derivation of such
constraints is straightforward from the results of the annihilation
scenario, so we limit our discussion on the matter to the comparison
with bounds from other experiments, in section \ref{D:Exp:Sec}.

\begin{figure}[t!]
  \centering
  \vspace{-5pt}\hspace{5pt}
  \includegraphics[trim=0 15 0 0,clip=true,width=0.45\textwidth]{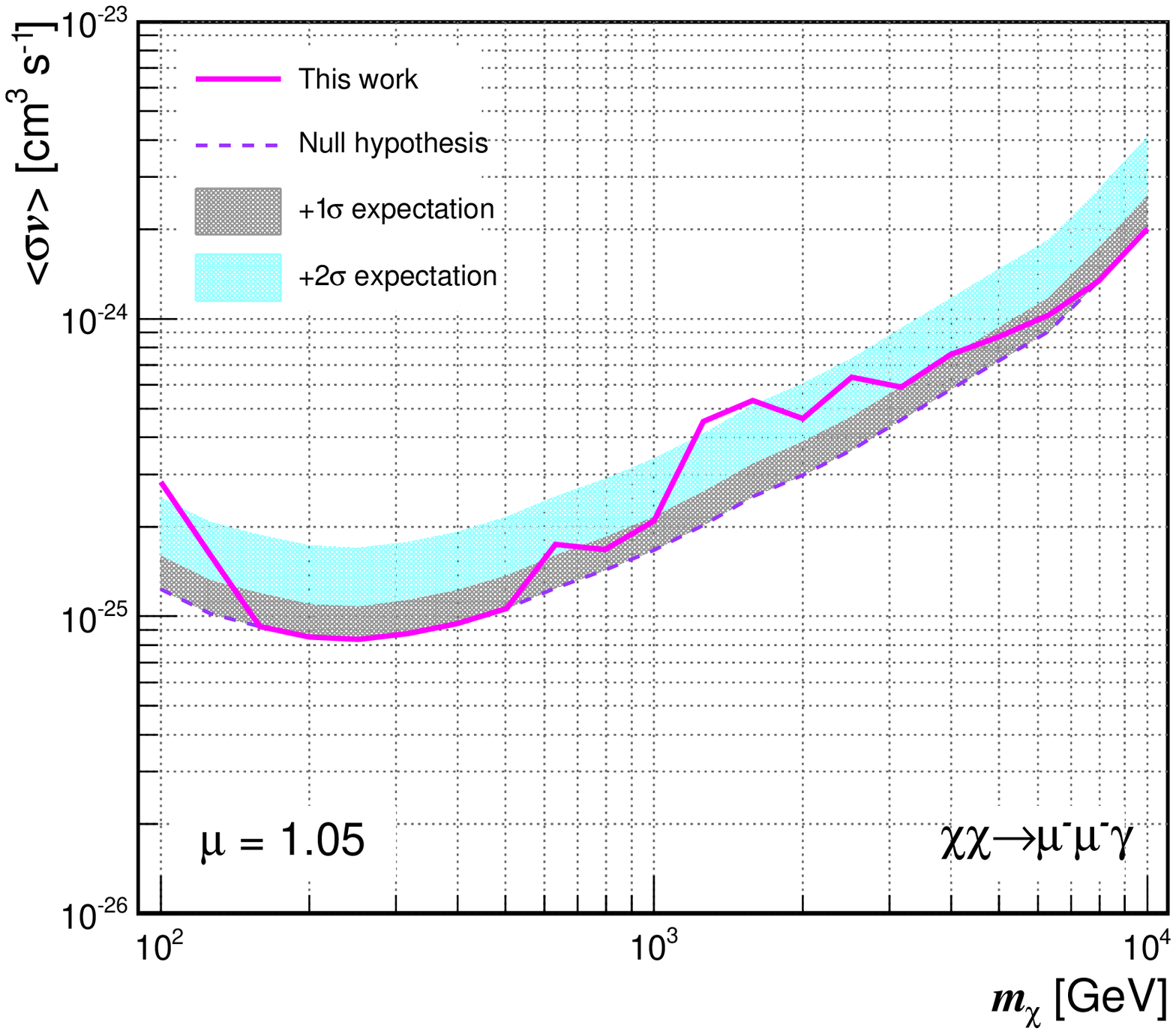}
  \includegraphics[trim=0 15 0 0,clip=true,width=0.45\textwidth]{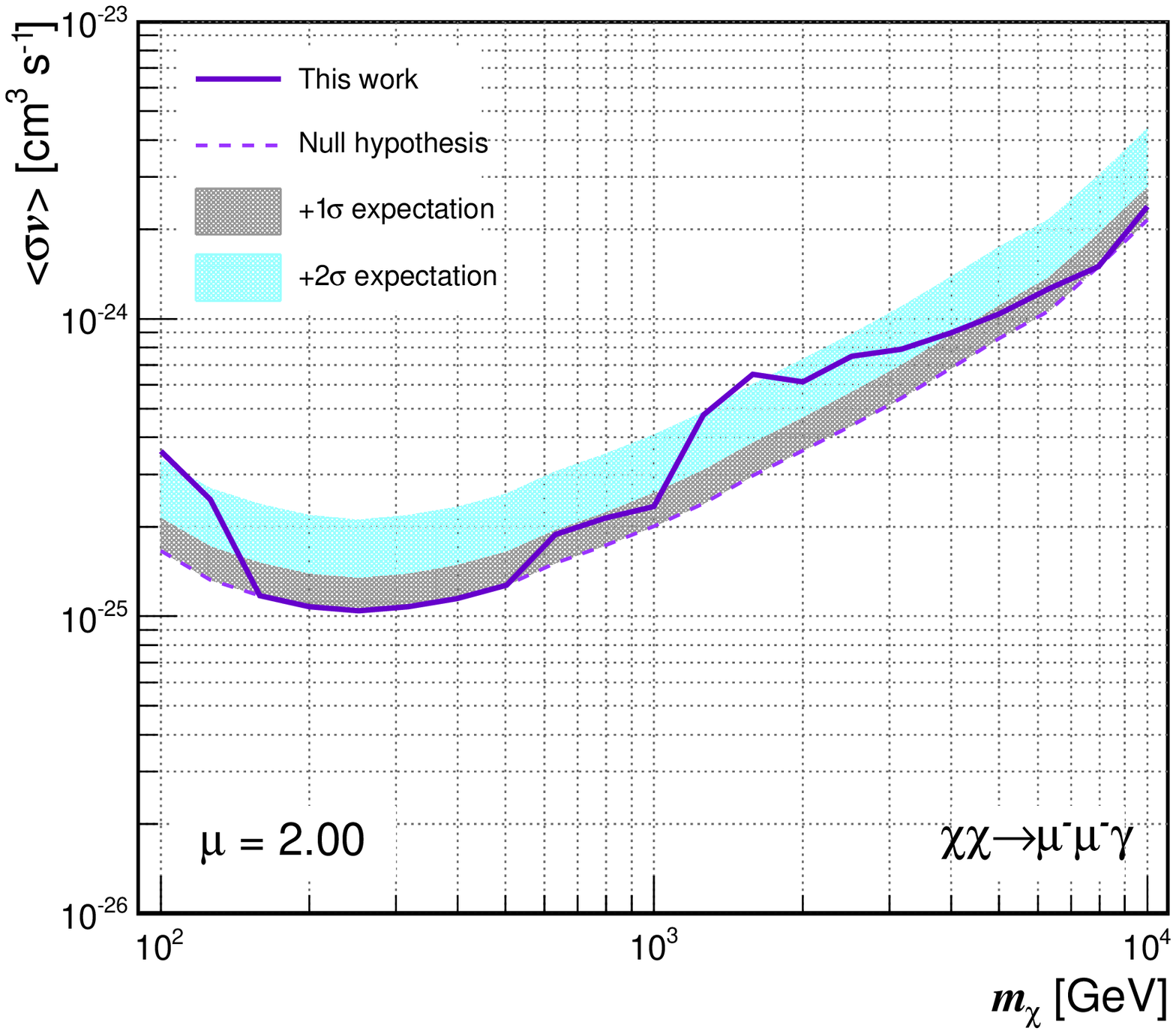}
  \vspace{0pt}
  \caption{Upper limits on $\sv$ for the $\mu^{+}\mu^{-}\gamma$ channel as
  a function of $\m$, from the Segue~1 observations with MAGIC (solid 
  line), and as expected for the case of no signal (dashed line), or
  for a signal of 1$\sigma$ or 2$\sigma$ significance (gray and
  light blue shaded areas, respectively). The value of the mass
  splitting parameter $\mu$ is 1.05 (\bt{left}) and 2.00 (\bt{right}).}
  \label{Fig8}
  \centering
  \vspace{0pt}\hspace{5pt}
  \includegraphics[trim=0 15 0 0,clip=true,width=0.45\textwidth]{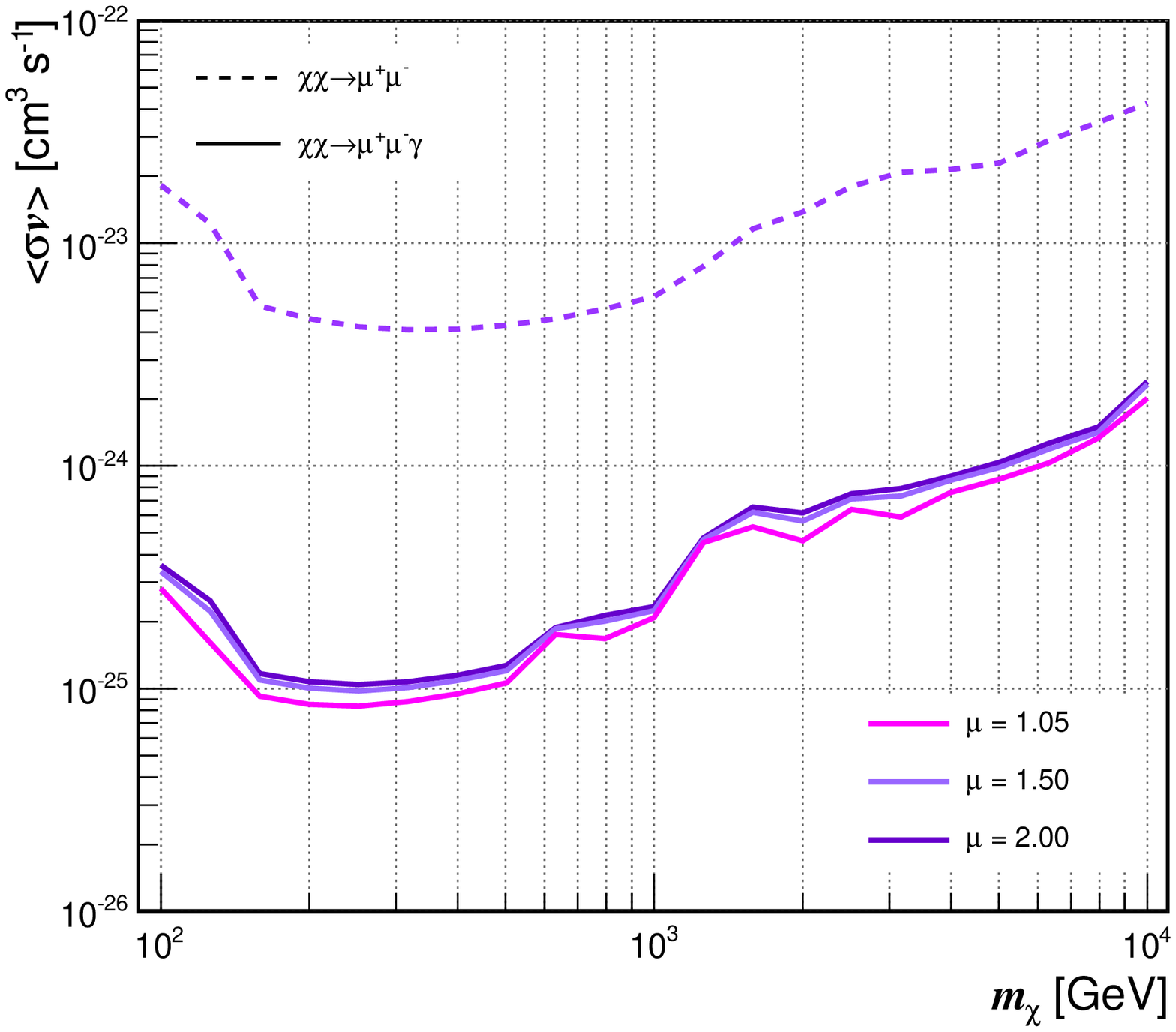}
  \includegraphics[trim=0 15 0 0,clip=true,width=0.45\textwidth]{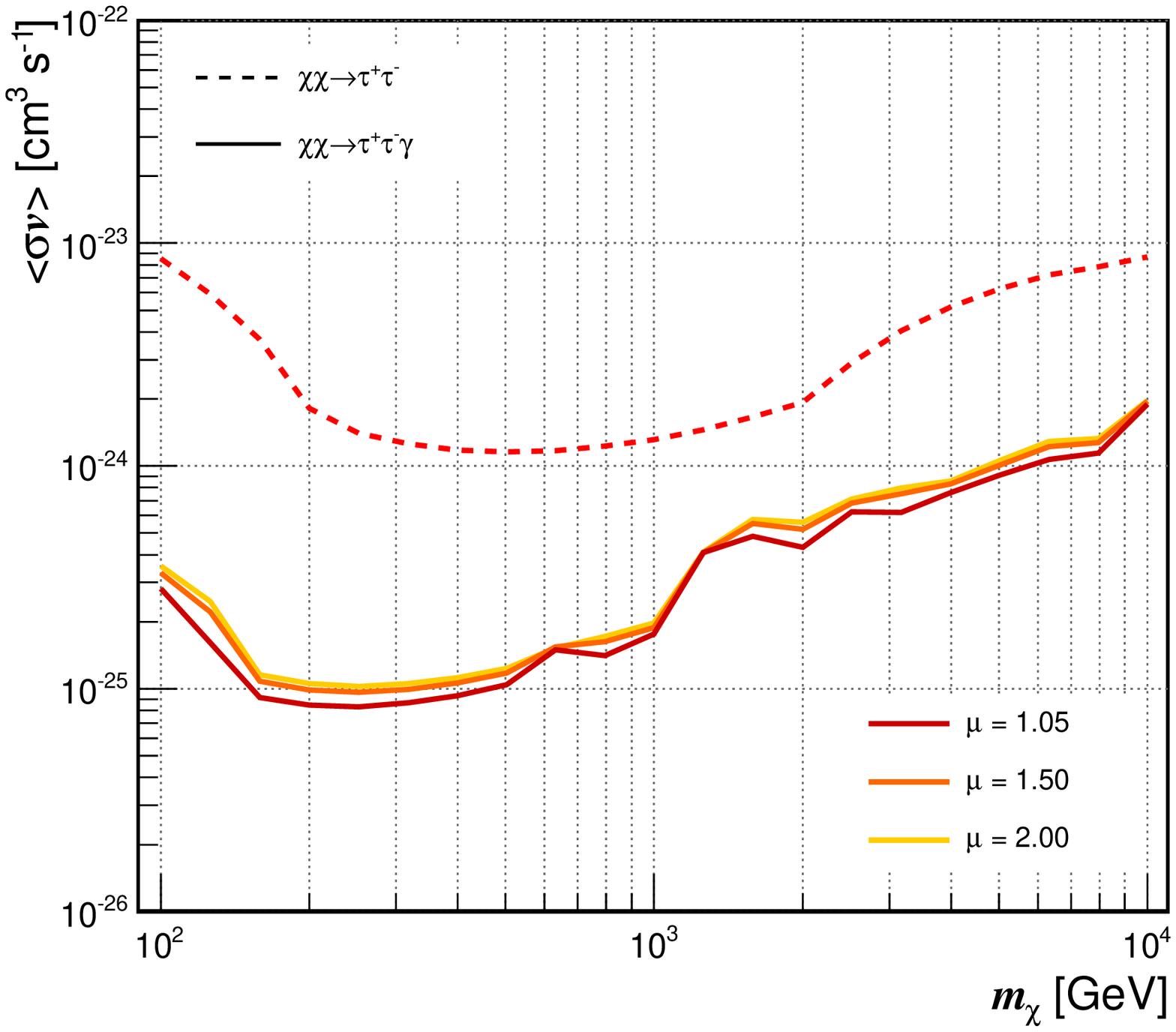}
  \vspace{0pt}
  \caption{Upper limits on $\sv$ for $\mu^{+}\mu^{-}\gamma$
    (\bt{left}) and $\tau^{+}\tau^{-}\gamma$ (\bt{right}) final
    states, as a function of $\m$, for different values of the mass
    splitting parameter $\mu$. Also shown are the exclusion curves for
    the annihilation without the VIB contribution (dashed line).}
  \label{Fig9}
\end{figure}
\begin{figure}[t!]
  \centering
  \vspace{-5pt}\hspace{5pt}
  \includegraphics[trim=0 15 0 0,clip=true,width=0.45\textwidth]{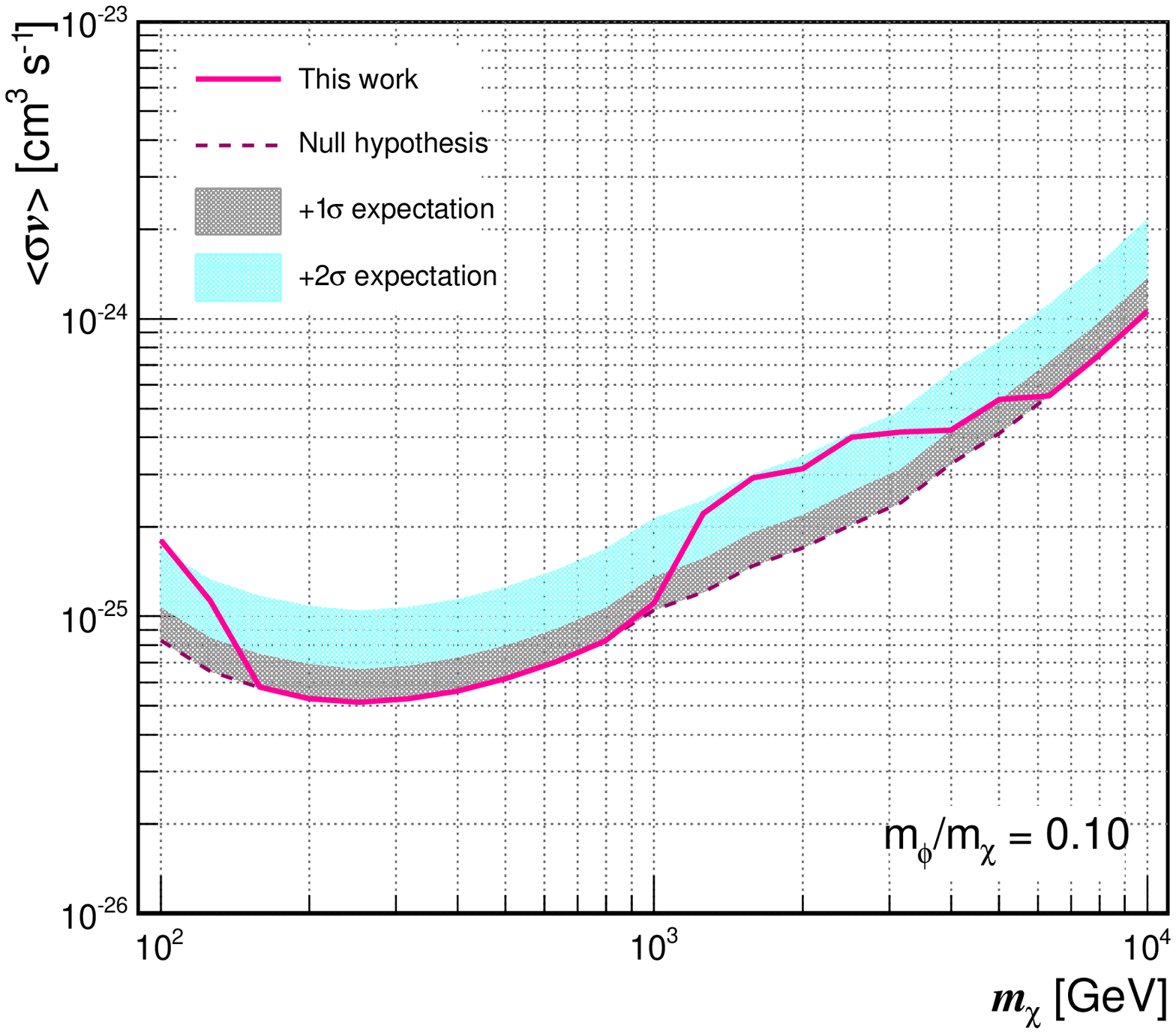}
  \includegraphics[trim=0 15 0 0,clip=true,width=0.45\textwidth]{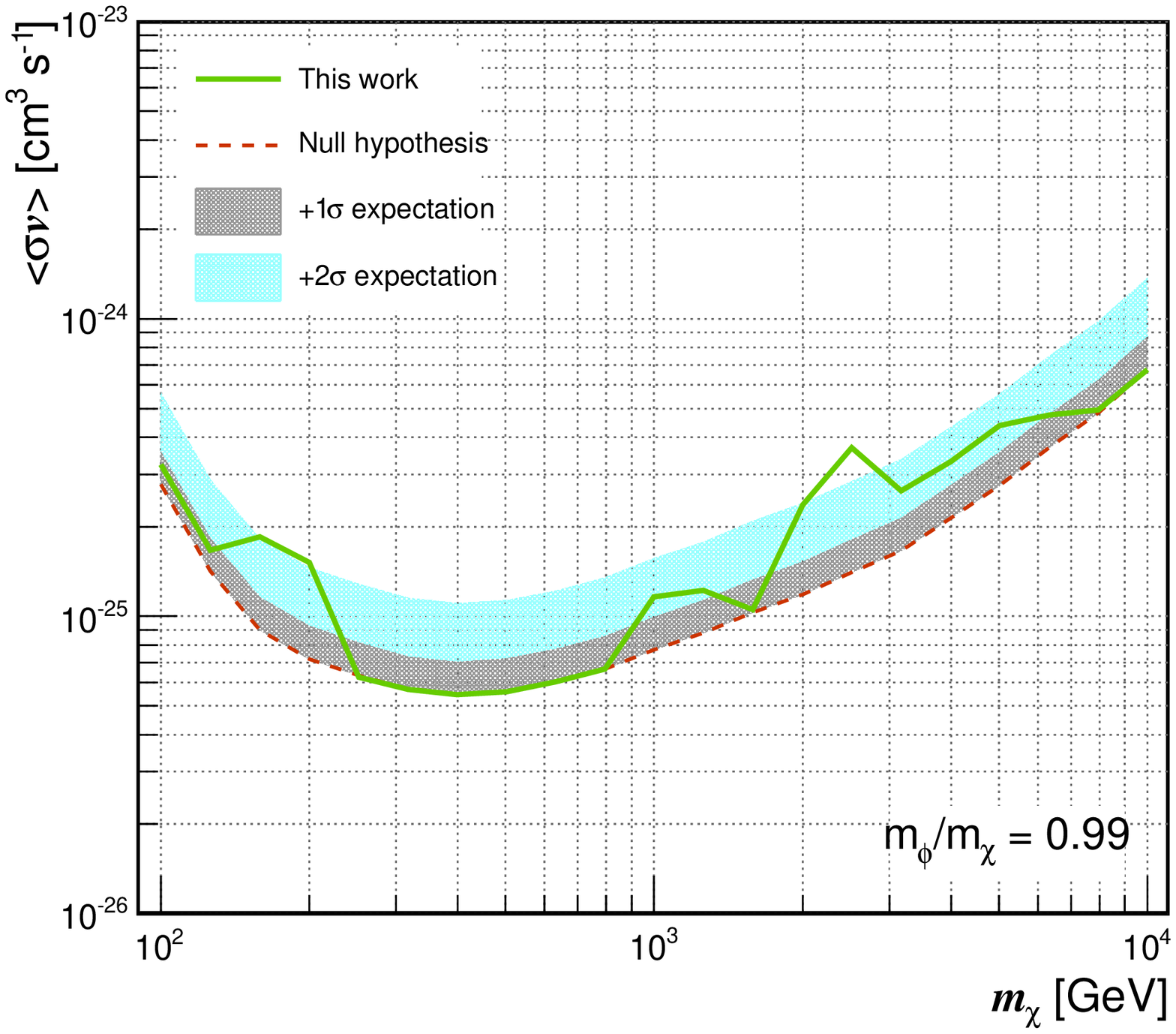}
  \vspace{0pt}
  \caption{Upper
    limits on $\sv$ for wide- ($m_{\phi}/\m$ = 0.1, \bt{left}) and
    narrow ($m_{\phi}/\m$ = 0.99, \bt{right})-box scenarios, as a
    function of $\m$, from the Segue~1 observations with MAGIC (solid 
    line), and as expected for the case of no signal (dashed line), as
    well as for the signal of 1$\sigma$ or 2$\sigma$ significance
    (gray and light blue shaded areas, respectively).}
  \label{Fig10}
  \centering
  \vspace{0pt}\hspace{5pt}
  \includegraphics[trim=0 15 0 0,clip=true,width=0.45\textwidth]{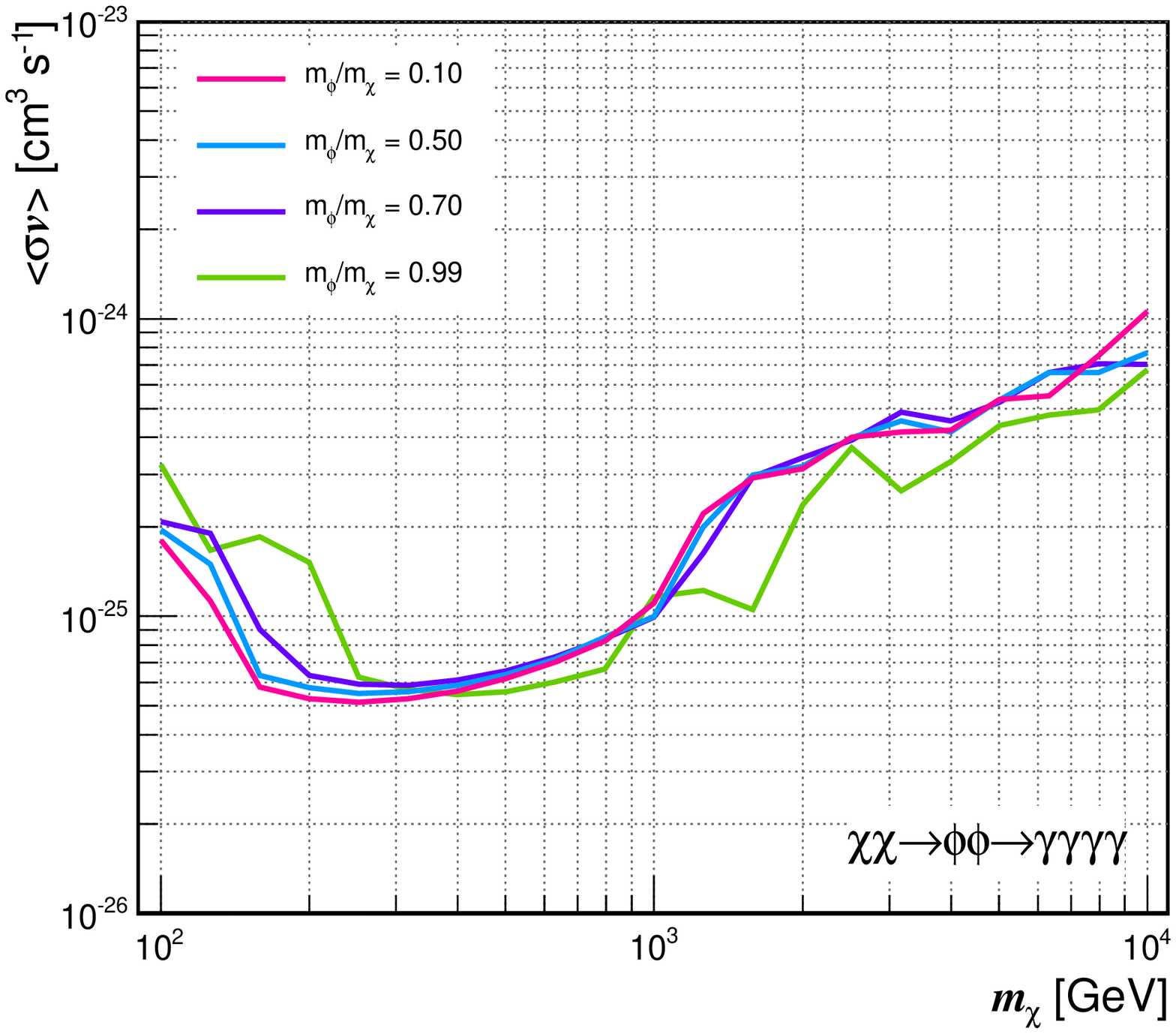}
  \includegraphics[trim=0 15 0 0,clip=true,width=0.45\textwidth]{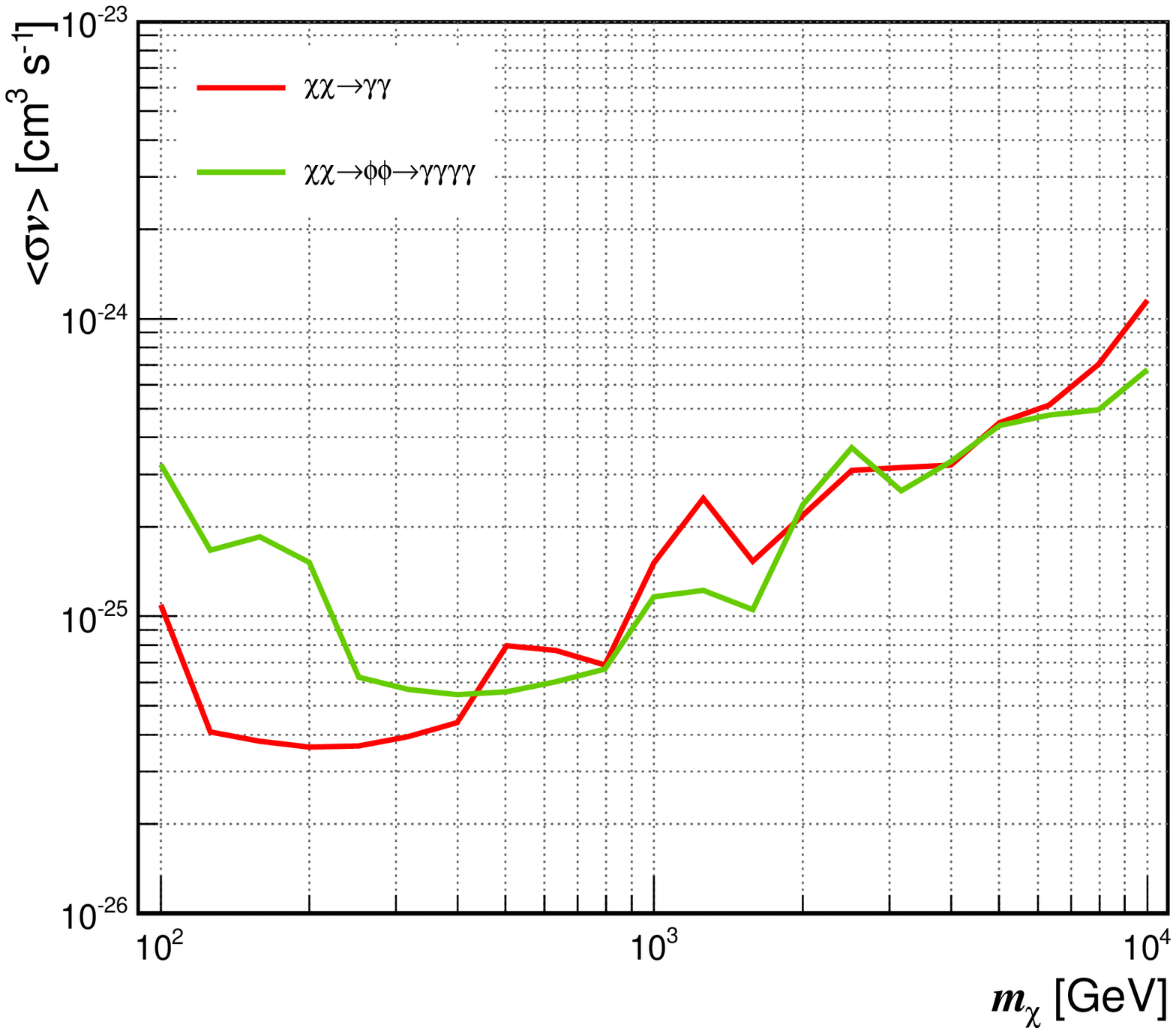}
  \vspace{0pt}
  \caption{Comparison of upper limits on $\sv$, as a function of $\m$,
    from the Segue~1 observations with MAGIC, for different ratios of
    scalar and DM particle masses (\bt{left}), and of the narrow box
    scenario ($m_{\phi}/\m$ = 0.99) with a monochromatic gamma-ray
    line (\bt{right}).}
  \label{Fig11}
\end{figure}
\subsection{Virtual internal bremsstrahlung contribution}
\label{R:VIB}

Here we consider the annihilation into leptonic channels with VIB
contribution, and set limits on the total 3-body annihilation cross
section. Figure \ref{Fig8} shows the $\sv$ upper limits for the
$\mu^{+}\mu^{-}\gamma$ channel, for different values of the mass
splitting parameter, chosen such that the VIB contribution is
significant with respect to the continuous one ($\mu$ = 1.05 and 2.00,
respectively). No positive detection can be claimed. 

The comparison of these exclusion curves is better illustrated in
figure \ref{Fig9}-left: we see that, for several mass-splitting values
($\mu$ = 1.05, 1.50 and 2.00), the obtained limits are rather similar.
This can be understood considering how the spectral shape of the VIB
signal is practically independent of $\mu$ for such small
mass-splitting parameter values. Still, it can be noticed that the
most degenerate case, $\mu$ = 1.05, provides the strongest limit, of
$\sv \sim 8.4\times 10^{-\sq{26}} \svu$ for $\m\sim$ 250 GeV. For
illustration purposes, on the same plot we also show the $\sv$
constraints calculated for the annihilation into the muon final state,
but without the VIB contribution. In this particular case, the
presence of VIB photons in the spectrum leads to almost two orders of
magnitude more stringent bounds.

Analogous conclusions are reached for annihilation into the 
$\tau^{+}\tau^{-}\gamma$ final state (figure \ref{Fig9}-right). The
strongest limit in this case corresponds to $\sv \sim
$ 8.3$\times$10$^{-{\sq{26}}} \svu$, for $\m \sim$ 250 GeV and $\mu$ =
1.05. The relative contribution of the VIB 'bump' in this case is less
significant than for the $\mu^{+}\mu^{-}\gamma$ channel, as the continuous
gamma-ray spectrum of $\tau$ leptons is of harder spectral slope. 


\subsection{Gamma-ray boxes}
\label{R:Box}

Here we consider the case of DM annihilation resulting in the
production of four photons
($\chi\chi\rightarrow\phi\phi\rightarrow\gamma\gamma\gamma\gamma$). As
for the DM decay scenarios, they are not covered here, given that the
transformation of an upper limit on $\sv$ to a lower limit on $\td$ is
trivial, by making the replacement $\sv/$8$\pi\m^{\sq{2}}$
$\rightarrow$ 1$/\m\td$ in eq. (\ref{eq2}) and $\m \rightarrow \m/$2
in eq. (\ref{eq3}).

Figure \ref{Fig10} shows the $\sv$ exclusion curves for extreme
degeneracies when $m_{\phi}/\m=0.1$ and $m_{\phi}/\m= 0.99$. In both
cases, the strongest constraints are similar, of the order $\sv\sim$
(5.1--5.4)$\times$10$^{-\sq{26}} \svu$, for $\m\sim$ 250 GeV and
$\sim$400 GeV, respectively. 

Figure \ref{Fig11}-left shows upper limits on $\sv$ for the already
mentioned extreme values of $m_{\phi}/\m$ (= 0.10, 0.99), as well as
for some intermediate cases ($m_{\phi}/\m$ = 0.50, 0.70). As can be
seen, with exception of the most narrow box scenario, all constraints
are essentially the same, and only a factor few weaker than for the
most degenerate configuration. This is understood given that the wide
boxes compensate the dimmer amplitudes (with respect to the
$m_{\phi}\approx\m$ cases) by extending to higher energies, where
the sensitivity of the telescopes is better.

For a more general view on the importance of box-shaped features,
figure \ref{Fig11}-right shows the upper limits on $\sv$ from the most
degenerate box model ($m_{\phi}/\m=$ 0.99) and from the line searches
previously shown (figure \ref{Fig7}). The obtained bounds are of the same
order of magnitude, although the direct comparison between the two
exclusion curves is not immediate: the line is centered at $E_{\gamma}
= \m$ and is normalized for 2 photons per annihilation, while the
box-shaped feature is centered at $E_{\gamma} = \m/2$ and is
normalized for 4 photons. This is reflected as a shift of the
exclusion curves in $x$ and $y$ coordinates.

For a more comprehensive overview, the most constraining bounds for
all of the final state channels presented in this section are
summarized in table \ref{Table2}.
\begin{table}[t!]
  \begin{center}
    \setlength{\extrarowheight}{1pt}
    {\small
      \begin{tabulary}{1.\textwidth}{ C |C C C C }
        \toprule[0.1em]
        \multicolumn{5}{c}{\bt{SECONDARY PHOTONS}}\\\midrule[0.1em]
        \multirow{7}{*}{\begin{sideways}{\bt{ANN~~~~}}\end{sideways}} &Final state & $\m$ [TeV] & $\sv$ [$\svu$] & Most constraining limit from dSphs \\\arrayrulecolor{gray}\hhline{~----}
        & $b\bar{b}$ & 1.78  & 5.44$\times$10$^{-{\sq{24}}}$ & $\m >$ 3.30 TeV \\
        & $t\bar{t}$ & 2.16 & 7.67$\times$10$^{-{\sq{24}}}$ &  no
        comparison possible\\
        & $\mu^{+}\mu^{-}$ & 0.33 & 4.09$\times$10$^{-{\sq{24}}}$ & $\m > $ 0.29 TeV \\
        & $\tau^{+}\tau^{-}$ & 0.50 & 1.16$\times$10$^{-{\sq{24}}}$ & $\m > $ 0.55 TeV\\
        & $W^{+}W^{-}$ & 1.35 & 4.52$\times$10$^{-{\sq{24}}}$& $\m >$ 2.77 TeV\\
        & $ZZ~$ & 1.53 & 4.94$\times$10$^{-{\sq{24}}}$&  no comparison
        possible \\\arrayrulecolor{black}\midrule[0.1em]
        \multirow{7}{*}{\begin{sideways}{\bt{DEC~~~~}}\end{sideways}}
        &Final state & $\m$ [TeV] & $\td$ [s] & Most constraining
        limit from dSphs \\\arrayrulecolor{gray}\hhline{~----}
        & $b\bar{b}$ & 20 & 2.54$\times$10$^{\sq{25}} $ & full range\\
        & $t\bar{t}$ & 20 & 1.97$\times$10$^{\sq{25}} $ &  no
        comparison possible\\
        & $\mu^{+}\mu^{-}$ & 20 & 5.89$\times$10$^{\sq{24}} $ & full range\\
        & $\tau^{+}\tau^{-}$ & 20 & 2.89$\times$10$^{\sq{25}} $ & full range\\
        & $W^{+}W^{-}$ & 20 & 2.45$\times$10$^{\sq{25}} $ & full range\\
        & $ZZ~$ & 20 & 2.58$\times$10$^{\sq{25}} $ &  no comparison possible\\
        \arrayrulecolor{black}\bottomrule[0.1em]
  
        \multicolumn{5}{c}{\bt{MONOCHROMATIC LINE}}\\\midrule[0.1em]
        \multirow{3}{*}{\begin{sideways}{\bt{ANN~~~}}\end{sideways}} &
        Final state & & $\m$ [TeV] & $\sv$ [$\svu$] \\\arrayrulecolor{gray}\hhline{~----}
        & $\gamma\gamma$ & & 0.21  & 3.61$\times$10$^{-{\sq{26}}}$ \\
        & $Z\gamma$ & & 0.25 & 7.82$\times$10$^{-{\sq{26}}}$ \\\arrayrulecolor{black}\midrule[0.1em]
        \multirow{2}{*}{\begin{sideways}{\bt{DEC~}}\end{sideways}} &
        Final state & & $\m$ [TeV] & $\td$ [s] \\\arrayrulecolor{gray}\hhline{~----}
        & $\gamma\nu$ & & 7.93 & 1.55$\times$10$^{\sq{26}} $ \\
        \arrayrulecolor{black}\bottomrule[0.10em]

        \multicolumn{5}{c}{\bt{VIRTUAL INTERNAL BREMSSTRAHLUNG}}\\\midrule[0.1em]
        \multirow{7}{*}{\begin{sideways}{\bt{ANN~~~~~}}\end{sideways}} & Final state & $\mu$ & $\m$ [TeV] & $\sv$ [$\svu$] \\\arrayrulecolor{gray}\hhline{~----}
        &\multirow{3}{*}{$\mu^{+}\mu^{-}$($\gamma$)} & 1.05  & 0.25 & 8.38$\times$10$^{-{\sq{26}}}$\\
        & & 1.50  & 0.25 & 9.78$\times$10$^{-{\sq{26}}}$\\
        & & 2.00  & 0.25 & 1.04$\times$10$^{-{\sq{25}}}$\\\hhline{~~---}
        &\multirow{3}{*}{$\tau^{+}\tau^{-}$($\gamma$)} & 1.05 & 0.25  & 8.28$\times$10$^{-{\sq{26}}}$\\
        & & 1.50  & 0.25 & 9.62$\times$10$^{-{\sq{26}}}$\\
        & & 2.00  & 0.25 & 1.02$\times$10$^{-{\sq{25}}}$\\
        \arrayrulecolor{black}\bottomrule[0.10em]
 
        \multicolumn{5}{c}{\bt{GAMMA-RAY BOXES}}\\\midrule[0.1em]
        \multirow{4}{*}{\begin{sideways}{\bt{ANN~~~~}}\end{sideways}} & Final state & $m_{\phi}/\m$ & $\m$ [TeV] & $\sv$ [$\svu$] \\\arrayrulecolor{gray}\hhline{~----}
        & & 0.10  & 0.25 & 5.13$\times$10$^{-{\sq{26}}}$\\
        & $\gamma\gamma\gamma\gamma$ & 0.50  & 0.25 & 5.51$\times$10$^{-{\sq{26}}}$\\
        & & 0.99  & 0.40 & 5.44$\times$10$^{-{\sq{26}}}$\\
        \arrayrulecolor{black}\bottomrule[0.15em]
      \end{tabulary}
    }
    \vspace{-10pt}
  \end{center}
  \caption[Summary of the results from this work.]
  {Summary of the strongest limits and corresponding 
    $\m$, obtained from the Segue~1 observations with 
    MAGIC, for various final states from DM 
    annihilation (ANN) and decay (DEC). When applicable, 
    it is stated for which range of considered $\m$ these 
    limits become the most constraining from dSph 
    observations, among the published results.}
\label{Table2}
\end{table}
\section{Discussion}
\label{D}

In this section, we discuss the advantages brought by the exploitation
of the full likelihood analysis method, compare our results with other
relevant experimental constraints and link them to the expectation
from theoretical models.

\subsection{Sensitivity gain from the full likelihood method}
\label{D:FL}

There are two aspects of the full likelihood analysis applied in this
work that carry advantages for DM searches with IACTs: $i)$ a
sensitivity improvement due to the use of specific spectral signatures
--- such as those coming from DM annihilation and decay, and, $ii)$
the combination of results from different data samples, e.g. obtained
under different experimental conditions, becomes a trivial operation.

To quantify the improvement in sensitivity, we compute the
\emph{improvement factors} as defined in \cite{I_FL}, i.e. the average
ratio of the widths of the confidence intervals for the signal
intensity, calculated by the full likelihood and conventional analysis
methods, respectively, assuming a common CL. The confidence intervals
are computed using fast Monte Carlo simulations of background events, with
the same statistics and PDF as in the actual experimental conditions.

\begin{table}[t!]
  \begin{center}
    \setlength{\extrarowheight}{1pt}
    \begin{tabulary}{0.99\textwidth}{C C C C C C C}
      \toprule[0.1em]
      $\m$ & \hspace{0.05cm} & \multicolumn{2}{c}{$b\bar{b}$} & \hspace{0.05cm} & \multicolumn{2}{c}{$\tau^{+}\tau^{-}$}\\\cmidrule{3-7}
      $[$GeV$]$   &  \hspace{0.05cm} &  no opt. & opt. &\hspace{0.05cm} &  no opt. & opt. \\\midrule[0.09em]
      100 & \hspace{0.05cm} & 2.6 & 2.6 & \hspace{0.05cm} & 2.1 & 2.1 \\
      1000 & \hspace{0.05cm} & 1.9 & 1.9 & \hspace{0.05cm} & 3.1 & 1.7 \\ 
      10000 & \hspace{0.05cm} & 3.3 & 1.9 & \hspace{0.05cm} & 13.9 & 2.4  \\
      \bottomrule[0.1em]
      \end{tabulary}
  \end{center}
  \caption{Sensitivity improvement factors 
    for the $b\bar{b}$ and $\tau^{+}\tau^{-}$ 
    annihilation channels from the use of 
    the full likelihood method over the 
    conventional one, when the latter is calculated for the full
    (no opt.) or optimized (opt.) energy integration range.}
  \label{Table3}
\end{table}
\begin{figure}[t!]
  \centering
  \includegraphics[trim=0 15 0 0,clip=true,width=0.55\textwidth]{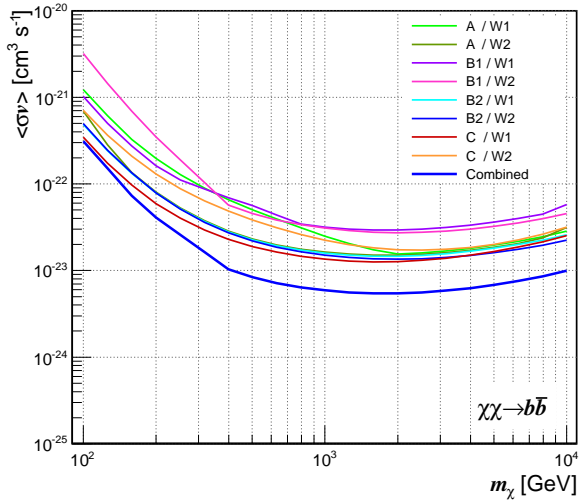}
  \vspace{0pt}
  \caption{Upper limits on $\sv$ for the $b\bar{b}$ annihilation channel,
    from individual wobble positions and different Segue~1
    observational periods. Also shown is the limit from the combined
    likelihood analysis.}
  \label{Fig12}
\end{figure}
Table \ref{Table3} shows the improvement factors obtained for signal
from the $b\bar{b}$ and $\tau^{+}\tau^{-}$ annihilation channels and
$\m$ of 100 GeV, 1 TeV and 10 TeV. The calculations with the
conventional method are done over both the optimized and the full
energy range. The obtained improvement factor values are in agreement
with the predictions made in \cite{I_FL}, and imply a very significant
boost in the sensitivity for DM searches: an improvement factor of $f$
is equivalent to $f^{\sq{2}}$ times more observation time.

Compared to our previous results of DM signals from Segue~1
\cite{I_MAGIC_Segue}, these results represent an overall improvement
of about a factor of 10. This has been accomplished by the increase in
the observation time of a factor $\sim$5.3 (i.e. a factor $\sim$2.3
better sensitivity), the operation of MAGIC as a stereoscopic system
(a factor $\sim$2 better sensitivity with respect to
single-telescope observations \cite{A_performance}), plus the
improvement factor coming from the full likelihood analysis.

Furthermore, the full likelihood method allows a trivial merger of
results obtained with different instruments or from different
observational targets. As discussed in section \ref{O}, our data
sample is divided into four periods with different instrumental
conditions. In addition, for each period we use two different pointing
positions, with slightly different background models. We have built
dedicated individual likelihood functions for each of the eight
different sub-periods, and merged them into a global likelihood
following eq. (5.1) in \cite{I_FL}, for our final results. The limits
on $\sv$ (assuming annihilation into the $b\bar{b}$ with Br = 100\%)
obtained by each of the eight considered sub-samples, compared to the
global limit, are shown in figure \ref{Fig12}.

\subsection{Comparison with the results from other gamma-ray experiments}
\label{D:Exp}

In the previous section we have estimated the $\sv$ upper limits and
$\td$ lower limits for various channels of DM
annihilation/decay. Here, we put those constraints in context and
compare them with the currently most stringent results from other
gamma-ray observatories. 

\subsubsection{Secondary photons from final state Standard Model particles}
\label{D:Exp:Sec}

\begin{figure}[t]
  \centering
  \includegraphics[trim=10 30 30 30,clip=true,width=0.45\textwidth]{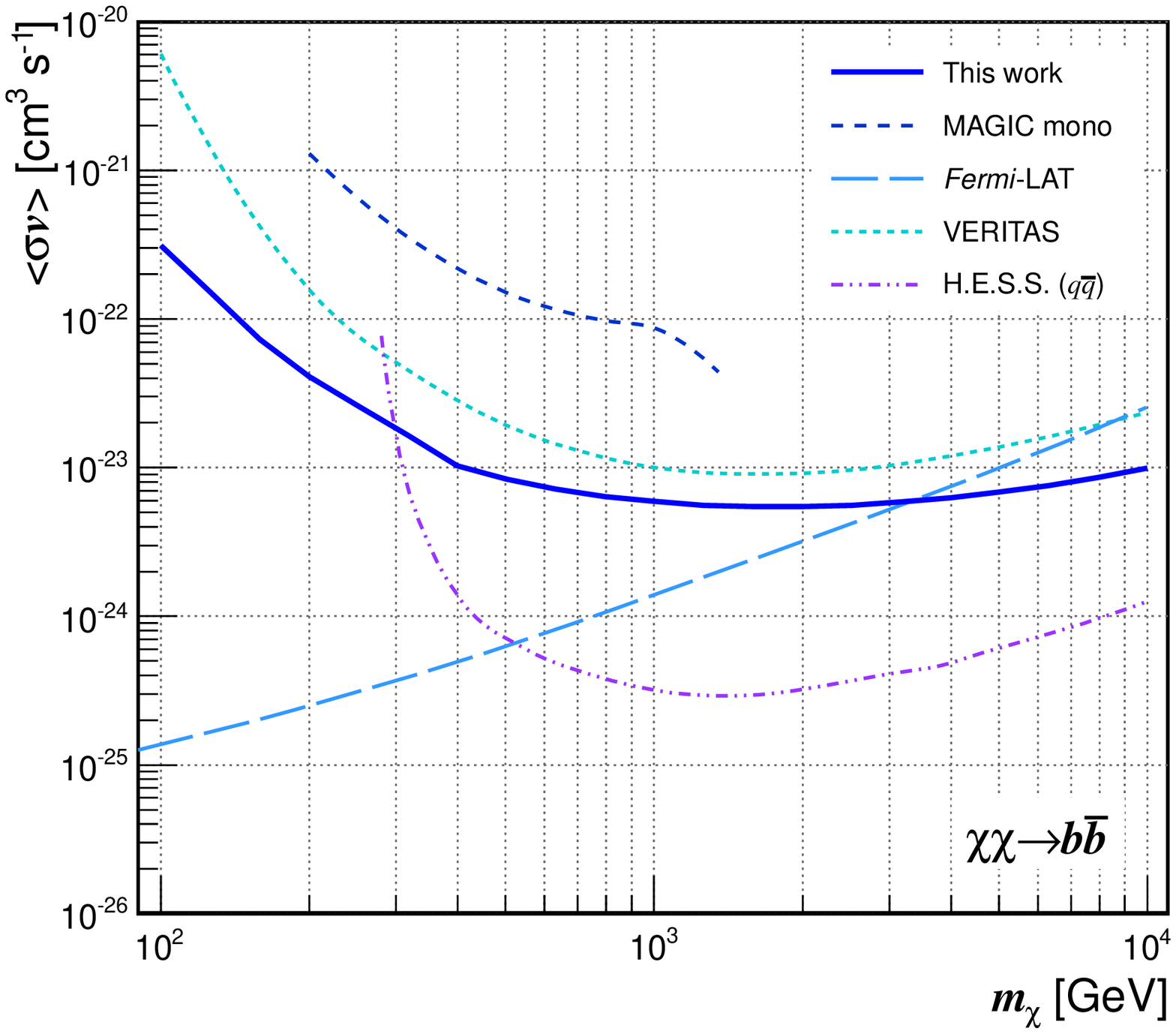}
  \includegraphics[trim=10 30 30 30,clip=true,width=0.45\textwidth]{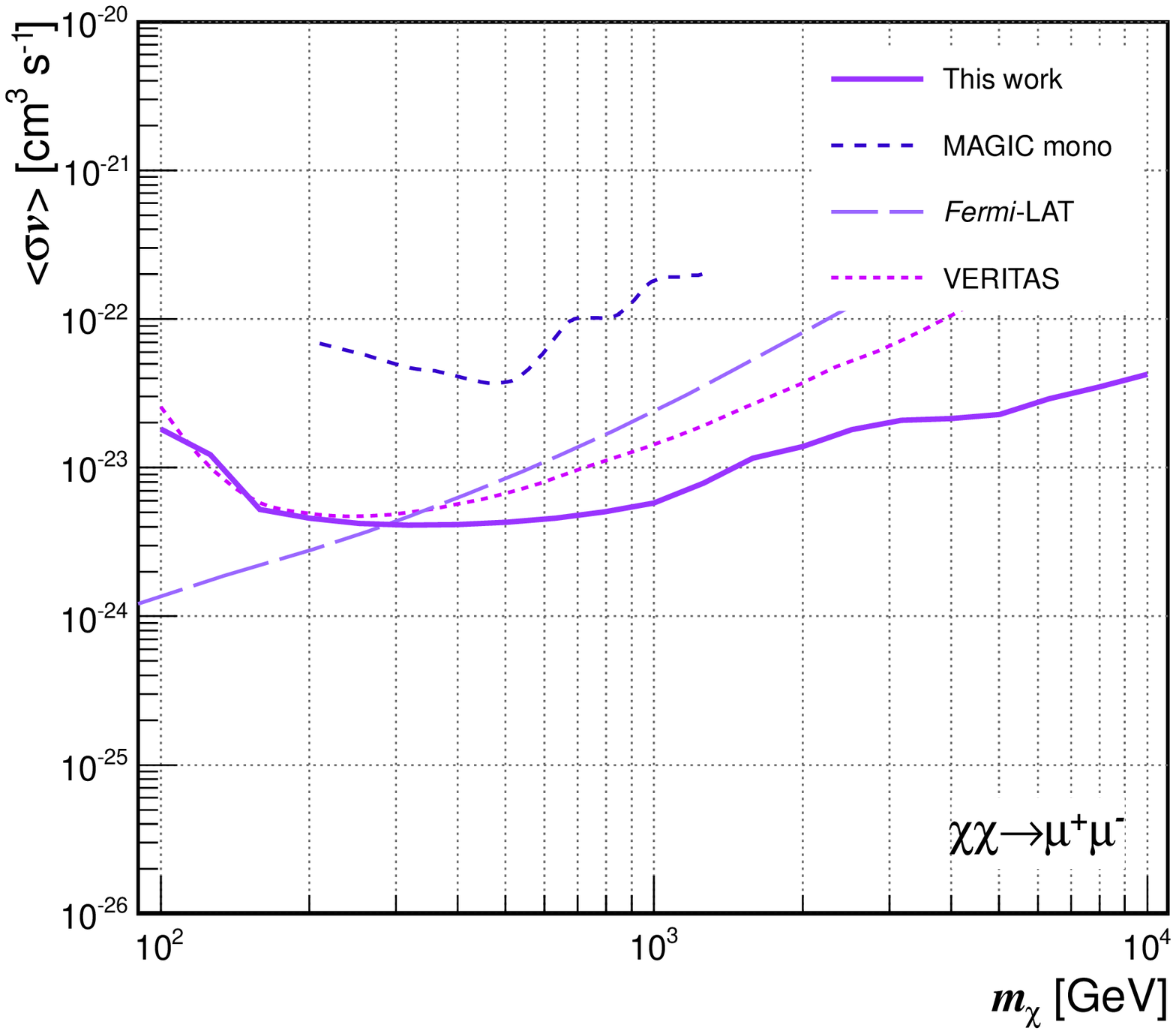}
  \includegraphics[trim=10 30 30 30,clip=true,width=0.45\textwidth]{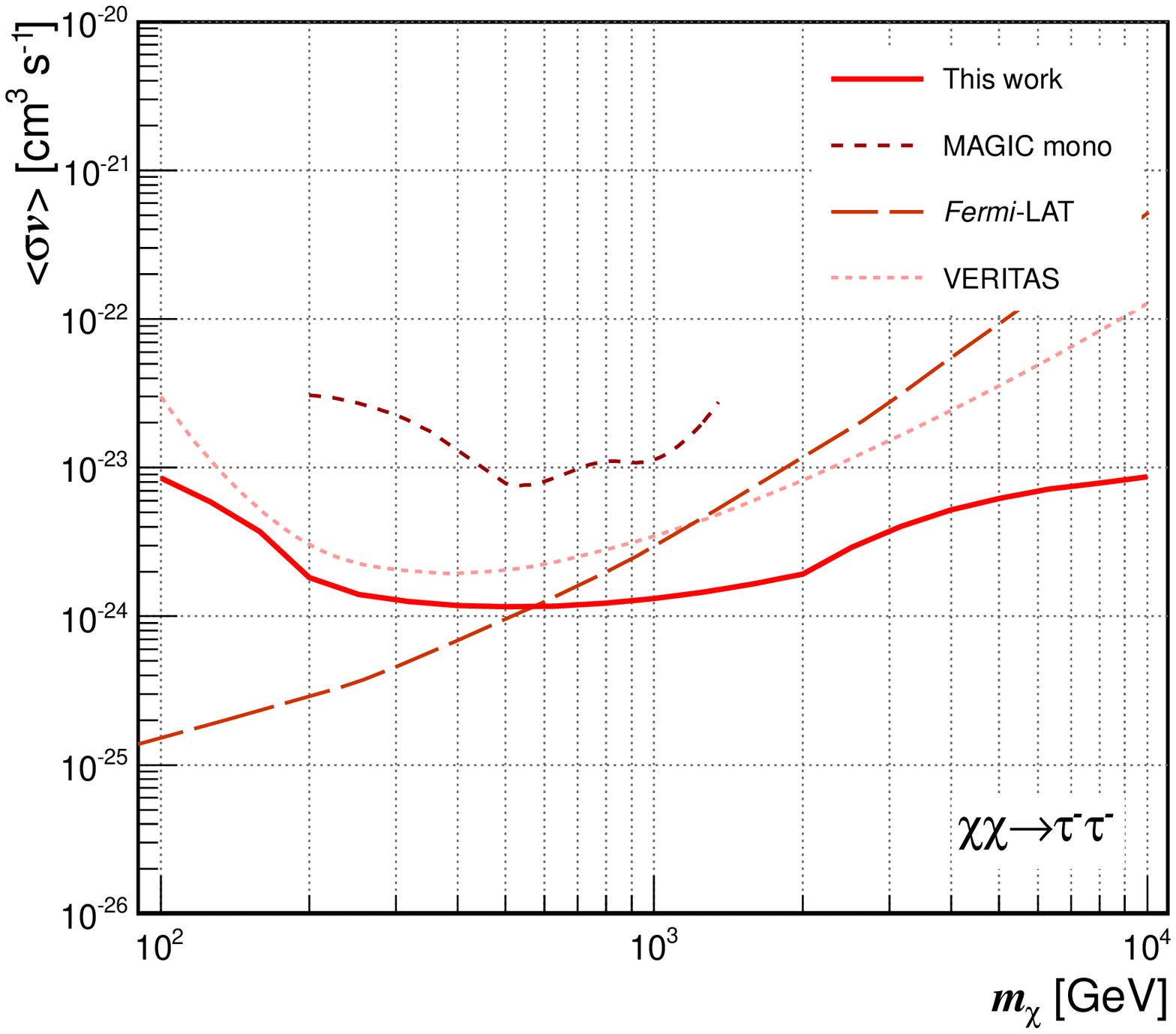}
  \includegraphics[trim=10 30 30 30,clip=true,width=0.45\textwidth]{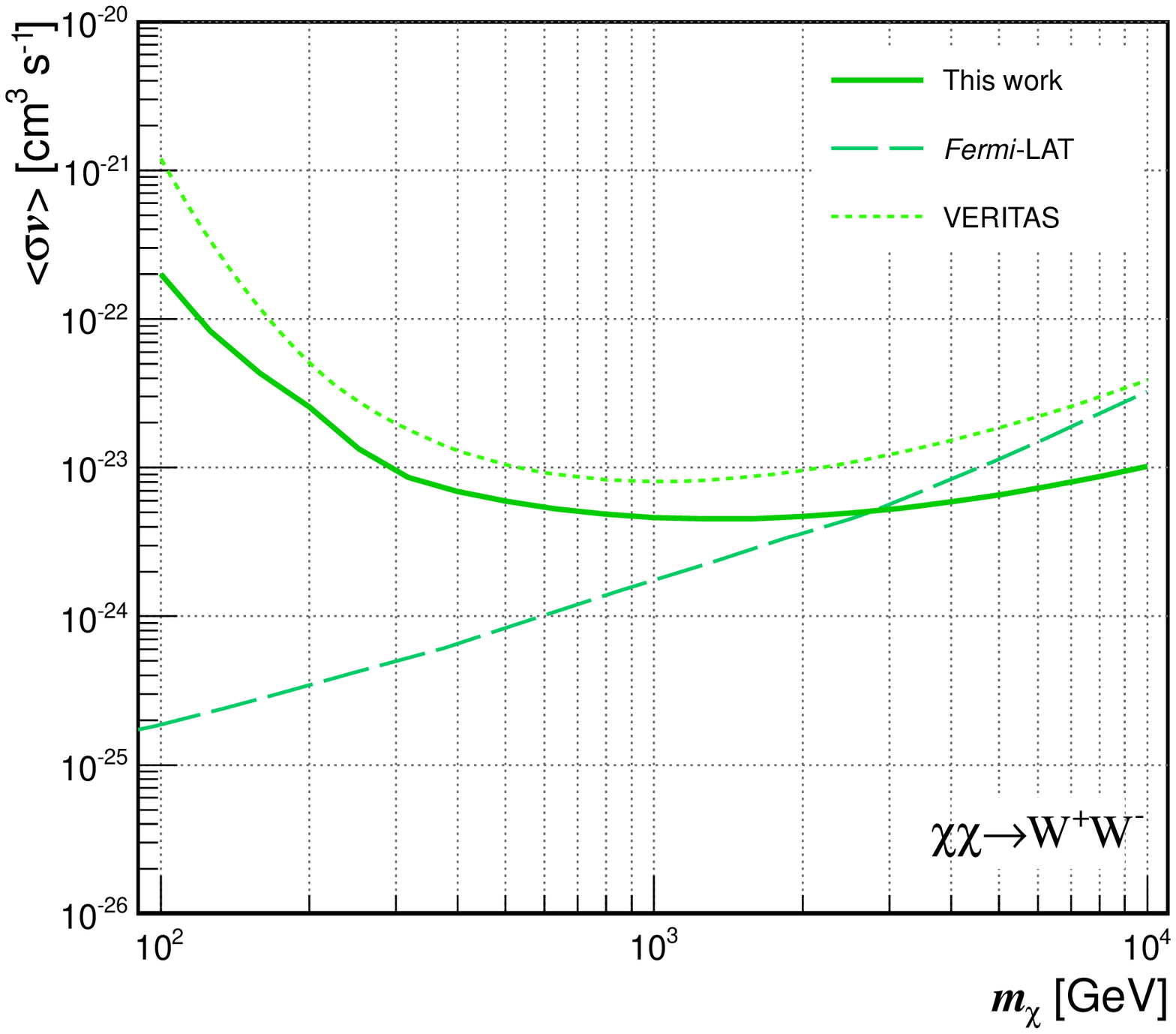}
  \caption{Upper limits on $\sv$ for different 
    final state channels (from top to bottom and left to right):
    $b\bar{b}$, 
    $\mu^{+}\mu^{-}$, $\tau^{+}\tau^{-}$ and $W^{+}W^{-}$, from the Segue
    1 observations with MAGIC (solid lines), compared with the
    exclusion curves from \tit{Fermi}-LAT (long-dashed lines, \cite{I_Fermi_dSph}),
    H.E.S.S. (dot-dashed line, \cite{D_HESS_GC}), VERITAS
    (dotted lines, \cite{I_VERITAS_Segue}) and MAGIC-I (dashed lines, \cite{I_MAGIC_Segue}).}
\label{Fig13}
\end{figure}

\paragraph{Annihilation} 
Figure \ref{Fig13} shows our $\sv$ upper limits from DM annihilation
into the $b\bar{b}$, $\mu^{+}\mu^{-}$, $\tau^{+}\tau^{-}$ and
$W^{+}W^{-}$ channels\footnote{We do not show the limits for the
  $t\bar{t}$ and $ZZ$ final states, as those channels are not
  discussed in the other mentioned studies.}, compared with the
corresponding constraints from $i$) joint analysis of 4 years of
observations of 15 dSphs by \tit{Fermi}-LAT \cite{I_Fermi_dSph}; $ii$)
112 hours of the GC Halo observations with H.E.S.S. (assuming generic
$q\bar{q}$ final state, \cite{D_HESS_GC}); and $iii$) $\sim$48 hours
of the Segue~1 observations with VERITAS \cite{I_VERITAS_Segue}. Note,
however, that the VERITAS results have been questioned in reference
\cite{I_FL}, where it is discussed why the VERITAS limits could be
over-constraining by a factor two or more. Lastly, we also show the
limits from $\sim$30 hours of the Segue~1 observations with MAGIC in
single-telescope mode \cite{I_MAGIC_Segue}.

As seen from figure \ref{Fig13}, for DM particles lighter than few
hundreds GeV (depending on the specific channel), the strongest limits
on $\sv$ come from \tit{Fermi}-LAT\footnote{Note that \tit{Fermi}-LAT
  results include each of the individual uncertainties on the
  astrophysical factors for the considered dSphs as nuisance
  parameters, whereas results by IACTs do not (since those limits
  scale trivially with $J$).}; for higher $\m$ values, the most
constraining bounds are derived from the H.E.S.S. observations of the
GC halo. We note, however, that the H.E.S.S. result heavily depends on
the assumed DM profile, as it is sensitive to the difference in the
DM-induced gamma-ray fluxes between the signal and background region,
rather than to the absolute flux. In fact, by using a
Navarro-Frenk-White (NFW) profile \cite{D_NFW} instead of the Einasto
one, the H.E.S.S. limit becomes a factor of $\sim$2 less constraining,
or even weaker for very cored profiles with similar fluxes from the
relatively close \st{ON} and \st{OFF} regions (figure 1 in
\cite{D_HESS_GC}). This is particularly relevant considering possible
uncertainties as, e.g., the effect of baryonic contraction in the GC
that could have an important effect on the final DM profile
\cite{D_DMprofile_baryonic}.

Our $\sv$ limits from 157.9 hours of the Segue~1 observations with
MAGIC are the strongest limits from the dSphs observations with IACTs,
and, for certain channels, also the most constraining bounds from dSph
observations in general (table \ref{Table2}). For annihilation into
the $b\bar{b}$ and $W^{+}W^{-}$ final states, the MAGIC constraints
complement those of the \tit{Fermi}-LAT dSphs observations for $\m >$
3.3 TeV and 2.8 TeV, respectively. For the leptonic channels, on the other
hand, our limits become the most constraining above $\m\sim$ 300 GeV
and $\sim$550 GeV, for annihilation into $\mu^{+}\mu^{-}$ and
$\tau^{+}\tau^{-}$, respectively.

\paragraph{Decay}
Over the last couple of years, a lot of attention has been given to
the decaying DM as a possible explanation of the flux excesses of
high-energy positrons and electrons measured by PAMELA
\cite{D_PAMELA}, \tit{Fermi}-LAT \cite{D_Fermi_electrons}, H.E.S.S
\cite{D_HESS_electrons} and AMS-02 \cite{D_AMS}. The needed DM
particle lifetime in such a case, $\td >$ 10$^{\sq{26}}$ s, is much
longer than the age of the Universe, so that the slow decay does not
significantly reduce the overall DM abundance and, therefore, there is
no contradiction with the astrophysical and cosmological observations.

Although DM decay has been disfavoured as the cause of the observed
flux excesses (compared to astrophysical explanations; see, e.g.,
\cite{D_Decay_Astro}), this scenario is an interesting subject by
itself: e.g., if the DM particle is a gravitino, and small R-parity
violation is allowed, the appropriate relic density through the
thermal production is achieved, naturally leading to a cosmological
history consistent with thermal leptogenesis and primordial
nucleosynthesis \cite{D_DecayTime}.
\begin{figure}[t]
  \centering
 \includegraphics[trim=10 30 30 30,clip=true,width=0.45\textwidth]{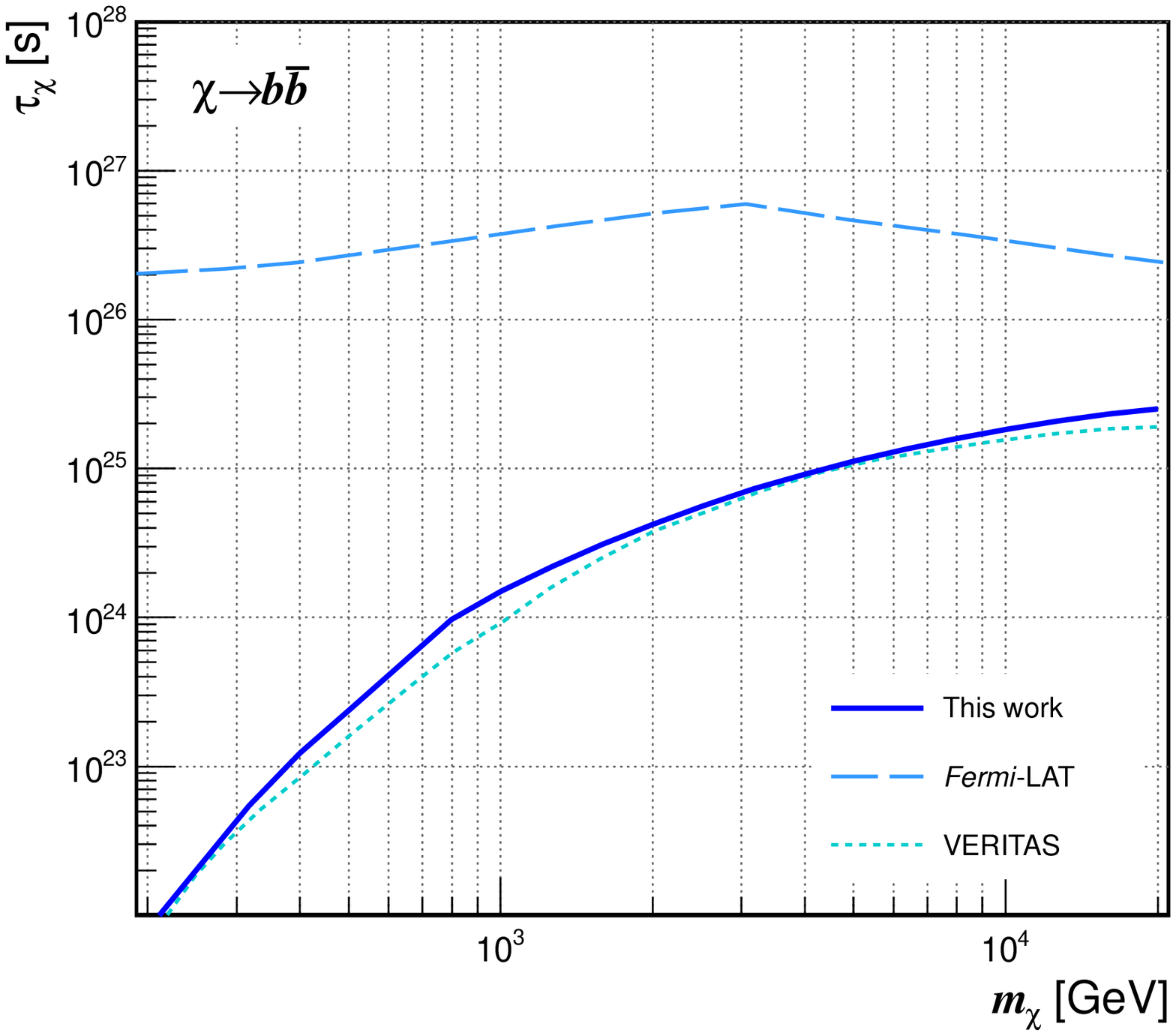}
  \includegraphics[trim=10 30 30 30,clip=true,width=0.45\textwidth]{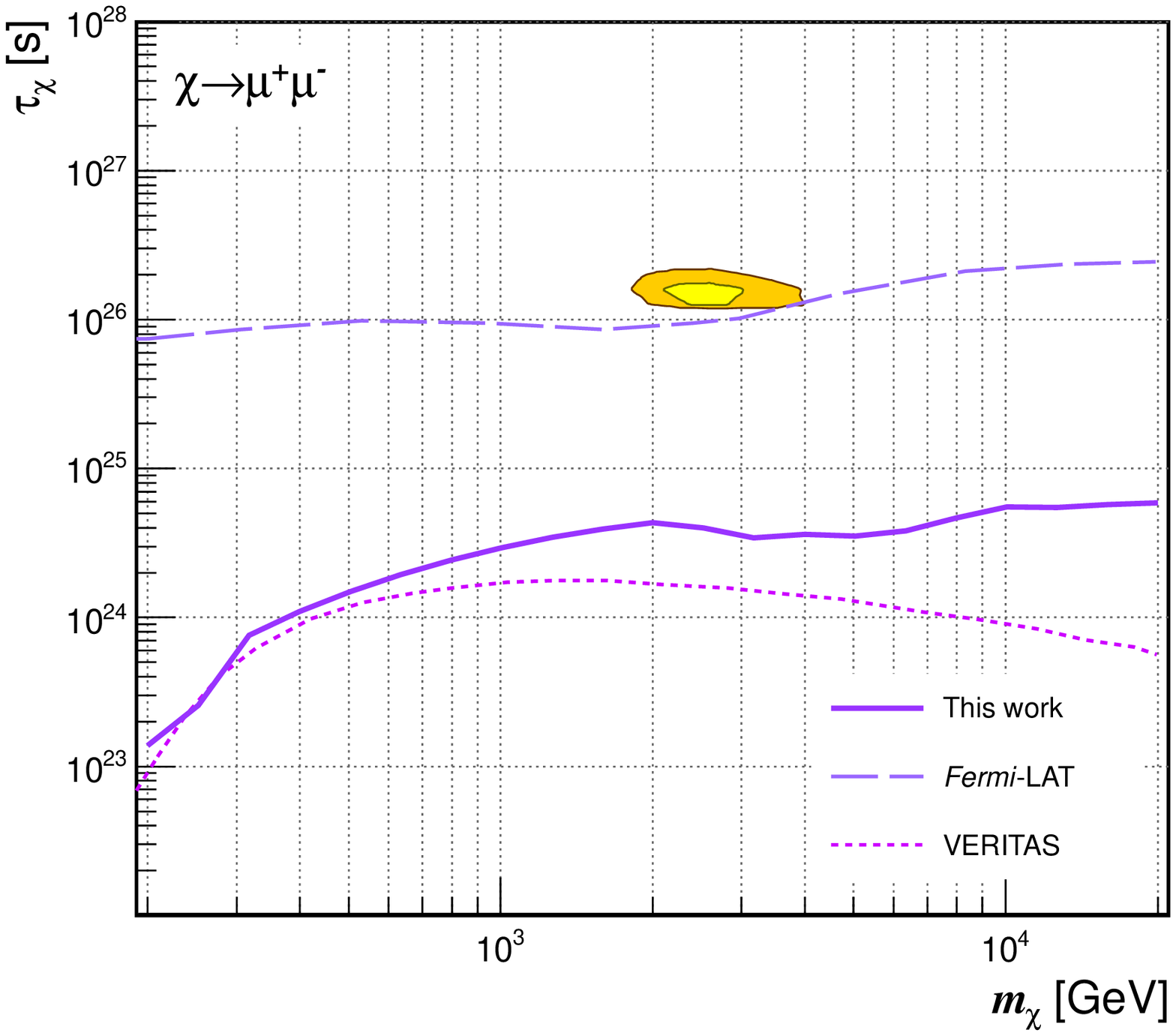}
  \includegraphics[trim=10 30 30 30,clip=true,width=0.45\textwidth]{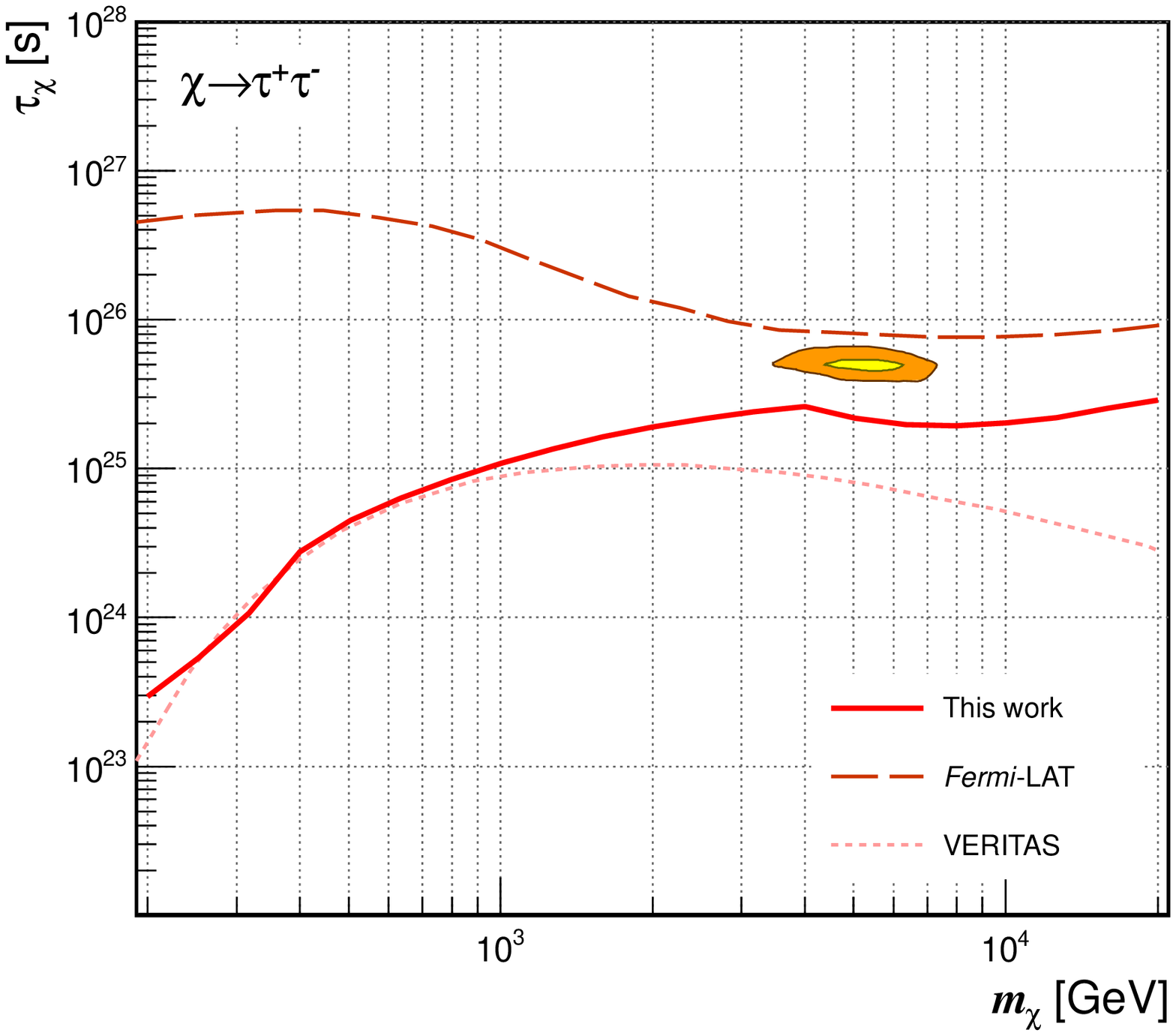}
  \includegraphics[trim=10 30 30 30,clip=true,width=0.45\textwidth]{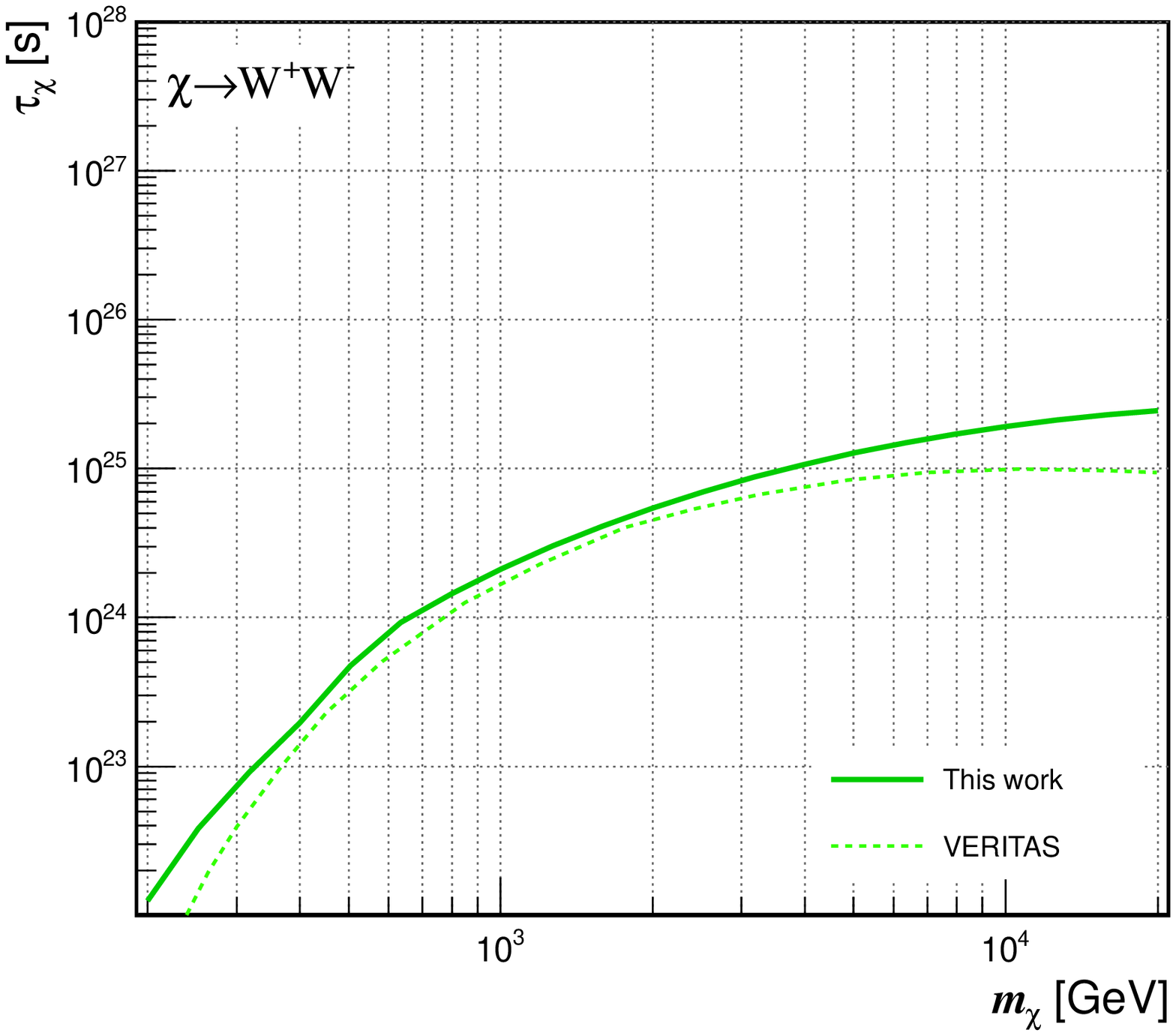}
  \caption{Lower limits on $\td$ for different final state channels
    (from top to bottom and left to right): $b\bar{b}$,
    $\mu^{+}\mu^{-}$, $\tau^{+}\tau^{-}$ and $W^{+}W^{-}$, from the
    Segue~1 observations with MAGIC (solid lines), compared with the
    exclusion curves from \tit{Fermi}-LAT (long-dashed lines,
    \cite{D_Fermi_Diffuse}), and VERITAS (dotted lines,
    \cite{I_VERITAS_Segue}). For the leptonic channels,
    $\mu^{+}\mu^{-}$ and $\tau^{+}\tau^{-}$, also shown are the
    regions that allow fit to the PAMELA, \tit{Fermi}-LAT and
    H.E.S.S. cosmic-ray excess measurements, with 95\% and 99.99\% CL
    (orange and red regions, respectively; taken from
    \cite{D_Pamela_bestFit}).}
\label{Fig14}
\end{figure}

The currently strongest constraints on $\td$ from gamma-ray
observatories are derived from the \tit{Fermi}-LAT diffuse gamma-ray
data: the 2-year long measurements at energies between $\sim$1 and 400
GeV \cite{D_Fermi_Diffuse} exclude decaying DM with lifetimes shorter
than 10$^{\sq{25}}$--10$^{\sq{26}}$ s (depending on the channel) for
$\m$ between 10 GeV and 10 TeV. VERITAS also provides lower limits to
$\td$ from $\sim$48 hours of Segue~1 observations
\cite{I_VERITAS_Segue} (albeit the already mentioned caveat regarding
reliability of those results applies also in this case), excluding
values below 10$^{\sq{24}}$--10$^{\sq{25}}$ s (depending on the
channel) for $\m\simeq$ 1--10 TeV. Contrary to the annihilation case,
the H.E.S.S. observations on the GC Halo are not competitive in the case
of decay, as the expected gamma-ray fluxes are very similar in the
signal and background regions. On the other hand, in \cite{D_Cirelli}
Cirelli et al. have shown that $\sim$15 hours of observations of the
Fornax cluster with H.E.S.S.  \cite{D_HESS_Fornax} could lead to the $\td$
lower limits of the order of $\sim$(10$^{\sq{24}}$--10$^{\sq{26}}$) s for
$\m$ between 1 and 20 TeV.

Figure \ref{Fig14} shows the results of this work, assuming the 2-body
decay of scalar DM particle into quark-antiquark, lepton-antilepton
and gauge boson pairs\footnote{For fermionic DM, decay channels such
  as $\chi\rightarrow l^{\pm}W^{\mp}$ and $Z\nu$ are possible, but
  these (in first approximation) can be analyzed as a combination of the
  corresponding 2-body scalar DM decay channels.}. Our strongest
limits correspond to $\m$ = 20 TeV, and are between
$\sim$5.9$\times$10$^{\sq{24}}$ s and
$\sim$2.9$\times$10$^{\sq{25}}$ s. Compared to the bounds from
\tit{Fermi}-LAT, for the lightest DM particles, the limits from this
work are three--four orders of magnitude weaker; on the other hand, for
more massive scenarios ($\m >$ 1 TeV), the MAGIC bounds are a factor
$\sim$30 to a factor $\sim$3 less constraining (for the muon and tau
lepton final states, respectively). In all of the considered scenarios
our results are more stringent than those of VERITAS (table
\ref{Table2}). For the leptonic channels, we also show the regions that
allow to fit the \tit{Fermi}-LAT, PAMELA and H.E.S.S. cosmic-ray
measurements \cite{D_Pamela_bestFit}, with 95\% and 99.99\% CL: our
exclusion curves are factors $\sim$30 and $\sim$2 away from
constraining those fits, for the $\mu^{+}\mu^{-}$ and $\tau^{+}\tau^{-}$
final states and masses $\m$ = 2.5 TeV and 5 TeV, respectively.

In general, observations of galaxy clusters are better suited than
dSph for constraining $\td$, due to the linear dependence of
$J_{\rt{dec}}$ with the density $\rho$ and the great amount of DM
present in this kind of objects. This is reflected in the fact that
the predicted limits for $\sim$15 hours observations of Fornax are of
the same order of the ones we obtain for $\sim$160 hours of Segue~1
data \cite{D_Cirelli}.

\subsubsection{Gamma-ray lines}
\label{D:Exp:Line}

The importance of the detection of gamma-ray lines from DM
annihilation or decay can not be overestimated: not only would a line
be a firm proof of DM existence, it would also reveal important
information about its nature. This is why this feature has been so
appealing, and many searches for a hint of it have been conducted so
far, in galaxy clusters \cite{D_Line_Cluster}, Milky Way dSph
satellites \cite{D_Line_dSph}, and in the GC and Halo
\cite{D_Fermi_Line_new, D_HESS_Line}. In addition, it is worth
mentioning that there is a recently claimed hint of a line-like signal
at $\sim$130 GeV in the \tit{Fermi}-LAT data of the GC region
\cite{DM_VIB, D_Line}: if the observed signal originates from direct
DM annihilation into two photons, the WIMP particle should have a mass
of $\m$ = 129$\pm$2.4$_{-{\sq{13}}}^{+\sq{7}}$ GeV and an annihilation
rate (assuming the Einasto profile) of $\sv _{\gamma\gamma}$ =
(1.27$\pm$ 0.32$_{-\sq{0.28}}^{+\sq{0.18}}$)$\times$10$^{-\sq{27}}
\svu$. Although this result could not be confirmed (nor disproved) by
the \tit{Fermi}-LAT Collaboration \cite{D_Fermi_Line_new}, the
potential presence of such a feature has stirred the scientific
community, and numerous explanations have appeared about its origin
(for a review, see \cite{D_Line_review}).

\paragraph{Annihilation}
The currently strongest upper limits on spectral lines from DM
annihilation are provided by the 3.7 years of observation of the
Galactic Halo by \tit{Fermi}-LAT \cite{D_Fermi_Line_new}, and 112
hours of the GC Halo region by H.E.S.S. \cite{D_HESS_Line}. The
\tit{Fermi}-LAT upper limits on $\sv$ extend from $\sim$10$^{-\sq{29}}
\svu$ at $\m$ = 10 GeV to $\sim$10$^{-\sq{27}} \svu$ at $\m$ = 300
GeV, while the H.E.S.S. bounds range between $\sim$10$^{-\sq{27}} \svu$ at
$\m$ = 500 GeV and $\sim$10$^{-\sq{26}} \svu$ at $m_\chi$ = 20 TeV. 
\begin{figure}[t]
  \centering
  \includegraphics[trim=10 20 30 30,clip=true,width=0.75\textwidth]{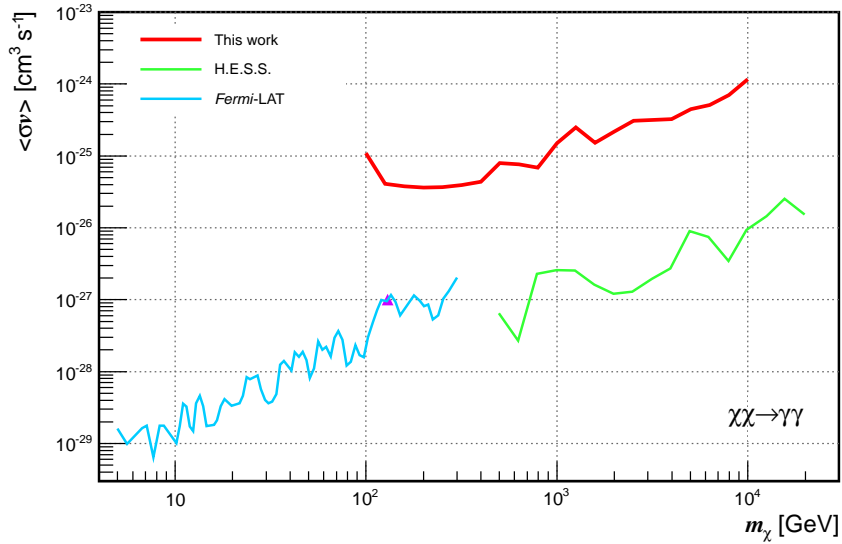}
  \caption{Upper limits on $\sv$ for direct DM annihilation into two
    photons, from the Segue~1 observations with MAGIC (red line), compared with the exclusion
    curves from the GC region observations from \tit{Fermi}-LAT (3.7
    years, blue line, \cite{D_Fermi_Line_new}) and H.E.S.S. (112 hours,
    green line, \cite{D_HESS_Line}). Also shown is the $\sv$ value
    corresponding to the 130 GeV gamma-ray line (violet triangle,
    \cite{D_Line}).}
\label{Fig15}
\end{figure}
\begin{figure}[t]
  \centering
  \includegraphics[trim=10 30 30 30,clip=true,width=0.55\textwidth]{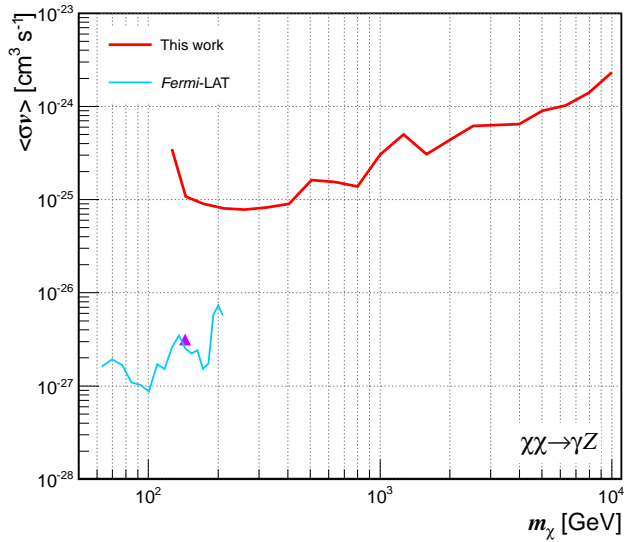}
  \caption{Upper limits on $\sv$ for DM annihilation into a photon and
    a $Z$ boson, from the Segue~1 observations with MAGIC (red line), compared with the
    exclusion curve from 2 years of the GC region observations with
    \tit{Fermi}-LAT (blue line, \cite{D_Fermi_Line_old}). Also show is
    the $\sv$ value corresponding to the 130 GeV gamma-ray line
    (violet triangle, \cite{D_Line}).}
\label{Fig16}
\end{figure}
\begin{figure}[t]
  \centering
  \includegraphics[trim=10 20 30 30,clip=true,width=0.75\textwidth]{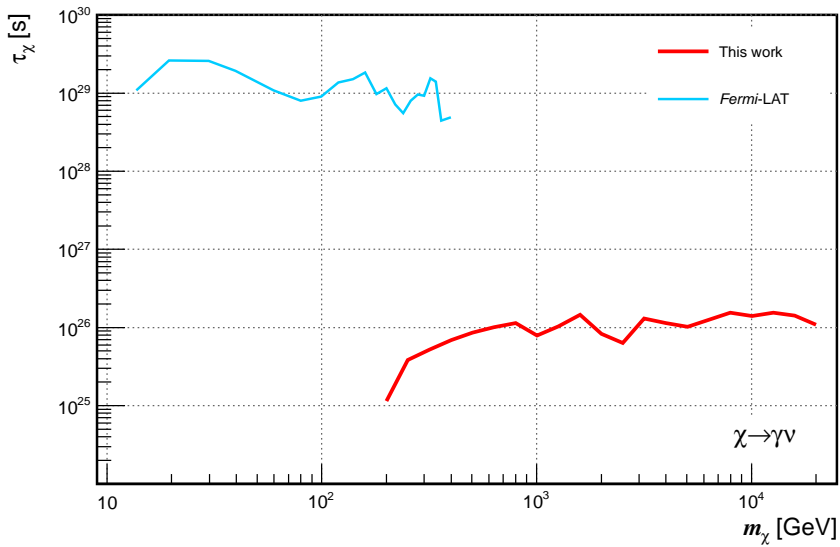}
  \caption{Lower limits on $\td$ for DM 
      decay into a neutrino and a photon, from the Segue~1 observations with MAGIC 
      (red line), compared with the exclusion curve 
      from 2 years of the GC region observations 
      with \tit{Fermi}-LAT (blue line, \cite{D_Fermi_Line_old}).}
\label{Fig17}
\end{figure}

Figure \ref{Fig15} shows our limits for the line search, assuming DM
annihilation into two photons, compared to the described bounds from
\tit{Fermi}-LAT and H.E.S.S. The strongest constraint from MAGIC is obtained
for $\m$ = 200 GeV, with $\sv \sim$ 3.6$\times$10$^{-26} \svu$, which
is about one order of magnitude higher than the \tit{Fermi}-LAT limit, and
a factor $\sim$30 above the sensitivity needed for testing the hint of
a line at 130 GeV. For higher $\m$ values, the H.E.S.S. limits are more
constraining than ours by a factor $\sim$50 (as expected). We note,
however, that similar considerations as those discussed in section
\ref{D:Exp:Sec} apply when comparing the results of line searches from
dSphs and Galactic Halo.


Results from line searches when DM particles annihilate into a photon
and a $Z$ boson are shown in figure \ref{Fig16}: the strongest bound
from this work corresponds to $\sv \sim$ 7.8$\times$10$^{-{\sq{26}}}
\svu$, for $\m\sim$ 250 GeV. Compared to the constraints from 2 years
of \tit{Fermi}-LAT observations of the GC region
\cite{D_Fermi_Line_old}, in the overlapping energy range, our limits
are one--two order(s) of magnitude weaker. Also shown is the $\sv$
estimate for the $\gamma Z$ explanation of the line-like feature at
130 GeV; the MAGIC upper limit is a factor $\sim$30 away from this value.

\paragraph{Decay}
We also use our search for spectral lines to constrain the properties
of decaying DM. If the DM particle is a gravitino in R-parity breaking
vacua, with $\td\sim$ 10$^{\sq{27}}$ s or larger, it can decay into a
photon and a neutralino, producing one monochromatic gamma-ray line at
$E_{\gamma}\simeq\m/$2. Searches for features of such origin have been
conducted by the \tit{Fermi}-LAT, in 2 years of observations of the GC
region \cite{D_Fermi_Line_old}, setting lower limits on $\td$ of
few$\times$10$^{\sq{29}}$ s up to $\m\sim$ 600 GeV, whereas, as
explained in section \ref{D:Exp:Sec}, H.E.S.S. observations on the GC
Halo are not sensitive for decaying DM searches.

Figure \ref{Fig17} shows our results compared to those from
\tit{Fermi}-LAT. Although the considered $\m$ range extends well
beyond the energies required for decay into $W$ or $Z$ bosons (that
would consequently fragment into photons with continuous spectrum),
here only the monochromatic emission is considered. MAGIC limits are
almost three orders of magnitude less constraining than those of
\tit{Fermi}-LAT, but cover a complementary range of masses. This is
expected, since (as discussed before), dSphs are suboptimal targets
for DM decay signals. Our strongest bound is of the order of $\td
\sim$ 1.5$\times$10$^{\sq{26}}$ s for $\m\sim$ 8 TeV. The case of the
decay of scalar DM into two photons is not considered here, as it is
trivial to derive the $\td$ lower limits for that scenario: the
gamma-ray signal would be the same as for the $\gamma\nu$ channel,
only twice as strong.

\subsection{Implications for models}
\label{D:Models}

Generating the correct WIMP relic density requires a DM annihilation
cross section at the time of the freeze-out of $\sv \simeq$
3$\times$10$^{-\sq{26}} \svu$. This value then constitutes a
model-independent upper limit on $\sv$ today, and sets the minimal
sensitivity required to observe a DM annihilation signal. The
constraints derived in this paper from deep observations of Segue~1
lie, in most cases, two orders of magnitude above the canonical
value. Nevertheless, for some channels --- notably in
$\chi\chi\rightarrow \tau^{+}\tau^{-}$ --- the very characteristic
photon spectrum allows us to derive more constraining bounds. In
particular, for $\m\sim$ 500 GeV the limit on $\sv$ for tau final
states lies just a factor of $\sim$40 away from the thermal cross
section. However, it should be borne in mind that a signal is expected
at $\sv \simeq$ 3$\times$10$^{-\sq{26}} \svu$ when the annihilation
cross section is $s$-wave dominated. Some well motivated DM scenarios
suggest the $p$-wave dominated annihilation (see, e.g.,
\cite{D_pwave}), and as such suppressed today by the velocity squared
of the DM particles. If this is the case, the expected $\sv$ can be
five--six orders of magnitude smaller than the canonical
value. Scenarios of this class include those where the DM particle is
a Majorana fermion that annihilates into a light fermion, for example
$\chi\chi\rightarrow\mu^+\mu^-$ or $\tau^+\tau^-$.

The expected rate of annihilations producing spectral features is also
typically smaller than the canonical value. The direct production of
two photons ($\chi\chi\rightarrow\gamma\gamma$) occurs at the one-loop
level, hence the expected rate for a thermally produced WIMP is
necessarily suppressed by a factor $\alpha^{\sq{2}}$, giving an
annihilitation cross section which is, in most scenarios,
$\lesssim{\cal O}$(10$^{-\sq{30}}$)$\svu$. Therefore, even if the
limits on $\sv$ for direct photon production are close to the thermal
value, a sensitivity increase of at least two--three orders of
magnitude is required in order to possibly observe a signal. For
annihilations with VIB contribution, the rate is suppressed compared
to the canonical value by the extra electromagnetic coupling and by
the three-body phase space, amounting to a factor of $\sim
\alpha/\pi$. Hence, observation of a gamma-ray signal from the
final states $\mu^+\mu^-\gamma$ or $\tau^+\tau^-\gamma$ requires a
sensitivity of $\sv \sim {\cal O}$(10$^{-\sq{28}}$)$\svu$, which is
three orders of magnitude below the limits obtained in this
paper. Lastly, the rate of annihilations producing gamma-ray boxes is
a priori unsuppressed, since $\sv$ for the process
$\chi\chi\rightarrow\phi\phi$ can be as large as the thermal value, and
the branching fraction $\phi\rightarrow\gamma\gamma$ can be sizable,
even 1. For this class of spectral features, the limits are then only a
factor of a few above the values for the cross section where a signal
might be expected.

It should be kept in mind, however, that these results are somewhat
conservative: no flux enhancements, due to possible boost factors,
have been considered. In general, the uncertainties entering the
expected fluxes are large enough so that potential surprises cannot be
excluded.
\section{Summary and conclusions}
\label{C}

We have reported the results on indirect DM searches obtained with the
MAGIC Telescopes using observations of the dSph galaxy Segue
1. The observations, carried out between January 2011 and February 2013,
resulted in 157.9 hours of good-quality data, thus making this the
deepest survey of any dSph by any IACT so far. In addition, this is
one of the longest observational campaigns ever, with MAGIC or any
other IACT, on a single, non-variable object. That imposes important
technical challenges on data analysis, for which suitable and
optimized solutions have been successfully designed and implemented.

The data have been analysed by means of the \tit{full likelihood method},
a dedicated analysis approach optimized for the recognition of
spectral features, like the ones expected from DM annihilation or
decay. Furthermore, with this method, the combination of data samples
obtained with different telescope configurations has been performed in
a straightforward manner. This has resulted in sensitivity
improvements by factors ranging between 1.7 and 2.6, depending on the
DM particle mass and the considered annihilation/decay channel.

No significant gamma-ray excess has been found in the Segue~1 sample.
Consequently, the observations have been used to set constraints on
the DM annihilation cross section and lifetime, assuming various final
state scenarios. In particular, we have computed limits for the
spectral shapes expected for secondary gamma-rays from annihilation
and decay into the SM pairs ($b\bar{b}$, $t\bar{t}$, $\mu^{+}\mu^{-}$,
$\tau^{+}\tau^{-}$, $W^{+}W^{-}$ and $ZZ$), for monochromatic
gamma-ray lines, for photons produced by the VIB processes and for the
spectral features from annihilation to gamma-rays via intermediate
scalars. The calculations have been done in a model-independent way,
by assuming a branching ratio of 100\% to each of the considered final
states. 95\% CL limits to $\sv$ and $\td$ have been obtained for
$\m$ in the 100 GeV $-$ 10 TeV range and 200 GeV $-$ 20 TeV, for
annihilation and decay scenarios, respectively. The most constraining
limits are obtained for DM annihilating or decaying purely into
$\tau^{+}\tau^{-}$ pairs: $\sv < 1.2\times 10^{-24} \svu$ for $\m
\simeq 500$ GeV and $\td > 3 \times 10^{25}$ s for $\m \simeq 2$
TeV. 

Studying different targets is of particular importance for indirect DM
searches. On one hand, a certain confirmation of the DM signal,
especially if it is a featureless one, can only come from observations
of at least two sources. On the other hand, diversity among
observational targets is necessary, as searches in different objects
are affected by different uncertainties. For instance, although most
aspects of the general cold DM halo structure are resolvable from
numerical approaches, the current knowledge and predictive power
regarding its behaviour are limited by the complex interplay between
the DM and baryonic components.  It is still a long way until the
effects baryons have on the DM distribution are fully perceived. This
is particularly relevant for targets like the GC and Halo, or galaxy
clusters, since their significant luminous content can influence the
evolution of the DM component. Furthermore, there are also
uncertainties coming from the presence of substructures in the halo,
and the possible enhancements of the cross-section due to the quantum
effects, that directly influence the value of the total expected
flux. These uncertainties are large ($\mathcal{O}$(10) or more) and
their impact on halos may be different on different scales. Thus,
diversification of the observational targets is the optimal strategy
for the discovery.
 
Altogether, the results from this work represent an important step
forward in the field of DM searches, significantly improving our
previous limits from dSph galaxies and complementing the existing
bounds from other targets.

\paragraph{Acknowledgments} 
\vspace{0.5cm}

We would like to thank the Instituto de Astrof\'{\i}sica de Canarias
for the excellent working conditions at the Observatorio del Roque de
los Muchachos in La Palma.  The support of the German BMBF and MPG,
the Italian INFN, the Swiss National Fund SNF, and the Spanish MICINN
is gratefully acknowledged. This work was also supported by the CPAN
CSD2007-00042 and MultiDark CSD2009-00064 projects of the Spanish
Consolider-Ingenio 2010 programme, by grant 127740 of the Academy of
Finland, by the DFG Cluster of Excellence ``Origin and Structure of
the Universe'', by the Croatian Science Foundation Project 09/176, by
the DFG Collaborative Research Centers SFB823/C4 and SFB876/C3, and by
the Polish MNiSzW grant 745/N-HESS-MAGIC/2010/0.

\appendix
\section{Flux upper limits}
\label{A}

As the Segue~1 observations with MAGIC have not resulted in a
detection, differential and integral upper limits are calculated for
the gamma-ray emission from the source, assuming a power law-shaped
spectra of slope $\Gamma$, and by relying on the conventional analysis
approach. The procedure and nomenclature used are those prescribed in
\cite{I_MAGIC_Segue}.

\begin{figure}[h!]
  \centering
  \vspace{-5pt}
  \includegraphics[trim=10 20 10 20,clip=true,width=0.75\textwidth]{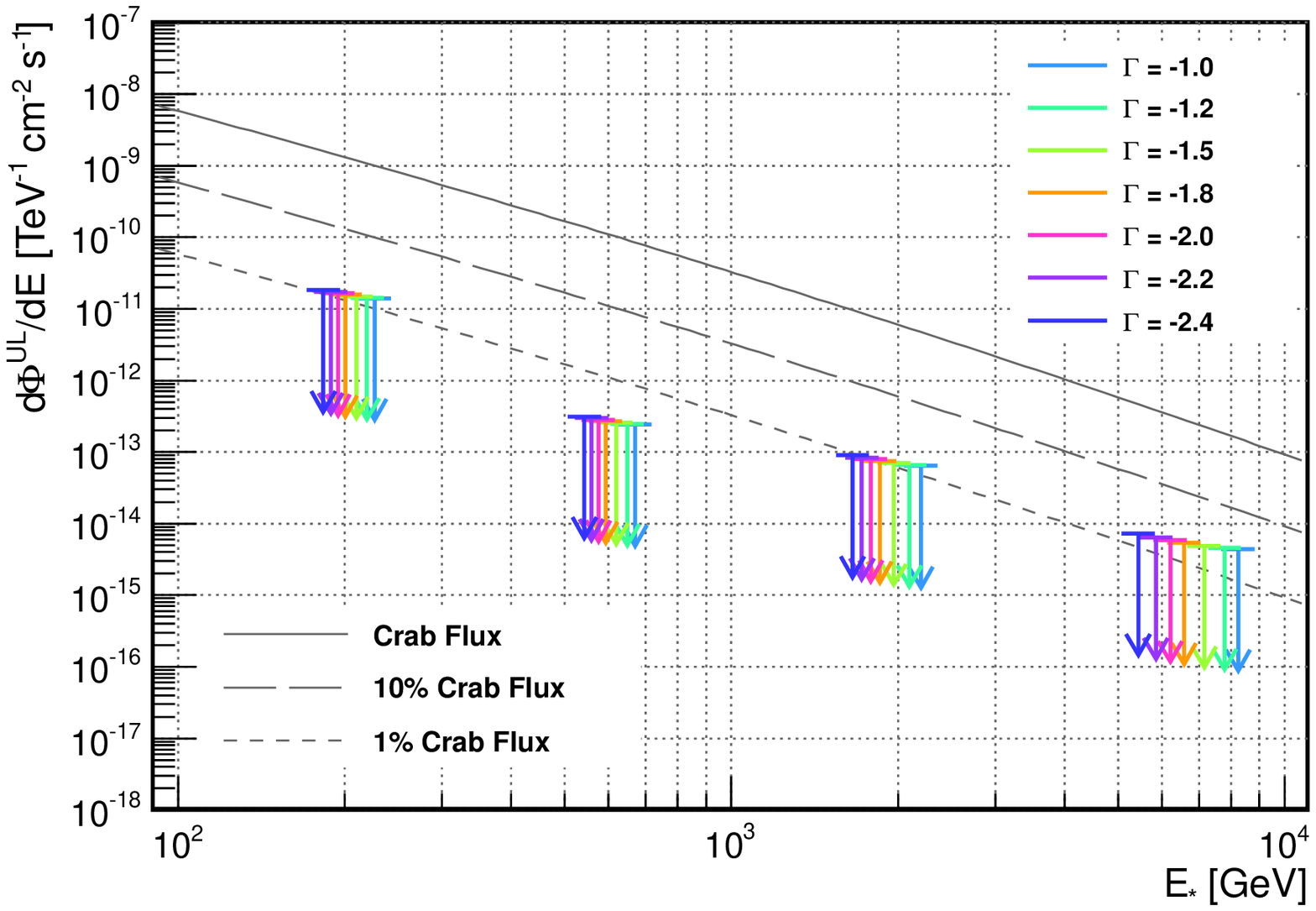}
  \vspace{0pt}
  \caption{Differential flux upper limits from 157.9 hours of the Segue~1
    observations with MAGIC, assuming a point-like source and a power law-shaped signal
    emission and different spectral slopes $\Gamma$. As a reference,
    the Crab Nebula differential flux (solid line,
    \cite{A_performance}) and its 10\% and 1\% fractions (long-dashed
    and dashed lines, respectively), are also drawn.}
  \label{A:Fig1}
  \vspace{1cm}
  \includegraphics[trim=10 20 10 20,clip=true,width=0.75\textwidth]{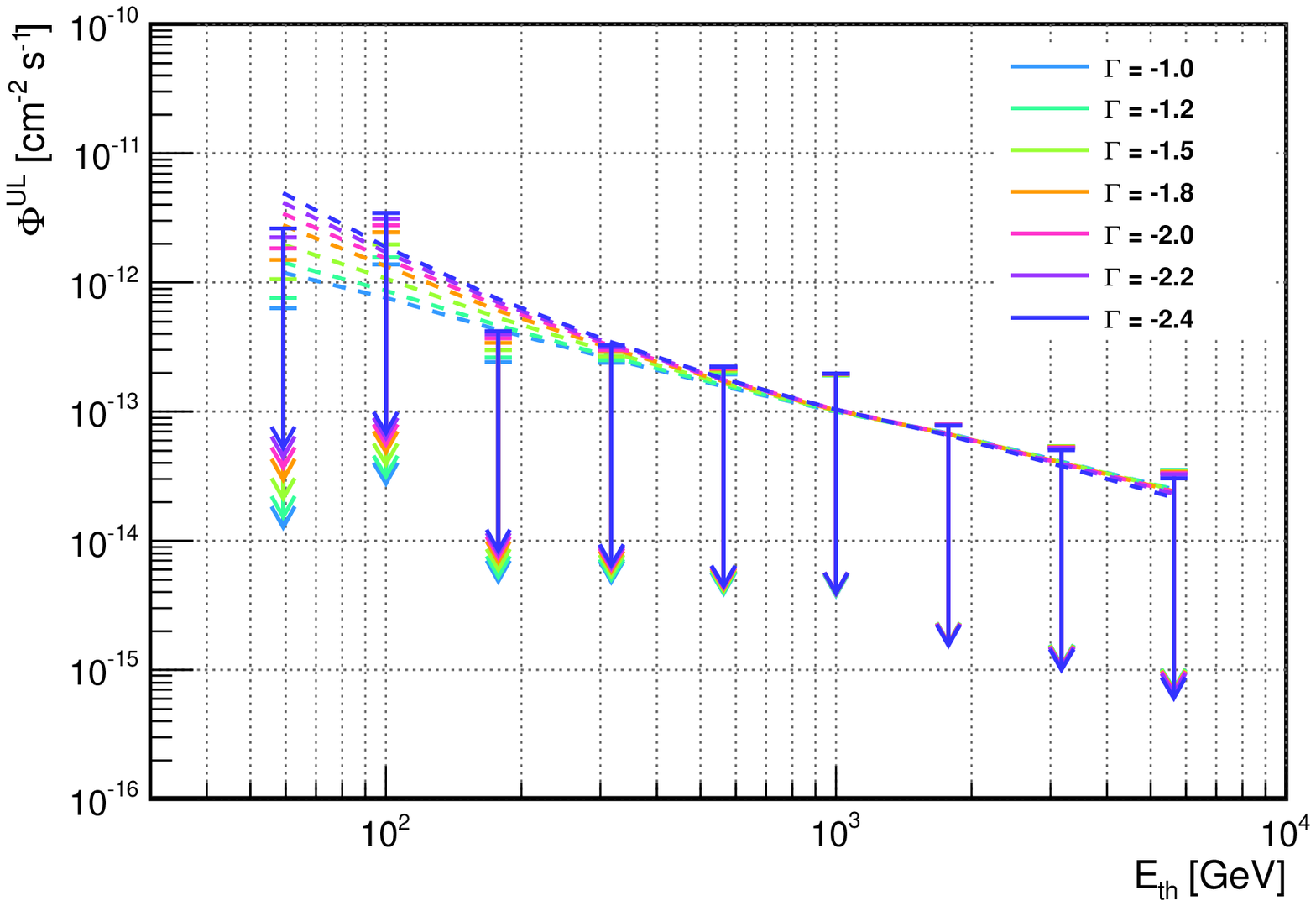}
  \vspace{0pt}
  \caption{Integral flux upper limits from 157.9 hours of the Segue~1
    observations with MAGIC, assuming a point-like source and a power law-shaped signal
    emission and different spectral slopes $\Gamma$. Dashed lines
    indicate the expectations for the null hypothesis case.}
  \label{A:Fig2}
\end{figure}

The differential flux upper limits are estimated in energy bins
$\Delta E$ as:
\begin{equation}
  \frac{d\Phi\UL}{dE}(E'_{\star}) = \frac{N_{\sq{ex}}\UL(\Delta
    E)}{\teff}\frac{1}{\int_{\sq{0}}^{\infty}~\Aeff(E'; \Delta E)
    S(E') dE'}, 
  \label{A:eq1}
\end{equation}
where $S(E') = (E'/E'_{\star})^{\Gamma}$ is the chosen power law
spectrum. $E'_{\star}$ is the pivot energy for the particular energy
bin, defined as:
\begin{equation}
  E'_{\star} = \frac{\int_{\sq{0}}^{\infty}~E' S(E') \Aeff(E'; \Delta
    E) dE'}{\int_{\sq{0}}^{\infty}~S(E') \Aeff(E'; \Delta E) dE'}.
  \label{A:eq2}
\end{equation}
$\Aeff (E';\Delta E)$ is the effective area for events with measured
energy within $\Delta E$, as a function of true energy $E'$;
$N_{\sq{ex}}\UL$ is computed using global $\Non$, $\Noff$ and $\tau$
values and the conventional method \cite{O_Rolke}, for a 95\% CL and
assuming a systematic uncertainty on the overall detection efficiency
of 30\%. $\Aeff(E'; \Delta E)$ is computed for the entire sample as
the weighted average of the effective areas of the four considered
data sets, with weights being the corresponding observation times,
i.e.:
\begin{equation}
  \Aeff(E'; \Delta E)  = \frac{\sum_{i}~\big({\Aeff}_{i}~{\teff}_{i}\big)}{\teff},
  \label{A:eq3}
\end{equation}
where the index $i$ runs over the different samples. Table
\ref{A:Table1} summarizes the upper limits in four (estimated) energy
logarithmic bins, between 100 GeV and 10 TeV, for different assumed
slopes of the power law-shaped signal emission ($\Gamma = -$1.0,
$-$1.2, $-$1.5, $-$1.8, $-$2.0, $-$2.2 and $-$2.4). The results are
also shown in figure \ref{A:Fig1}.

The integral flux upper limits are calculated above a certain energy
threshold $\Eth$ as:
\begin{equation}
  \Phi\UL(>\Eth) =
  \frac{N_{\sq{ex}}\UL(E>\Eth)}{\teff}
  \frac{\int_{\Eth}^{\infty} S(E') dE'}
  {\int_{\sq{0}}^{\infty}~\Aeff(E'; \Delta E) S(E') dE'}, 
  \label{A:eq4}
\end{equation}
The values of $N_{\sq{ex}}\UL (E>E_{\sq{th}})$ and $\Aeff(E';
E>\Eth)$ are obtained in a similar manner as described for the
differential upper limits. Again, several power law spectra are
assumed, with different slopes $\Gamma$. The results are summarized in
table \ref{A:Table2} and also shown in figure \ref{A:Fig2}, where they
are also compared with the values expected for the null hypothesis
case, i.e. $N_{\sq{ex}} (E>\Eth)$ = 0.
\begin{sidewaystable}
  \begin{center}
    \setlength{\extrarowheight}{1pt}
    {\small
      \begin{tabulary}{1.\textwidth}{ C C C C C C C C C C C C C }
        \toprule[0.1em] & & & & &
        \multicolumn{7}{c}{$d\Phi^{\sq{UL}}/dE$ [TeV$^{-\sq{1}}$
          cm$^{\sq{-2}}$
          s$^{\sq{-1}}$]\vspace{0.2cm}}\\\cmidrule{6-12} $\Delta E$
        [GeV] & $\Non$ & $\Noff$ & $\tau$ & $N_{\rt{ex}}^{\sq{UL}}$ &
        $\Gamma = -$1.0 & $\Gamma = -$1.2 & $\Gamma = -$1.5 & $\Gamma
        = -$1.8 & $\Gamma = -$2.0 & $\Gamma = -$2.2 & $\Gamma =
        -$2.4\\\midrule[0.1em] 
        100$-$320 & 10996 & 10871 & 1.004 & 652.4 &
        1.41$\times$10$^{-\sq{11}}$ & 1.44$\times$10$^{-\sq{11}}$ &
        1.49$\times$10$^{-\sq{11}}$ & 1.57$\times$10$^{-\sq{11}}$ &
        1.64$\times$10$^{-\sq{11}}$ & 1.72$\times$10$^{-\sq{11}}$ &
        1.82$\times$10$^{-\sq{11}}$ \\
        320$-$1000 & 1606 & 1608 & 0.976 & 80.6 & 2.43$\times$10$^{-\sq{13}}$
        & 2.48$\times$10$^{-\sq{13}}$ & 2.58$\times$10$^{-\sq{13}}$ &
        2.71$\times$10$^{-\sq{13}}$ & 2.83$\times$10$^{-\sq{13}}$ &
        2.96$\times$10$^{-\sq{13}}$ & 3.12$\times$10$^{-\sq{13}}$\\
        1000$-$3200 & 341 & 314 & 1.001 & 109.2 & 6.40$\times$10$^{-\sq{14}}$ &
        6.60$\times$10$^{-\sq{14}}$ & 6.99$\times$10$^{-\sq{14}}$ &
        7.48$\times$10$^{-\sq{14}}$ & 7.89$\times$10$^{-\sq{14}}$ &
        8.38$\times$10$^{-\sq{14}}$ & 8.97$\times$10$^{-\sq{14}}$ \\
        3200$-$10000 & 63 & 59 &  1.034 & 36.6 & 4.42$\times$10$^{-\sq{15}}$ &
        4.57$\times$10$^{-\sq{15}}$ & 4.92$\times$10$^{-\sq{15}}$ &
        5.42$\times$10$^{-\sq{15}}$ & 5.86$\times$10$^{-\sq{15}}$ &
        6.44$\times$10$^{-\sq{15}}$ & 7.31$\times$10$^{-\sq{15}}$ \\
        \bottomrule[0.1em]
      \end{tabulary}
    }
    \vspace{-10pt}
  \end{center}
  \caption{Differential flux upper limits 
    from 157.9 hours of Segue~1 observations 
    with MAGIC, in four energy bins and for 
    several power law-shaped spectra.}
  \label{A:Table1}
  \vspace{1cm}
  \begin{center}
    \setlength{\extrarowheight}{1pt}
    {\small
      \begin{tabulary}{1.\textwidth}{ C C C C C C C C C C C C C }
        \toprule[0.1em] & & & & & \multicolumn{7}{c}{$\Phi^{\sq{UL}}$
          [cm$^{\sq{-2}}$
          s$^{\sq{-1}}$]\vspace{0.2cm}}\\\cmidrule{6-12} $E_{\sq{th}}$
        [GeV] & $\Non$ & $\Noff$ & $\tau$ & $N_{\rt{ex}}^{\sq{UL}}$ &
        $\Gamma = -$1.0 & $\Gamma = -$1.2 & $\Gamma = -$1.5 & $\Gamma
        = -$1.8 & $\Gamma = -$2.0 & $\Gamma = -$2.2 & $\Gamma =
        -$2.4\\\midrule[0.1em] 59.2 & 25804 & 26018 & 1.001 & 297.2 &
        6.37$\times$10$^{-\sq{13}}$ & 7.66$\times$10$^{-\sq{13}}$ &
        1.06$\times$10$^{-\sq{12}}$ & 1.49$\times$10$^{-\sq{12}}$ &
        1.84$\times$10$^{-\sq{12}}$ & 2.23$\times$10$^{-\sq{12}}$ &
        2.63$\times$10$^{-\sq{12}}$ \\
        100.0 & 13006 & 12832 & 1.001 & 708.0 & 1.39$\times$10$^{-\sq{12}}$
        & 1.57$\times$10$^{-\sq{12}}$ & 1.96$\times$10$^{-\sq{12}}$ &
        2.44$\times$10$^{-\sq{12}}$ & 2.78$\times$10$^{-\sq{12}}$ &
        3.12$\times$10$^{-\sq{12}}$ & 3.44$\times$10$^{-\sq{12}}$\\
        177.8 & 4869 & 4913 & 0.994 & 135.8 & 2.44$\times$10$^{-\sq{13}}$ &
        2.63$\times$10$^{-\sq{13}}$ & 3.01$\times$10$^{-\sq{13}}$ &
        3.43$\times$10$^{-\sq{13}}$ & 3.71$\times$10$^{-\sq{13}}$ &
        3.98$\times$10$^{-\sq{13}}$ & 4.22$\times$10$^{-\sq{13}}$ \\
        317.2 & 2034 & 2007 & 0.983 & 144.9 & 2.41$\times$10$^{-\sq{13}}$ &
        2.51$\times$10$^{-\sq{13}}$ & 2.70$\times$10$^{-\sq{13}}$ &
        2.90$\times$10$^{-\sq{13}}$ & 3.03$\times$10$^{-\sq{13}}$ &
        3.14$\times$10$^{-\sq{13}}$ & 3.25$\times$10$^{-\sq{13}}$ \\
        562.3 & 878 & 855 & 0.990 & 125.2 & 1.96$\times$10$^{-\sq{13}}$   &
        1.99$\times$10$^{-\sq{13}}$ & 2.06$\times$10$^{-\sq{13}}$ &
        2.13$\times$10$^{-\sq{13}}$ & 2.17$\times$10$^{-\sq{13}}$ &
        2.21$\times$10$^{-\sq{13}}$ & 2.24$\times$10$^{-\sq{13}}$ \\
        1000.0 & 404 & 373 & 1.009 & 126.5 & 1.90$\times$10$^{-\sq{13}}$ &
        1.91$\times$10$^{-\sq{13}}$ & 1.93$\times$10$^{-\sq{13}}$ &
        1.96$\times$10$^{-\sq{13}}$ & 1.97$\times$10$^{-\sq{13}}$ &
        1.97$\times$10$^{-\sq{13}}$ & 1.97$\times$10$^{-\sq{13}}$ \\
        1778.3 & 175 & 173 & 1.021  & 54.2 & 8.02$\times$10$^{-\sq{14}}$ &
        8.02$\times$10$^{-\sq{14}}$ & 8.01$\times$10$^{-\sq{14}}$ &
        7.99$\times$10$^{-\sq{14}}$ & 7.96$\times$10$^{-\sq{14}}$ &
        7.90$\times$10$^{-\sq{14}}$ & 7.78$\times$10$^{-\sq{14}}$ \\
        3172.3 & 64 & 60 & 1.026 & 36.2 & 5.44$\times$10$^{-\sq{14}}$ &
        5.42$\times$10$^{-\sq{14}}$ & 5.39$\times$10$^{-\sq{14}}$ &
        5.34$\times$10$^{-\sq{14}}$ & 5.28$\times$10$^{-\sq{14}}$ &
        5.19$\times$10$^{-\sq{14}}$ & 5.04$\times$10$^{-\sq{14}}$\\
        5623.4 & 20 & 19 & 1.225 & 22.0 & 3.54$\times$10$^{-\sq{14}}$ &
        3.52$\times$10$^{-\sq{14}}$ & 3.48$\times$10$^{-\sq{14}}$ &
        3.42$\times$10$^{-\sq{14}}$ & 3.36$\times$10$^{-\sq{14}}$ &
        3.25$\times$10$^{-\sq{14}}$ & 3.03$\times$10$^{-\sq{14}}$ \\
        \bottomrule[0.1em]
      \end{tabulary}
    } 
    \vspace{-10pt}
  \end{center}
  \caption{Integral flux upper limits from 
    157.9 hours of Segue~1 observations with 
    MAGIC, for different energy thresholds and 
    several power law-shaped spectra.}
    \label{A:Table2}
    \vspace{-5pt}
  \end{sidewaystable}

\end{document}